\documentclass[a4paper,11pt]{article}
\pdfoutput=1

\usepackage{jheppub}
\usepackage[T1]{fontenc} 
\usepackage{dsfont}
\usepackage{mathrsfs}

\usepackage{comment}
\usepackage{stackengine,xcolor}
\usepackage{todonotes}
\usepackage{subcaption}
\usepackage{tabularx}
\usepackage{mathbbol}
\usepackage{booktabs}

\graphicspath{{\main/Pictures/}{Pictures/}}

\DeclareMathOperator{\arcsinh}{arcsinh}

\newcommand{\ket}[1]{\left| #1 \right>} 
\newcommand{\bra}[1]{\left< #1 \right|} 
\newcommand{\e}{\mathrm{e}}

\newcommand{\ud}{{\mathrm{d}}}
\newcommand{\be}{\begin{equation}}
\newcommand{\ee}{\end{equation}}

\newcommand{\tidal}[1]{{\textcolor{blue}{#1}}}
\title{Gravitational Bound Waveforms from Amplitudes}
\date{\today}

\author[1]{Tim Adamo,}
\author[2]{Riccardo Gonzo}
\author[2]{\& Anton Ilderton}  

\affiliation[1]{School of Mathematics and Maxwell Institute for Mathematical Sciences \\ University of Edinburgh, EH9 3FD, UK}
\affiliation[2]{Higgs Centre for Theoretical Physics, School of Physics and Astronomy \\ University of Edinburgh, EH9 3FD, UK}

\emailAdd{t.adamo@ed.ac.uk}\emailAdd{rgonzo@ed.ac.uk}\emailAdd{anton.ilderton@ed.ac.uk}

\abstract{With the aim of computing bound waveforms from scattering amplitudes, we explore gravitational two-body dynamics using the Schwinger-Dyson equations and Bethe-Salpeter recursion. We show that the tree-level scattering waveform admits a natural analytic continuation, in rapidity, to the bound waveform, which we confirm from an independent calculation, in the Post-Newtonian expansion, of the time-domain multipoles at large eccentricity. We demonstrate consistency of this scattering-to-bound map with the Damour-Deruelle prescription for orbital elements in the quasi-Keplerian parametrization (which enters into the evaluation of the multipoles) and with the analytic continuation, in the binding energy, of radiated energy and angular momentum at 3PM.}

\begin{document}

\maketitle

\section{Introduction}
\label{sec:intro}

In recent years the LIGO-Virgo-KAGRA collaboration has successfully detected numerous gravitational wave signals, primarily attributed to the binary mergers of compact objects. This remarkable achievement was possible thanks to a set of template banks used for the detection, which were produced using both analytic and numerical methods in general relativity. Numerical relativity techniques -- while generically more accurate -- are computationally expensive, especially beyond a few orbits. On the other hand, traditional analytical tools like the effective-one-body (EOB)~\cite{Buonanno:1998gg,Buonanno:2000ef,Damour:2008yg} formalism or the gravitational self-force (GSF)~\cite{Barack:2009ux,Poisson:2011nh,Barack:2018yvs,Pound:2021qin} approach not only allow waveforms for a wider range of parameter values, but also provide a connection with the underlying two-body dynamics.

For the inspiral part of the binary evolution, the waveform is determined by the long-distance gravitational interaction between two massive compact bodies can be studied using effective field theory tools for point particles~\cite{Goldberger:2004jt,Foffa:2013qca,Porto:2016pyg}. Recently developed amplitude techniques~\cite{Bern:2019crd,Buonanno:2022pgc,Kosower:2022yvp,DiVecchia:2023frv} and worldline methods~\cite{Mogull:2020sak,Kalin:2020mvi,Shi:2021qsb} provide a new perspective on this problem, at least for scattering orbits, bringing efficient computational methods from quantum field theory to gravitational wave physics. On one hand, this enables extraction of the scattering angle (and therefore the Hamiltonian) up to very high orders in the Post-Minkowskian (PM) expansion \cite{Cheung:2018wkq,DiVecchia:2021bdo,DiVecchia:2021ndb,Bjerrum-Bohr:2021din,Bern:2019nnu,Bern:2021dqo,Bern:2021yeh,Dlapa:2021vgp,Cho:2023kux,Kalin:2020fhe,Dlapa:2022lmu,Jakobsen:2023ndj,Jakobsen:2023hig,Damgaard:2023ttc,Jakobsen:2022zsx,FebresCordero:2022jts} . On the other hand, observables like the waveform~\cite{Jakobsen:2021smu,Jakobsen:2021lvp,Cristofoli:2021jas,Bautista:2021inx,Mougiakakos:2022sic,Riva:2022fru,Jakobsen:2022psy,Brandhuber:2023hhy,Elkhidir:2023dco,Herderschee:2023fxh,Georgoudis:2023lgf,DeAngelis:2023lvf,Brandhuber:2023hhl,Aoude:2023dui,Georgoudis:2023ozp,Bohnenblust:2023qmy,Bhattacharyya:2024aeq} or radiated energy and angular momentum~\cite{Parra-Martinez:2020dzs,Herrmann:2021lqe,Herrmann:2021tct,Riva:2021vnj,Alessio:2022kwv,Manohar:2022dea,Dlapa:2022lmu,Jakobsen:2023hig,Damgaard:2023ttc} can be computed from scattering amplitudes using the KMOC formalism~\cite{Kosower:2018adc,Maybee:2019jus,Cristofoli:2021vyo}, or with generalizations of the in-in formalism for worldline methods~\cite{Jakobsen:2022psy,Kalin:2022hph,Damgaard:2023vnx}. These results have also been partially verified, in the overlapping region of validity, with traditional Post-Newtonian (PN) methods applied to the case of eccentric hyperbolic orbits \cite{Peters:1963ux,Peters:1964zz,Bini:2019nra,Bini:2020nsb,Bini:2020wpo,Bini:2021gat,Bini:2020hmy,Bini:2020rzn,Bini:2022enm,Bini:2023mdz,Bini:2023fiz}.

The major drawback of the Post-Minkowskian calculations is that they are intrinsically defined for the scattering scenario, while the physical situation corresponds to (elliptic-type) bound motion. This becomes particularly relevant when trying to import results such as the Hamiltonian and fluxes into a framework like EOB \cite{Antonelli:2019ytb,Khalil:2022ylj} or GSF, which also incorporates (beyond the inspiral) the merging and ringdown phases that are necessary for the generation of complete waveform templates. Luckily, the gravitational dynamics of classical particles is entirely described by a set of universal (differential) equations of motion, where only the boundary conditions distinguish between scattering and bound configurations. Such universality means that, at the level of observables, contributions which depend locally on the trajectory for scattering orbits should have a direct relation with the corresponding ones for bound orbits. 

A relation between scattering and bound dynamics was first observed by Damour and DeRuelle in 1985~\cite{Damour:1985}, using the orbital elements of the quasi-Keplerian parametrization in the PN approach. This parametrization describes the relative motion of two spinless bodies in the center of mass frame, which generalizes the Kepler motion in the PN expansion, and admits a (gauge-dependent) analytic continuation in terms of the eccentricity of the orbit. A new, gauge-invariant map -- deemed the ``boundary-to-bound'' (B2B) dictionary -- was recently proposed by K\"alin and Porto~\cite{Kalin:2019rwq,Kalin:2019inp,Cho:2021arx} for the planar dynamics, including aligned-spin configurations. This directly links scattering observables (e.g., scattering angle) and bound ones (e.g., periastron advance) through the energy and angular momentum. In the probe limit, but at all orders in the perturbative expansion, this has been further extended to include precession for generic Kerr orbits in~\cite{Gonzo:2023goe}. Moreover, an extension has been proposed to study the boundary-to-bound map for radiative observables like the total radiated energy and angular momentum~\cite{Saketh:2021sri,Cho:2021arx}.

Recently, a new approach has been proposed to study classical bound states via the Bethe-Salpeter recursion~\cite{Adamo:2022ooq} (see also \cite{Khalaf:2023ozy}), where the amplitude-action relation~\cite{Bern:2021dqo,Kol:2021jjc,Damgaard:2021ipf} was expressed in terms of the conservative two-massive particle irreducible (2MPI) kernel to perform the analytic continuation. Moreover, the scattering wavefunction was obtained from the amplitude, and its structure in the complex energy plane was studied to make contact with the classical binding energy, showing in particular the need to choose a single branch cut prescription for the analytic continuation in the energy variable~\cite{Adamo:2022ooq}. 

\medskip

In this paper, we build on the these recent developments to establish the boundary-to-bound dictionary for the gravitational (tree-level) waveform and the corresponding 3PM fluxes. In Section~\ref{sec:bound_wavefunction}, we consider the explicit toy example of scattering and bound state scalar wavefunctions on the linearized Schwarzschild background; these are prototypes of a `scalar waveform.' In the partial wave basis, we confirm that the single branch cut prescription for analytic continuation relates the scattering and bound wavefunctions, where one is required to take the residue of the bound state pole after analytic continuation to obtain the properly normalized bound wavefunction. This prompts us to revisit the superclassical iterations arising in the 2-body problem in Section~\ref{sec:theory-intro}. We begin with the Schwinger-Dyson equations in an effective one-body model before moving on to the generic two-body case, defining matrix elements for scattering and bound states by taking the residue on the corresponding poles. Scattering amplitudes provide a good computational approach here, allowing for full resummation in impact parameter space, in terms of two-massive-particle-irreducible conservative and radiative kernels. This bypasses the need to consider superclassical iterations and the corresponding residues, enabling the analytic continuation to be carried out directly at the level of the classical radiative kernel.

Consequently, in Section~\ref{sec:waveform_B2B} we study the tree-level scattering waveform in the time domain, derived from the radiative kernel, and conjecture an analytic continuation to its bound counterpart. This analytic continuation is entirely in terms of the binding energy (or, equivalently, the rapidity). We test this conjecture against the direct calculation of PN multipoles using the quasi-Keplerian parametrization. We also consider the total radiated energy and angular momentum at 3PM order in Section~\ref{sec:KMOC-bound}, showing how the corresponding scattering-to-bound map can be derived by studying the integration over the retarded time of the fluxes, emphasizing the new analytic continuation in terms of the binding energy. We conclude in Section~\ref{sec:conclusion}, summarizing the current status of relations between scattering and bound observables.

\subsection{Setup, notation and conventions}
We study the gravitational two-body dynamics of two minimally coupled massive scalar fields $\phi_1, \phi_2$, with masses $m_1$ and $m_2$, as described by the action
\begin{equation}
\label{eq:the-action}
   S = -\frac{1}{16 \pi G_N} \int \mathrm{d}^4 x \sqrt{-g}\, R + S_{\mathrm{GF}} +\sum_{j=1,2} \frac{1}{2} \int \mathrm{d}^4 x \sqrt{-g} \left( g^{\mu \nu} \partial_\mu \phi_j \partial_\nu \phi_j- m_j^2 \phi_j^2\right) \;,
\end{equation}
in which the first term is the Einstein-Hilbert action, with $G_N$ the Newton constant, while $S_{\mathrm{GF}}$ is a gauge fixing term. Defining $\kappa := \sqrt{32 \pi G_N}$, we expand the metric $g_{\mu \nu}$ in terms of the linearized graviton field $h_{\mu \nu}$ as $g_{\mu \nu} = \eta_{\mu \nu} + \kappa h_{\mu \nu}$,  which allows us to study perturbative graviton-scalar scattering on Minkowski spacetime.

For $(4+N)$-point amplitudes $\mathcal{M}_{4 + N}(p_1,p_2;p_1^{\prime},p_2^{\prime},k_1^{\prime},\dots,k_N^{\prime})$, we will label the momenta of the incoming (resp. outgoing) massive legs with $p_1^{\mu}$, $p_2^{\mu} = P^{\mu} - p_1^{\mu}$ (resp. $p_1^{\prime \, \mu}$, $p_2^{\prime \, \mu} = P^{\prime \, \mu} - p_1^{\prime \, \mu}$), while the $N$ outgoing gravitons will have massless momenta $k_1^{\prime \, \mu}, \dots, k_N^{\prime \, \mu}$. Therefore, $P^2 = E$ is identified with the incoming center of mass energy and the momentum collectively radiated into gravitons is $\sum_{j=1}^N k_{j}^{\prime \, \mu} = P^{\mu}-P^{\prime \, \mu}$. We further define the momentum transfers  $q_j^{\mu} = \hbar \, \bar{q}_j^{\mu}$ with $j=1,2$ and the classical 4-velocities $v_A^{\mu} = p_A^{\mu} /m_A$, $v_B^{\mu} = p_B^{\mu} /m_B$ from
\begin{align}
\label{eq:HEFT_parametrization}
\hspace{-8pt}p_1^{\mu} = m_A v_A^{\mu} + \hbar \frac{\bar{q}_1^{\mu}}{2}\,, 
    \,\,\,\,
    p_1^{\prime \, \mu} = m_A v_A^{\mu} - \hbar \frac{\bar{q}_1^{\mu}}{2}\,, \,\,\,\,
    p_2^{\mu} = m_B v_B^{\mu} - \hbar \frac{\bar{q}_2^{\mu}}{2}\,, \,\,\,\,
    p_2^{\prime \, \mu} = m_B v_B^{\mu} + \hbar \frac{\bar{q}_2^{\mu}}{2}\,.
\end{align}
As a consequence, the masses of classical particles are defined as $m_A^2 := p_A^2 = m_1^2 - q_1^2/4$ and $m_B^2 := p_B^2 = m_2^2 - q_2^2/4$ with $m_A \sim m_1$, $m_B \sim m_2$ in the $\hbar \to 0$ limit. 

Finally, we use the convention that $A_{(\mu} B_{\nu)} = A_{\mu} B_{\nu} -  A_{\nu} B_{\mu}$ and adopt the shorthand notation $\hat{\delta}(\cdot) := (2 \pi) \delta(\cdot)$ and $\hat{\mathrm{d}}^4 q := \mathrm{d}^4 q / (2 \pi)^4$ for delta functions and integral measures, respectively. We also work in `mostly minus' metric conventions.

\section{Probe wavefunctions from scattering to bound}
\label{sec:bound_wavefunction}

At leading order in the PM expansion, the two-body problem can be mapped, in the regime $m_A/m_B \ll 1$ relevant to large mass ratio inspirals, to the problem of a scalar probe, mass $m_A$, moving in the linearized Schwarzschild metric sourced by the mass $m_B$ (see for example \cite{Kabat:1992tb,Damour:2016gwp,Adamo:2022ooq}). We begin by working in this background-field approximation as a warm up, proving explicitly the analytic continuation between the (one-body) scattering and bound state wavefunctions of the probe.

\subsection{Warm up: linearized Schwarzschild}
\label{sec:lin-schwarz}
In the background field approximation, and neglecting radiation, we drop the field $\phi_2$ which generates the background from (\ref{eq:the-action}) and we fix $g_{\mu\nu}\to {\bar g}_{\mu\nu}$ to be the linearized Schwarzschild metric generated by a static source of mass $m_B$,
\begin{align}
   {\bar g}_{\mu\nu}= \eta_{\mu \nu} - \frac{4 G_N m_B}{r} \bigg(v_{B \mu} v_{B \nu} - \frac{1}{2} \eta_{\mu \nu}\bigg) \,, \qquad v_B^\mu = (1,{\bf 0}) \;.
    \label{eq:linearized_Schw}
\end{align}
A scalar field $\phi_1$ of mass $m_A$ propagating on this spacetime obeys the Klein-Gordon equation
\begin{align}
\bigg(
\frac{1}{\sqrt{-{{\bar g}}}} \partial_\mu \sqrt{-{\bar g}}\, {\bar g}^{\mu\nu}\partial_\nu + \frac{m_A^2}{\hbar^2} \bigg) \Psi(x)= 0\,,
\label{eq:deq_EOBwavefunctions}
\end{align}
the solution of which has been discussed many times (e.g.,~\cite{Kabat:1992tb,Messiah,Gottfried2003,Kol:2021jjc,Adamo:2022ooq}), so we will be brief. First, (\ref{eq:deq_EOBwavefunctions}) must be supplemented with appropriate boundary conditions. For both scattering and bound solutions, we impose regularity at the origin\footnote{In a fully non-linear Schwarzschild background, one imposes boundary conditions (e.g., regularity, no outgoing wave, etc.) at the event horizon, rather than the origin. Of course, the linearized Schwarzschild metric relevant for the leading PM expansion we consider here has no event horizon.}; this is natural for making contact with perturbative scattering amplitudes and for bound solutions. For all solutions we make the separation of variables ansatz
${\Psi}(x)= \e^{-i E t / \hbar}\, {\Psi}(\mathbf{x})$. 

With this, and working to order~$G_N$, (\ref{eq:deq_EOBwavefunctions}) takes a form explicitly equivalent to the Coulomb equation, 
\begin{align}
    \left[\hbar^2 \boldsymbol{\nabla}^2+|\mathbf{p}|^2+\frac{2 \hbar |\mathbf{p}| \zeta}{r}  \right] {\Psi}(\mathbf{x}) =0 \,,
    \label{eq:Coulomb-eq}
\end{align}
where $\boldsymbol{\nabla}^2 := {\delta^{i j}} \partial_i \partial_j$ and we \emph{define} variables $|{\bf p}|$ and $\zeta$ by
\begin{align}
    \label{eq:alpha-def}
   |{\bf p}| := \sqrt{E^2-m_A^2} \;, \qquad  \zeta:= {\frac{G_N m_B}{\hbar} \frac{(2 E^2-m_A^2)}{\sqrt{E^2-m_A^2}}} \,.
\end{align} 
When solving (\ref{eq:Coulomb-eq}) we look for scattering solutions which behave like plane waves in the asymptotic past, and bound solutions which are exponentially suppressed for large $r$. In the scattering case, the wavefunction which is regular at the origin and represents an incoming plane wave of momentum $\mathbf{p}$ in the asymptotic past, call it $\Psi^{>}_{\mathbf p}(x)$, is (cf., Chapter 33 of \cite{NIST:DLMF})
\begin{align}
   {\Psi}_{\mathbf{p}}^{>}(x)
   =
   \e^{\pi \zeta / 2}\, \Gamma(1-i \zeta) \, { }_1F_1\!\bigg(i \zeta; 1; \frac{i(|\mathbf{p}| r-\mathbf{p} \cdot \mathbf{r})}{\hbar}\bigg)
   \e^{-i p\cdot x/\hbar}
   \,,
\label{eq:phiscatteringwavefunctionfull}
\end{align} 
where ${}_1F_1$ is the generalized hypergeometric function. (The superscript $>$ (resp. $<$) labels scattering (resp. bound) state quantities throughout.) Of course, the energy appearing here is now on-shell, $E = \sqrt{{\bf p}^2+m_A^2}$, so $|{\bf p}|$ in (\ref{eq:Coulomb-eq}) really is the modulus of the momentum. The normalization of the wavefunction is determined by the standard Klein-Gordon inner product:
\be
\label{eq:KG-normalization}
\langle {\Psi}^{>}_{\mathbf{p}^{\prime}} | {\Psi}^{>}_{\mathbf{p}}\rangle  =
i \int\!\ud^3 x\,  {\Psi}^{*>}_{\mathbf{p}^{\prime}}(x)\, \overleftrightarrow{\partial_t} {\Psi}^{>}_{\mathbf{p}}(x)  = 2 E_{\mathbf{p}}\,  \hat{\delta}^3(\mathbf{p}' - \mathbf{p})\,.
\ee
While the continuum of scattering solutions to (\ref{eq:Coulomb-eq}) obey $E^2>m_A^2$, bound state solutions have $E_n^2<m_A^2$ as a consequence of the quantization condition~\cite{Kabat:1992tb,Messiah,Gottfried2003}
\begin{align}
\label{eq:quantization}
\frac{G_N m_B}{\hbar}
\frac{2 E_n^2-m_A^2}{\sqrt{m_A^2-E_n^2}}
\equiv i \zeta  =n \,, \qquad n \in \mathbb{N}^+ \;.
\end{align}
Going from scattering to bound therefore corresponds to analytically continuing $|{\bf p}|$ from the positive real to positive imaginary axis, while~$\zeta$ goes from positive real to negative imaginary. 
Under this continuation the bound state wavefunctions arise as poles in the scattering wavefunctions; to see this it is convenient to expand~\eqref{eq:phiscatteringwavefunctionfull} into partial waves as
\begin{align}
   {\Psi}_{\mathbf{p}}^{>}(x) &= \sum_{\ell = 0}^{\infty} \sum_{m = -\ell}^{\ell} \, R^{>}_{{\bf p}\ell}(r) Y_{\ell m}(\hat{\mathbf{r}}) Y^*_{\ell m}(\hat{\mathbf{p}}) \,, \\
R^{>}_{{\bf p}\ell}(r) &=
4\pi\,
\e^{\pi \zeta/2}
\frac{\Gamma(\ell+1 -i \zeta)}{(2 \ell+1) !}
\bigg(\frac{2i|\mathbf{p}| r}{\hbar}\bigg)^{\ell}
\e^{i |\mathbf{p}| r/\hbar}
{}_1 F_1\!\bigg(\ell+1 -i \zeta ; 2 \ell+2 ; - \frac{2 i |\mathbf{p}| r}{\hbar}\bigg) \,,\label{eq:Rpl-explicit} 
\end{align}
in which the $Y_{\ell m}$ are the usual spherical harmonics. Now, taking
\be
\label{eq:continuation-of-parameters}
    |{\bf p}|\to i \lambda/n \;,
    \qquad  \zeta \to -in, 
\ee
for $\lambda := n\sqrt{m_A^2 - E_n^2}>0$, the $\Gamma$-function in \eqref{eq:Rpl-explicit} diverges due to a simple pole. Taking the residue of this simple pole in $\zeta$ we find:
 \begin{align}
R^{<}_{n\ell}(r) := \text{Res}_{\zeta = -i n} R^{>}_{\bf p \ell}(r) = 
-\frac{4 \pi  i^{n+1}}{\Gamma (n+\ell+1)}
\e^{-\frac{\lambda  r}{\hbar n}} \left(\frac{2 \lambda  r}{\hbar n}\right)^\ell L_{n-\ell-1}^{(2 \ell+1)}\!\left(\frac{2 r \lambda }{\hbar n}\right) \;,
\end{align} 
for $L_n^{(\alpha)}$ the associated Laguerre polynomials. These residues are precisely the radial profiles entering the bound state solutions ${\Psi}_{n \ell m}^{<}(x)$ of (\ref{eq:Coulomb-eq}) which are given by\footnote{This basis of wavefunctions, up to boundary conditions and normalization, has been discussed in~\cite{Baumann:2018vus,Baumann:2019eav,Baumann:2021fkf}.}~\cite{Messiah}
\be
	{\Psi}_{n \ell m}^{<}(x) := \e^{-i E_n t / \hbar}\, R^{<}_{n \ell}(r)\, Y_{\ell m}(\theta,\phi)\,,
	 \label{eq:ansatz_boundpsi}
\ee
with a normalisation here inherited from the Klein-Gordon inner product (\ref{eq:KG-normalization}).

Of course similar arguments apply to the full scattering solution (\ref{eq:phiscatteringwavefunctionfull}); we can continue~$|{\bf p}|$ into the complex plane and take the residue on a bound state pole \cite{Gottfried2003}. The result is a normalizable wavefunction built from the bound state solutions (\ref{eq:ansatz_boundpsi}).  In general, then, this analytic continuation can be expressed in terms of rapidity $y=E/m_A$ as:
\begin{align}
	\Psi^{<}_{n}(x,\sqrt{1-y^2}) = \text{Res}_{\zeta = -i n} \bigg[
	\Psi^{>}(x,\sqrt{y^2-1} \to i\sqrt{1-y^2}) \bigg] \;.
   \label{eq:B2B-wavefunction2}
\end{align}
This analytic continuation is an exact feature of solutions to the wave equation. It therefore holds, in particular, in the large distance and classical limits appropriate for the calculation of on-shell observables relevant to gravitational wave physics.  For example, the large distance expansion of the scattering wavefunction  (\ref{eq:phiscatteringwavefunctionfull}) follows from the asymptotic expansion of the hypergeometric function for large (positive imaginary) argument (cf.~eq.(4.4) of~\cite{Kabat:1992tb}):
\begin{align}
  {\Psi}_{\mathbf{p}}^{>}(x) &\sim  
  \e^{-i p\cdot x/\hbar 
  - i \zeta \log(|\mathbf{p}| r-\mathbf{p} \cdot \mathbf{r}|/\hbar)} + \frac{f^{>}_{\mathbf{p}}(\theta)}{r}\, 
\e^{-i (E t - |{\bf p}|r)/\hbar + i \zeta \log (2 |\mathbf{p}| r/\hbar)}\,,
\label{eq:phiscatteringwavefunction}
\end{align}
in which the scattering amplitude $f^{>}_{\bf p}(\theta)$ is
\begin{align}
f^{>}_{\mathbf{p}}(\theta) = \frac{\hbar\, \zeta}{2 |\mathbf{p}|}\, \frac{\Gamma(1-i \zeta) }{\Gamma(1 + i \zeta)}\, \frac{1}{(\sin^2(\theta/2))^{1-i \zeta}} \,,
\label{eq:scalar_wavefunction}
\end{align}
and we have chosen coordinates such that $|\mathbf{p}| r-\mathbf{p} \cdot \mathbf{r} = 2 |\mathbf{p}| r \sin^2(\theta/2)$.  We observe from (\ref{eq:phiscatteringwavefunction}) and (\ref{eq:scalar_wavefunction}) that the residue of the wavefunction as $\zeta\to -in$ comes from the factor $\Gamma(1-i\zeta)$: it lies entirely in the \emph{scattering amplitude}. We are therefore prompted to make contact with \emph{on-shell} tools.

\subsection{On-shell derivation of classical wavefunctions via KMOC}
\label{sec:on-shell}

We now turn to the full two-body problem, with the initial two-particle state
\be\label{initial}
   \ket{\text{in}}= \int\!\ud\Phi(p_1,p_2)\varphi_1(p_1)\varphi_2(p_2)\ket{p_1}\ket{p_2} \equiv \ket{\text{in}}_2\ket{\text{in}}_1\;, 
\ee
in which $\varphi_1(p_1)$ and $\varphi_2(p_2)$ are wavepackets peaked around classical momenta $p_1 \sim m_A v_A$ and $p_2 \sim m_B v_B$ \cite{Kosower:2018adc}. Neglecting radiation for now (i.e., restricting to the conservative sector of the theory), we evolve (\ref{initial}) to asymptotic late time using the $S$-matrix, and ask how to extract the wavefunction of particle $1$ from the final state $S\ket{\text{in}}$. This cannot be the expectation value of the field operator $\langle\phi_1(x)\rangle$ because, in the classical limit, number-changing contributions for massive particles are suppressed\footnote{This is unlike the situation for massless fields, which are described by coherent states in the classical limit, thus the \emph{waveform} -- that is, the expectation value of the massless field operator~\cite{Kosower:2018adc,Cristofoli:2021jas} -- is nonzero.}, so $S\ket{\text{in}}$ is a one-particle state for each scalar; the expectation value $\langle\phi_1(x)\rangle$ is zero in such states.

Recall instead that the usual definition of a one-particle wavefunction is, essentially, $ \phi_p(x) = \bra{0}\phi(x)\ket{p}$, where $\phi$ is the relevant field and $\ket{p}$ is a one-particle state. 
Given this we observe that the overlap
\be
\label{eq:wavefunction-projection}
    \bra{\text{in}}_2 \bra{0}_1\phi_1(x)\ket{\text{in}}= \int\!\ud\Phi(p_1)\,\varphi_1(p_1)\,\e^{-ip_1\cdot x / \hbar}\;,
\ee
extracts the wavefunction of particle $1$ in the initial state -- we have used here the asympotically free mode expansion of the scalar field. We therefore take the same overlap to obtain the wavefunction at asymptotically late times, $\bra{\text{in}}_2\bra{0}_1 \phi_1(x) \, S \ket{\text{in}}$ (and we will further justify this below). Subtracting (\ref{eq:wavefunction-projection}) then gives the change in the wavefunction between asymptotically early and late times, call this $\Delta \phi_1(x)$, which we expect to contain the elastic scattering amplitude. Writing $S=1+iT$ and inserting a complete set of states we find
\be
\label{eq:deltas}
\begin{split}\Delta\phi_1(x) &=
     \int\!\ud\Phi(p'_2,
     p'_1,
     p_2,p_1)\,
     \varphi_2(p'_2) \varphi_2(p_2)\varphi_1(p_1)\,
     \e^{-ip'_1\cdot x  / \hbar}\bra{p'_2 p'_1}i T \ket{p_1 p_2} \;, \\
    &\bra{p'_2 p'_1}i T \ket{p_1 p_2}  = i\,{\hat \delta}^4(p_2'+p_1'-p_1-p_2)\, \mathcal{M}_4(p_1, p_2; p'_1, p'_2) \;,
\end{split}
\ee
where $\mathcal{M}_4$ is the $2\to2$ scattering amplitude.

To enforce the classical limit, we parameterise the particle momenta as \eqref{eq:HEFT_parametrization} where $v_B^{\mu} = (1,{\bf 0})$ again. Following~\cite{Cristofoli:2021jas} we ignore shifts of order $\hbar$ in wavepackets, which allows us to factorise and drop the $p_2$ integral and $|\varphi_2(p_2)|^2$. The wavepacket $\varphi_1$, on the other hand, is inherited by $\Delta\phi_1(x)$ as it should be. Assuming an incoming plane wave as in (\ref{eq:phiscatteringwavefunctionfull}), though, we can also drop $\varphi_1$ and the $p_1$ integral.  So, \emph{in the classical limit}, we identify
\be
\label{eq:KMOC-scalar-back1}
\begin{split}
     \Delta\phi_1(x) &\sim i\,\int\!\ud\Phi(p'_2)\ud\Phi(p'_1)\, \e^{-ip'_1\cdot x  / \hbar}{\hat \delta}^4(p_1'+p_2'-p_1-p_2) \, \mathcal{M}_4(p_1, p_2; p'_1, p'_2) \;.
\end{split}
\ee
To make contact with the linearized Schwarzschild background we assume that the momentum transfer $\hbar {\bar q}$ is small compared to the mass of particle 2, expanding in $(\hbar {\bar q})^2/m_B^2$.  In this limit particle $1$ propagates as a probe on the linearized Schwarzschild background generated by particle $2$, and $\mathcal{M}_4$ becomes equal to the exponential resummation of ladder and cross-ladder diagrams, also known as the leading eikonal approximation~\cite{Kabat:1992tb,Adamo:2021rfq}. 

Using the delta-functions in (\ref{eq:KMOC-scalar-back1}) to integrate out the momentum transfer ${\bar q}$ leaves 
\be
\label{eq:KMOC-scalar-back}
    \Delta\phi_1(x) \sim i\,\int\!\ud\Phi(p'_1)\, \frac{\e^{-ip'_1\cdot x  / \hbar}}{2m_B}\, {\hat \delta}(E_{{\bf p}'_1}-E_{{\bf p}_1}) \, \mathcal{M}_4(p_1 , p_B; p_1', p_B)
    \;.
\ee
Finally, since we are interested in the scalar profile at large distances we can use the saddle-point approximation
\begin{align}
\label{eq:asymptotic-exp}
\e^{-i p \cdot x / \hbar} \stackrel{r \rightarrow \infty}{\sim}
\frac{2\pi\hbar}{i |\mathbf{p}| r}\,
\e^{-i (E_{\mathbf{p}} t- |{\mathbf p}|r)/\hbar}\,
\delta_\Omega(\hat{\mathbf{x}}-\hat{\mathbf{p}}) \;,
\end{align}
and thus all remaining integrals in (\ref{eq:KMOC-scalar-back}) can be evaluated. In the considered limit, and upon using the the delta functions above, we have the relation~\cite{Kabat:1992tb,Adamo:2021rfq,Adamo:2022ooq}.
\begin{align}
\mathcal{M}_4(p_1 , p_B; p_1', p_B) = \frac{8\pi m_B}{\hbar} \bigg(\frac{\mu}{|{\bf p}|}\bigg)^{-2i\zeta} f_{{\bf p}}^{>}(\theta) \;,
\label{eq:matching}
\end{align}
in which $\mu$ is an IR regulator. We thus obtain the final result
\be
    \Delta \phi_1(x) \sim \frac{f^>_{\bf p}(\theta)}{r}\,
    \e^{-i (E_{\mathbf{p}_1} t - |{\bf p}_1|r) / \hbar -2i \zeta \log(\mu/ |{\bf p}_1|)} \;.
\ee
which recovers (\ref{eq:phiscatteringwavefunction}) up to (constant phases and) the Coulomb-type phase. This is unsurprising, as KMOC assumes a free asymptotic mode expansion, and it is well known in the Coulomb/Schwarzschild case that this misses a phase, but gives the correct cross section.

We have now seen that the 4-point amplitude determines the wavefunction (as suggested in~\cite{Fried:1981fd} by matching a gauge-invariant observable like the cross-section and later generalized in~\cite{Kabat:1992tb,Kalin:2019rwq, Adamo:2021rfq,Adamo:2022ooq}). Given the identification~\eqref{eq:matching} and the analytic continuation~\eqref{eq:B2B-wavefunction2}, it follows that, for the 4-point amplitude, an analytic continuation in energy is the same analytic continuation as for the wavefunction, confirming the suggestion in~\cite{Adamo:2022ooq}.

\begin{figure}[t!]
\centering
\includegraphics[scale=0.95]{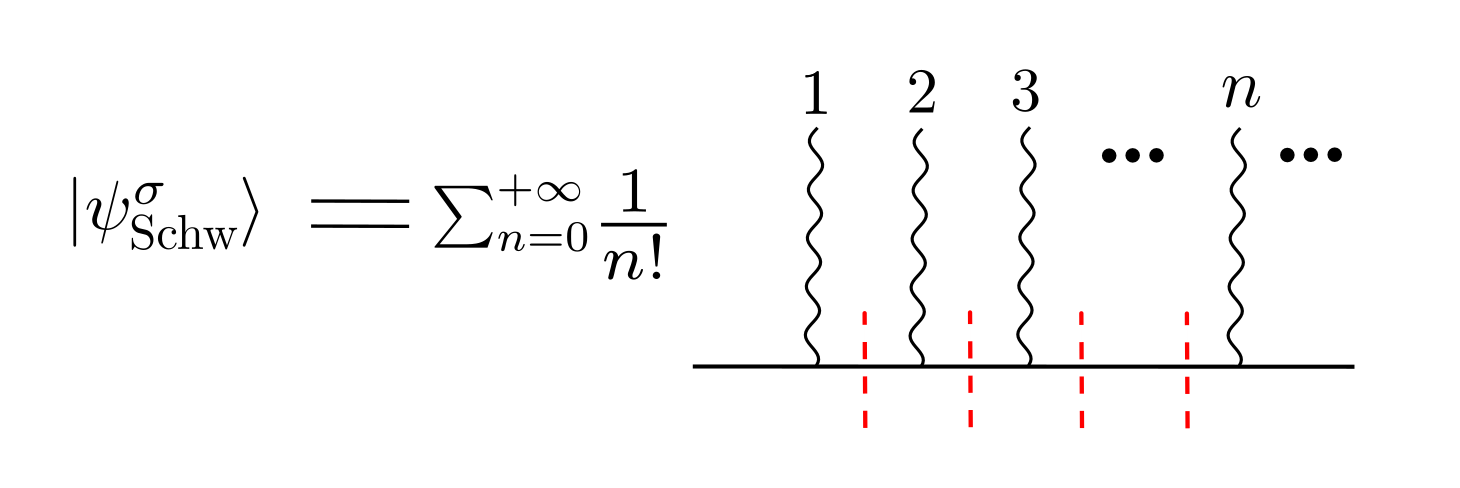}
\caption{The linearized Schwarzschild state $\ket{\psi_{\mathrm{Schw}}^{\sigma}}$ is generated by a coherent state of virtual gravitons. The dashed red lines indicate that the massive propagators are cut.}
\label{fig:linearized_Schw}
\end{figure}

\medskip

An alternative perspective on the above is given by considering the linearized Schwarz- schild background as a coherent state of \emph{virtual} gravitons determined by a massive particle at large distances\footnote{Coherent states of \emph{real} gravitons generate solutions of the vacuum equations, i.e.,~source-free solutions (cf., \cite{Cristofoli:2021jas,Cristofoli:2021vyo})}. Generating classical solutions from off-shell gravitons and scattering amplitudes was first explored in~\cite{Cristofoli:2020hnk}, with the approach based on off-shell coherent states developed subsequently in~\cite{Monteiro:2020plf} and~\cite{Adamo:2022ooq}. 

Borrowing notation from~\cite{Adamo:2022ooq}, we define a state (see Fig.\ref{fig:linearized_Schw})
\begin{align}
\label{eq:Schw_state}
\left|\psi_{\mathrm{Schw}}^\sigma\right\rangle=\frac{1}{\mathcal{N}} \int \mathrm{d}\Phi(p_2) \varphi_{2}(p_2) 
   \exp \left[i \int \frac{\hat{\mathrm{d}}^4 l}{l^2+i \epsilon} \hat{\delta}\left(2 p_2 \cdot l\right)\, \mathcal{M}_3^{(0) \mathrm{cl}}\left(p_2, l^\sigma\right) A_\sigma^{\dagger}(l)\right]
   \ket{p_2} \;,
\end{align}
\begin{figure}[t!]
\centering
\includegraphics[scale=0.9]{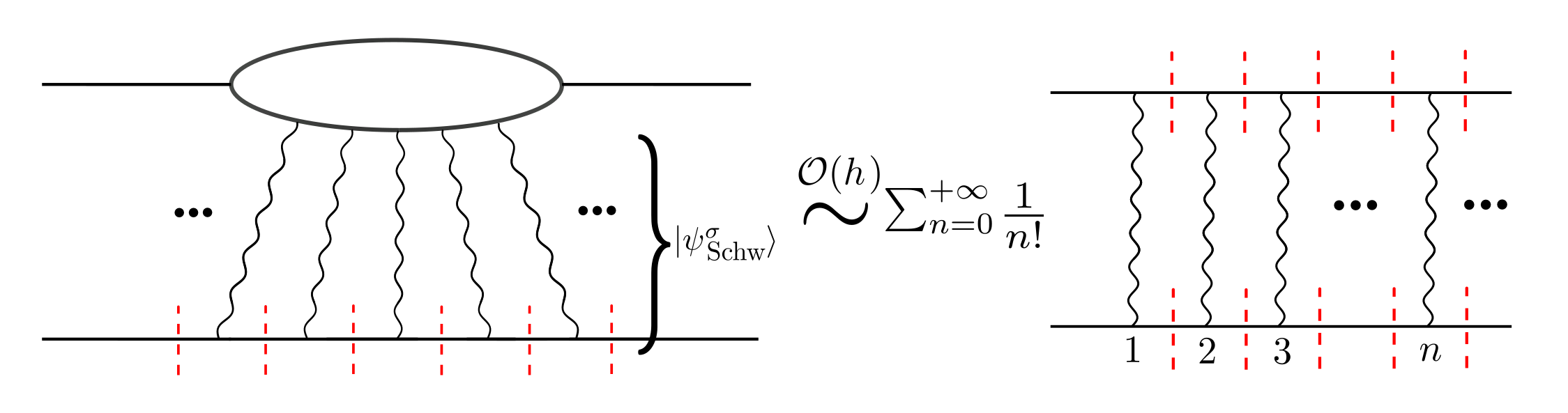}
\caption{A covariant description for the scattering of a massive probe on a linearized Schwarzschild background can be given by sewing the classical amplitude with graviton emissions from a massive particle line with a coherent state of virtual gravitons.}
\label{fig:background_scalar}
\end{figure}
%
where $\ket{p_2}$ is a momentum state for the massive particle, $\varphi_{2}(p_2)$ is a localized wavepacket for the massive particle of 4-velocity $v_B^{\mu}$, $A_\sigma^{\dagger}(l)$ is a placeholder for some operator which creates a `quanta' of virtual gravitons, and $\mathcal{M}_3^{(0) \mathrm{cl}}\left(p, l^\sigma\right)$ is the 3-pt `amplitude' for the emission of an graviton of \emph{off-shell} momentum $l^{\mu}$ and polarization vector $\varepsilon^{\mu \nu}_\sigma(l) = \varepsilon^{\mu}_\sigma(l) \varepsilon^{\nu}_\sigma(l)$, i.e.
\begin{align}
\mathcal{M}_3^{(0) \mathrm{cl}}\left(p, l^\sigma\right)=-\kappa \left(p\cdot \varepsilon_\sigma(l)\right)^2 \,. 
\end{align}
The expectation value of the graviton field operator in this coherent state is\footnote{{Note that at leading order the exponential coherent state effectively linearizes, and this way of obtaining the linear classical solution coincides with the off-shell prescription of~\cite{Cristofoli:2020hnk}.}} \cite{Monteiro:2020plf,Adamo:2022ooq}
\begin{align}
	h_{\mu \nu}^{\mathrm{cl}}(x)=\left\langle\psi_{\mathrm{Schw}}^\sigma\left|h_{\mu \nu}(x)\right| \psi_{\mathrm{Schw}}^\sigma\right\rangle=-\frac{4 G_N M}{r} P_{\mu \nu \alpha \beta} v_B^\alpha v_B^\beta  \,,
\end{align}
such that we recover the linearized Schwarzschild metric~\eqref{eq:linearized_Schw}. In the considered limit, the probe particle propagates on the Schwarzschild background as represented by (\ref{eq:Schw_state}), while particle $2$ becomes a \emph{spectator}, effectively decoupling from the theory. As such the overlap $\bra{0}_1 S\ket{\text{in}}$ returns the wavefunction of particle 1, multiplied by the unchanged initial state of particle 2, and the additional projections in (\ref{eq:wavefunction-projection}) simply remove this spectator state. 

\section{Scattering and bound matrix elements from Schwinger-Dyson}
\label{sec:theory-intro}
We have reviewed how bound state wavefunctions arise through poles in scattering wavefunctions, which may themselves be seen (in the appropriate large distance limit) as observables calculated from scattering amplitudes in the full two-body problem. The bound state poles are generated by an all-orders resummation of corrections to the Born amplitude; this is what yields the $\Gamma$ function in (\ref{eq:scalar_wavefunction}). This prompts us to reconsider classical Bethe-Salpeter recursion, with the idea that performing the resummation of superclassical iterations should simplify the map.

There are three different physical scenarios for the initial and final pair of massive particles interacting via the gravitational field and emitting radiation. The binding energy $\mathcal{E}$ distinguishes between these scenarios:
\begin{align}
    \mathcal{E} = \frac{E - m_A - m_B}{\mu} \,,\qquad \mu = \frac{m_A m_B}{m_A + m_B}\,,
    \label{eq:binding_energy}
\end{align}
where $E$ is the total incoming energy of the two-body system and $\mu$ is the reduced mass. In the first scenario of hyperbolic scattering, an incoming scattering state of two scalars evolves to an outgoing scattering state of two scalars (plus radiation), and the binding energy is positive, $\mathcal{E}>0$, at all times. In the second scenario of evolution from a two-body bound state to another two-body bound state plus radiation, $\mathcal{E}<0$ at all times. The third possibility is \emph{bound state formation}, where an incoming scattering configuration of two scalars evolves to a bound state and we pass dynamically from $\mathcal{E}>0$ to $\mathcal{E}<0$. For illustrative purposes we focus on the last case in sections (\ref{sec:bound_matrixelement}) and (\ref{sec:Schwinger-Dyson}).
Analogous results for the other scenarios will then be obvious.

\subsection{One-body approach}
\label{sec:bound_matrixelement}
As in Section~\ref{sec:bound_wavefunction}, we begin in a simplified one-body (OB) approximation, before returning to the full two-body problem. We will be somewhat agnostic about the nature of this approximation, but we use the OB action
\begin{align}
        \hspace{-10pt} S^{\text{\tiny OB}} = -\frac{1}{16 \pi G_N} \int \mathrm{d}^4 x \sqrt{-g}\, R + \frac{1}{2} \int \mathrm{d}^4 x \sqrt{-g} \left(g^{\mu \nu} \partial_{\mu} \Phi\, \partial_{\nu} \Phi - \mu^2\,\Phi^2\right) \;,
        \label{eq:scalarGR-EOB}
\end{align}
where we expand $g_{\mu\nu} = {g}^{\text{\tiny OB}}_{\mu\nu} + \bar{h}_{\mu\nu}$, for ${g}^{\text{\tiny OB}}$ a fixed metric which may be either a chosen background metric, as before, or the metric of the effective-one-body (EOB) formalism~\cite{Buonanno:1998gg,Damour:2008yg}, and $\bar{h}_{\mu\nu}$ is a the graviton fluctuation around ${g}^{\text{\tiny OB}}$. 

\begin{figure}[t!]
\centering
\includegraphics[scale=1]{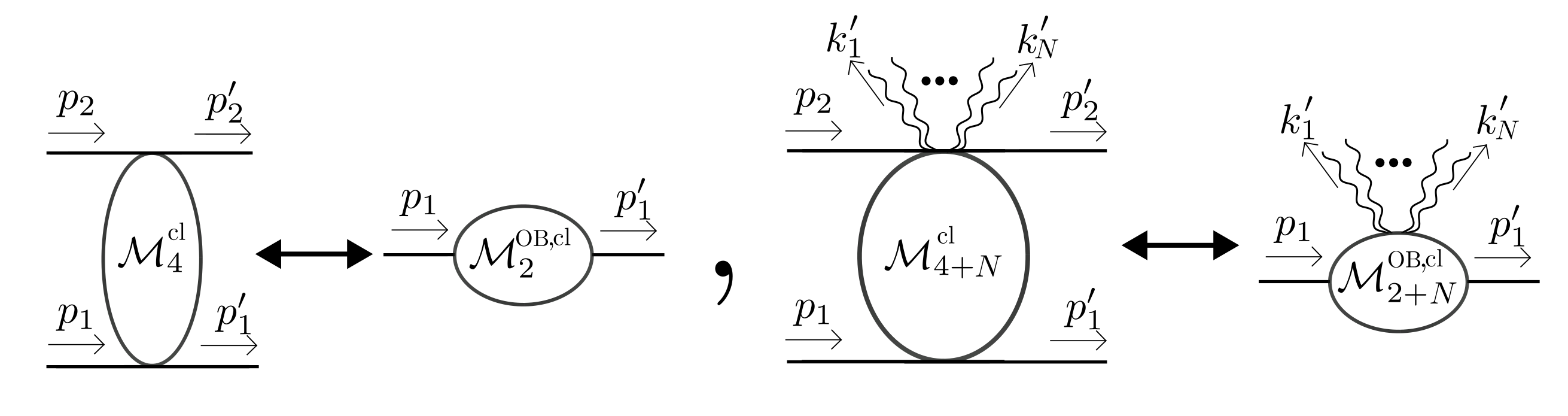}
\caption{Matching of on-shell amplitudes between the OB model and the full two-body problem. Left: conservative sector, right: radiative sector.}
\label{fig:EOBmatching}
\end{figure}

In the EOB approach, the two-body system is mapped \emph{exactly} to the effective motion of a single particle in a curved background determined by the original scattering configuration. The effective metric can -- in principle -- be determined perturbatively, by matching $4+N$-point amplitudes in the full theory with $2+N$-point amplitudes in the OB model, see Fig.\ref{fig:EOBmatching}. It is known that all conservative effects can be perturbatively mapped to an unambiguous `potential' piece of the metric up to 3PM (see, for example,~\cite{Damgaard:2021rnk}), but the inclusion of radiative and recoil effects is not yet fully understood and might imply that the map becomes ambiguous at 4PM~\cite{Bini:2020nsb,Bini:2020hmy}. Our working assumption, then, is that our OB model applies only when a Hamiltonian (and thus the corresponding scattering and bound wavefunctions to be discussed immediately below) are well-defined; that is, in background field theory or up to 3PM in the EOB approach.

We begin by introducing a complete set of one-body scattering and bound states for our probe particle (with a `bar' on a quantity reminding us that we are in the OB setup):
\begin{align}
\label{eq:states-in-EOB}
\ket{\mathcal{\bar{U}}_{\mathbf{P}}} &=  \text{unbound state, total momentum} \, \mathbf{P}  \,, \\
\ket{\mathcal{\bar{B}}_{ \{n\}}} &=  \text{bound state, quantum numbers $\{n\}$} \,, \nonumber 
\end{align}
obeying the completeness relation
\begin{align}
1 =
\sum_{\{n\}} \, \ket{\mathcal{\bar{B}}_{\{n\}}} \bra{\mathcal{\bar{B}}_{\{n\}}}
+ \int \!
\frac{\mathrm{\hat{d}}^3 P}{2 E^{>}_{\mathbf{P}}} \ket{\mathcal{\bar{U}}_{\mathbf{P}}} \bra{\mathcal{\bar{U}}_{\mathbf{P}}} \,.
\label{eq:completeness-4pt-back-earlier}
\end{align}
Such a decomposition\footnote{The \emph{existence} of such a decomposition of the Hilbert space is a consequence of the completeness of eigenfunctions for a self-adjoint Hamiltonian operator, as first discussed in the non-relativistic setup in~\cite{Newton:2004} and later generalized in~\cite{BERGGREN1968}. Essentially, the proof relies on analyticity of the resolvent (the operator $(E-\mathcal{\hat{H}})^{-1}$ for a local Hamiltonian $\mathcal{\hat{H}}$) in the complex energy plane, which determines a decomposition of the Green's function in terms of scattering and bound states (and possibly resonances \cite{Hernandez1984}).} may be realised concretely for a linearized Schwarzschild background by exploiting the map to the Coulomb problem; the completeness of the Coulomb wavefunctions has been demonstrated in e.g.~in~\cite{Hostler,Mukunda:1978} (see also~\cite{Mukhamedzhanov:2006cs}) in the form 
\begin{align}
\sum_{l=0}^{\infty} \sum_{m=-l}^{l} \left[\sum_{n=l+1}^{\infty} \Psi^{<}_{n l m}(\mathbf{x}) \Psi^{<*}_{n l m}\left(\mathbf{x}^{\prime}\right)+\int_0^{\infty} \!\mathrm{d}E_p\,  \Psi^{>}_{\mathbf{p} l m}(\mathbf{x}) \Psi^{>*}_{\mathbf{p} l m}\left(\mathbf{x}^{\prime}\right) \right] =\delta^3\left(\mathbf{x}-\mathbf{x}^{\prime}\right) \,,
\label{eq:scattering_bound_decomp}
\end{align}
in which the $\Psi^{>}$ and $\Psi^{<}$ are the scattering and bound wavefunctions introduced earlier, but where the latter are now normalised directly by the Klein-Gordon inner product. For a interesting recent discussion of the relativistic wavefunctions see~\cite{Lippstreu:2023vvg}.

The states (\ref{eq:states-in-EOB}) and (position-space) wavefunctions are related by
\begin{align}
\label{eq:EOB-wavefunctions-earlier}
\bar{\Psi}^{<}_{\{n\}}(y) &= 
\left\langle\Omega^{\text{\tiny OB}}\left|
\bar{\Phi}(y) \right| \mathcal{\bar{B}}_{\{n\}}\right\rangle \,,  \\
\bar{\Psi}^{* >}_{\mathbf{p}}(x) &=
\langle\mathcal{\bar{U}}_{\mathbf{p}}|
\bar{{\Phi}}^{\dagger}(x)| \Omega^{\text{\tiny OB}}\rangle \,, \nonumber
\end{align}
where here $\Omega^{\text{\tiny OB}}$ is an (asymptotically flat) background state determined by $g^{\text{\tiny OB}}$ (different from the standard Poincar\'e invariant vacuum $\Omega$) . An additional ingredient in the OB approach is that, due to the nonlinear interaction of gravity, the emitted gravitons have a non-trivial wavefunctions,
\begin{align}\label{eq:EOB-graviton-earlier}
    \bar{H}^{* >}_{\mathbf{K}^{\prime},  \mu \nu}
    (z) =
    \big\langle\Omega^{\text{\tiny OB}}\left|
\bar{h}_{\mu \nu}(z) \right| \bar{h}_{\mathbf{K}^{\prime}} \big\rangle \,,
\end{align}
which enter when we consider \emph{radiative} transitions between the states (\ref{eq:EOB-wavefunctions-earlier}), as encoded in the $(2+N)$-point Green's functions for $N\geq 1$. 

The Green's functions possess energy poles corresponding to both scattering states and bound states; the traditional Bethe-Salpeter approach to radiative transitions (in the two-body problem)~\cite{Mandelstam:1955sd,Faustov:1970hn,Faustov:1974qt,Gross:1987bu} is based on isolating, in the relevant momentum-space Green's functions, the poles corresponding to the incoming scattering state, the outgoing bound state and any emitted, outgoing gravitons.  For the case of single graviton emission in the OB model, we need the 3-point Green's function.
\begin{align}
G^{\text{\tiny OB}}_{\mu_1 \nu_1}(x_1; y_1,z_1) = \bra{\Omega^{\text{\tiny OB}}} T \bar{h}_{\mu_1 \nu_1}(z_1) \bar{\Phi}(y_1) \bar{\Phi}^{\dagger}(x_1)  \ket{\Omega^{\text{\tiny OB}}} \,. 
\end{align}
We show in appendix~\ref{sec:appendixA} that, performing LSZ reduction on the 3-point function and isolating the scattering and bound state poles, the matrix element for bound state formation is
\begin{align}
\left\langle\mathcal{\bar{B}}_{\{n\}} ; \bar{h}_{\mathbf{K}^{\prime}} \big|S\big| \mathcal{\bar{U}}_{\mathbf{p}}\right\rangle &= \int
\hat{\mathrm{d}}^4 r\,
\hat{\mathrm{d}}^4 s\,
\hat{\mathrm{d}}^4 l\, 
\bar{H}^{* >}_{\mathbf{K}^{\prime}\, {\mu} \,{\nu}}(l)  \bar{\Psi}_{\{n\}}^{* <}(r) \bar{\Psi}^{>}_{ \mathbf{p}}(s)  \bar{M}^{{\mu} \,{\nu}}_{3} (r, l; s) \,,
\label{eq:master-eq-earlier}
\end{align}
where $\bar{M}^{\mu\nu}_{3}(r,l;s)$ is the 3-point $S$-matrix element in momentum-space for \emph{generic}, i.e.~\emph{off-shell} momenta. This is simply convoluted with the momentum space wavefunctions. 

Note that (\ref{eq:master-eq-earlier}) is fully consistent with the perturbiner approach to scattering: transition elements are obtained by evaluating multi-linear pieces of the on-shell action on the asymptotic solutions of the field equations~\cite{Arefeva:1974jv,Shrauner:1977sk,Abbott:1983zw,Jevicki:1987ax,Selivanov:1999as,Lee:2022aiu,Adamo:2023cfp,Kim:2023qbl,Jain:2023fxc}. Here, the novelty is that these asymptotic solutions can represent bound -- rather than scattering -- states. As such bound-bound and scattering-scattering  transition elements are obtained simply by substituting the wavefunctions by their appropriate bound or scattering counterparts.

\subsection{Bethe-Salpeter and Schwinger-Dyson in the two-body problem}
\label{sec:Schwinger-Dyson}
We now turn to scattering-to-bound transitions in the full 2-body problem. We will consider both the relevant `asymptotic' states and, in some detail, the Green's functions contributing to radiative transitions, as these are a key ingredient of the waveforms to be discussed in Sect.~\ref{sec:waveform_B2B}. To begin, we recall the position space Green's functions of the scalar and graviton fields, defined as usual by
\begin{align}
\label{eq:green-intro}
G_{\mu_1 \nu_1 \dots \mu_N \nu_N}&(x_1,x_2; y_1,y_2,z_1,\dots,z_N) \\
& := \bra{\Omega} T h_{\mu_1 \nu_1}(z_1) \dots h_{\mu_N \nu_N}(z_N) \phi_1(y_1) \phi_2(y_2) \phi_1^{\dagger}(x_1) \phi^{\dagger}_2(x_2) \ket{\Omega} \,, \nonumber 
\end{align}
where $\ket{\Omega}$ is the Minkowski space vacuum, $T$ denotes time-ordering, and we have specialised to the 4-scalar, $N$-graviton case. We use $x$, $y$, $z$, resp.~$p$, $p'$, $k'$, for the position resp.~momentum-space arguments of $\phi^\dagger$, $\phi$, $h$, throughout, so for $(4+N)$-point amplitudes the incoming (outgoing) massive legs have momenta $p_1$, $p_2$ ($p_1'$, $p_2'$) and outgoing gravitons have momenta $k_1^{\prime} \dots k_n^{\prime}$.

As in the one-body setup, matrix element for e.g.~bound state formation are found by isolating the relevant poles in the relevant momentum-space Green's functions. To do this one first goes to centre-of-mass ($X$, $Y$) and relative ($x$, $y$) coordinates, which are convenient for describing both scattering and bound states: 
\begin{align}
\label{eq:coord_variables}
X = \frac{1}{2} (x_1 + x_2) \,,\quad Y = \frac{1}{2} (y_1 + y_2)\,,
\qquad
x = x_1 - x_2 \,, \quad y = y_1 - y_2\,,
\end{align}
along with their respective conjugate momenta $P$, $P^{\prime}$, $Q$, $Q^{\prime}$. In particular, $P = p_1 + p_2$ and $P' = p_1^{\prime} + p_2^{\prime}$ are the total momenta of the initial and final scattering/bound states, while $Q$ and $Q'$ are the relative momenta of the \emph{constituents} of those states. Further, $P^2 = s$ is identified with the center of mass energy and the momentum collectively radiated into gravitons is given by $\sum_{j=1}^N k_{j}^{\prime} = P-P^{\prime}$. Following~\cite{Petraki:2015hla}, we then define a basis of \emph{two-body} scattering and bound states in the conservative Hilbert space, given by
\begin{align}
\label{eq:2-body_states}
\ket{\mathcal{U}_{\mathbf{P},\mathbf{Q}}} &= \phi_1-\phi_2 \,\text{scattering state, total momentum} \, \mathbf{P} \, \text{and relative momentum} \, \mathbf{Q} \,, \nonumber  \\
\ket{\mathcal{B}_{
\{n\}}} &= \phi_1-\phi_2 \,\text{bound state, 
quantum numbers} \, \{n\} \,. 
\end{align}
These states obey the completeness relation
\begin{align}
1 = 
\sum_{\{n\}}  \, \ket{\mathcal{B}_{\{n\}}} \bra{\mathcal{B}_{\{n\}}}
+ \int \frac{\mathrm{\hat{d}}^3 P}{2 E^{>}_{\mathbf{P},\mathbf{Q}}} 
\frac{\mathrm{\hat{d}}^3 Q}{2 \epsilon^{>}_{\mathbf{P},\mathbf{Q}}} \ket{\mathcal{U}_{\mathbf{P},\mathbf{Q}}} \bra{\mathcal{U}_{\mathbf{P},\mathbf{Q}}} \,,
\label{eq:completeness-4pt}
\end{align}
where $E^{<}_{\{n\}}$ is the two-body bound state energy, while $E^{>}_{\mathbf{P},\mathbf{Q}}$ and $\epsilon^{>}_{\mathbf{P},\mathbf{Q}}$ are the scattering state energies.   In the \emph{free} theory ($\kappa\to 0$) these behave as
\begin{align}
    E^{<}_{\{n\}} \to 0\,, \qquad E^{>}_{\mathbf{P},\mathbf{Q}} \, \epsilon^{>}_{\mathbf{P},\mathbf{Q}} \to E_{\mathbf{p}_1} E_{\mathbf{p}_2}\,, \quad\text{where}\quad  E_{\mathbf{p}_j} = \sqrt{|\mathbf{p}_j|^2 + m_j^2} \,.
\end{align}
From the states (\ref{eq:2-body_states}) we define \emph{two-body} scattering and bound state wavefunctions
\begin{align}
\Psi_{\mathbf{P}, \mathbf{Q}}^{* >}(s) &= \int \mathrm{d}^4 x \, \e^{-i s\cdot x / \hbar} \, \langle\mathcal{U}_{\mathbf{P}, \mathbf{Q}}|T\phi_1^{\dagger}(x/2)\phi_2^{\dagger}(-x/2)|\Omega\rangle \;, \nonumber \\
 \Psi_{\{n\}}^{<}(r) &= \int \mathrm{d}^4 y \, \e^{i r\cdot y / \hbar} \, \left\langle\Omega\left|T \phi_1\!\left(y/2\right) \phi_2\!\left(-y/2\right)\right| \mathcal{B}_{\{n\}}\right\rangle \,, 
\label{eq:two-bodywavefunction}
\end{align}
by Fourier transforming the time-ordered correlators. 

With this, we turn to the Green's functions themselves, written in centre-of-mass and relative variables. Focusing on the 5-point case (single graviton emission), the momentum space Green's function is
\begin{align}\label{eq:Fouriertransform-5pt-v2-earlier}
G^{\mu \nu}&\big(\tfrac{P}{2} + Q, \tfrac{P}{2} - Q; \tfrac{P^{\prime}}{2} + Q^{\prime}, \tfrac{P^{\prime}}{2} - Q^{\prime},  k_1^{\prime} \big)=\int \mathrm{d}^4 (X, Y, x, y, z_1) \, \nonumber \\
&\quad \times \e^{-i (P\cdot X + Q\cdot x -P^{\prime}\cdot Y^{\prime} - Q^{\prime}\cdot y^{\prime} - k_1^{\prime}\cdot z_1) / \hbar}\,  G^{\mu \nu }\big( X+\tfrac{x}{2}, X-\tfrac{x}{2}; Y+\tfrac{y}{2}, Y-\tfrac{y}{2},z_1\big) \,.  
\end{align}
We use the separation (\ref{eq:completeness-4pt}) to isolate the matrix element for bound state formation and then perform LSZ reduction, isolating the poles at
\begin{equation*}
P^0 \to E^{>}_{\mathbf{P},\mathbf{Q}} \,\, \text{(scattering)}\,,
\quad
P^{\prime 0} \to E^{<}_{\{n\}} \,\, \text{(bound)}\,,\quad 
k_1^{\prime 0} \to E_{\mathbf{k}_1^{\prime}} \equiv |\mathbf{k}_1^{\prime}| \quad \text{(graviton)} \,,
\end{equation*}
to obtain the transition element for bound state formation (see also~\cite{Petraki:2015hla})
\begin{align}
\label{eq:full-2-body-5-pt-result-earlier}
\left\langle\mathcal{B}_{\mathbf{P'}, \{n\}} ; h^{\sigma}_{\mathbf{k}_1^{\prime}} \big|S\big| \mathcal{U}_{\mathbf{P}, \mathbf{Q}}\right\rangle &= \varepsilon^{*\sigma }_{{\mu} {\nu}}(k_1^{\prime}) \int \hat{\mathrm{d}}^4 r \, \hat{\mathrm{d}}^4 s\, \Psi_{\{n\}}^{* <}(r)\, \Psi^{>}_{\mathbf{P}, \mathbf{Q}}(s)\, \hat{\delta}^4(P^{\prime} +k_1^{\prime} - P) \nonumber \\
&  \times \mathcal{M}^{{\mu} {\nu}}_5 \Big(\tfrac{P}{2} + s, \tfrac{P}{2} - s; \tfrac{P^{\prime}}{2} + r, \tfrac{P^{\prime}}{2} - r,  k_1^{\prime}\Big) \,.
\end{align}
Here, $\varepsilon^{\sigma}_{{\mu} {\nu}}(k_1^{\prime})$ is the helicity-$\sigma$ polarization tensor of the emitted on-shell graviton with momentum $k_1^{\prime}$ and $\mathcal{M}_5$ is the 5-point $S$-matrix element for \emph{generic} massive momenta and one on-shell graviton.

This result allows us, in principle, to determine the bound state formation transition element in a general quantum field theory setup.
The matrix element $\mathcal{M}_5$ can computed with standard off-shell Feynman diagrams techniques, but, unfortunately, the computation of the two-body wavefunctions is more difficult, as they obey complicated higher-order differential equations~\cite{Wick:1954eu}. While these equations admit a simple solution in the non-relativistic limit~\cite{Petraki:2015hla}, solving the relativistic equations is a formidable task\footnote{This is due to various, not unrelated, difficulties: 1) higher order differential equations~\cite{Wick:1954eu}, 2) absence of a probe limit result~\cite{Cutkosky:1954ru}, 3) highly non-trivial frame-dependence due to Poincar\'e covariance (in particular boost symmetries, for unequal times)~\cite{Jarvinen:2004pi,Hoyer:2006xg}.}. The advantage of the OB setup is that the wavefunctions obey far simpler differential equations and, as we saw in Section~\ref{sec:lin-schwarz}, there are solvable examples.

\subsection{Classical recursion relations for amplitudes and the 2MPI kernels}
\label{sec:2MPI_kernels}

\begin{figure}[t!]
\centering
\includegraphics[scale=0.14]{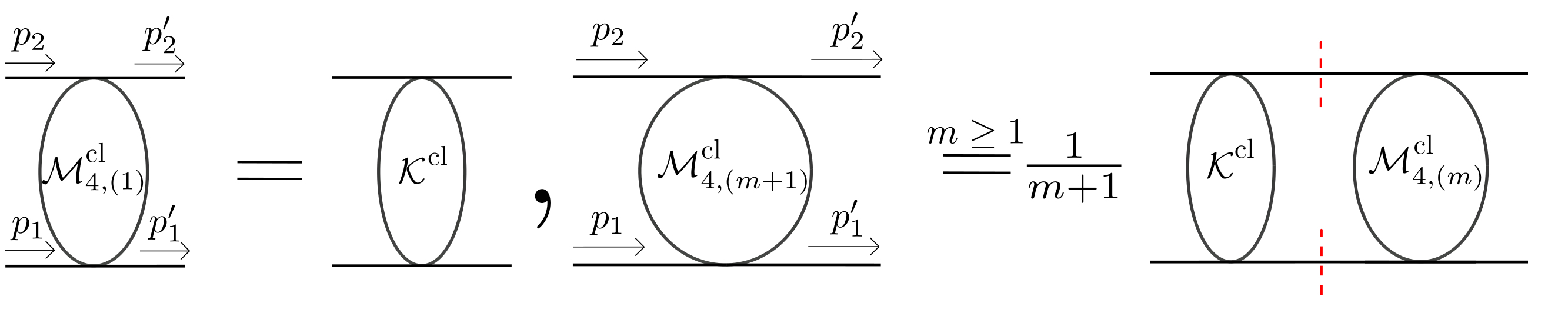}
\caption{The classical 4-pt amplitude recursion relation (\ref{eq:classical-BS}).}
\label{fig:recursion4pt-cl}
\end{figure}

Because of the general difficulties in explicity computing the wavefunctions, we return again to the Green's functions (\ref{eq:green-intro}): the recursion relations obtained from the Schwinger-Dyson equations of the theory can also be turn into (solvable) amplitude recursion relations using the standard LSZ reduction procedure in the classical limit. 
To illustrate, consider first the \emph{conservative} sector of the two body problem. This is controlled by the 4-point Green's function of the scalar fields. As discussed in Appendix~\ref{app:SD_radiative}, the Schwinger-Dyson equations lead in this case to the following recursion relation for the 4-point scattering amplitude $\mathcal{M}_4$ 
\begin{align}
\mathcal{M}_4(p_1, p_2; p_1^{\prime}, p_2^{\prime}) &=
\mathcal{K}(p_1, p_2; p_1^{\prime}, p_2^{\prime}) 
\nonumber \\
&+ \int\! \hat{\mathrm{d}}^4 s_1 \,  
\mathcal{K}(p_1, p_2; s_1, s_2) \Delta(s_1,s_2) \mathcal{M}_4(s_1, s_2; p_1^{\prime}, p_2^{\prime})\,,
\label{eq:BS-equation-amplitude-earlier}
\end{align}
in which in which it is understood that $p_1^{\prime}+p_2^{\prime} = p_1 + p_2 = s_1+s_2$ by momentum conservation, $\mathcal{K}$ is the interaction kernel given by connected \emph{two-massive-particle irreducible} (2MPI) diagrams, and $\Delta(p_1, p_2) \equiv \Delta_1(p_1) \Delta_2(p_2)$ in which
\begin{equation}
	\Delta_j(p) = \frac{i}{p^2 - m_j^2 + i \epsilon} \;,    
\end{equation}
is simply the free disconnected Green's function for particle $j$.
\begin{figure}[t!]
\centering
\includegraphics[scale=0.9]{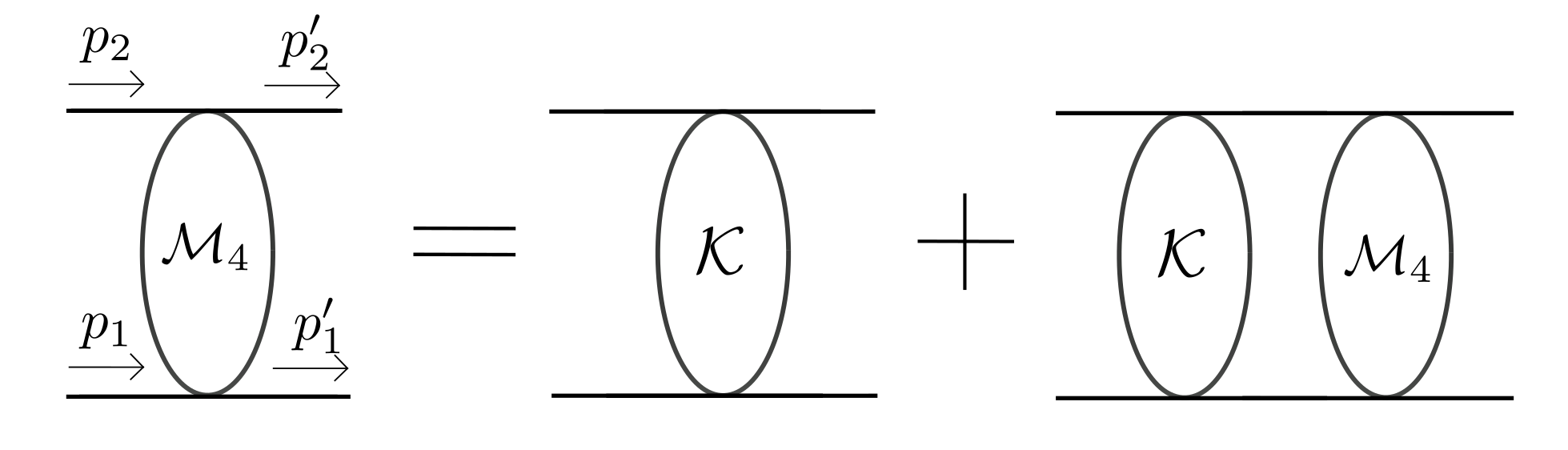}
\caption{The 4-pt amplitude recursion relation (\ref{eq:BS-equation-amplitude}) derived from the Schwinger-Dyson equations.}
\label{fig:recursion4pt-earlier}
\end{figure}
In~\cite{Adamo:2022ooq} it was shown, making use of the HEFT formalism developed in~\cite{Brandhuber:2021eyq,Brandhuber:2023hhy,Damgaard:2019lfh,Aoude:2020onz}, how to solve~\eqref{eq:BS-equation-amplitude-earlier} in the space of conservative classical amplitudes $\mathcal{H}_{4,\text{cl}}$, defined by quotienting $4$-point diagrams with external massive particles over the symmetrization of internal graviton exchanges. With this prescription and using the parametrization \eqref{eq:HEFT_parametrization}, we obtained the classical Bethe-Salpeter equation for the two-massive particle reducible amplitude $\mathcal{M}^{\text{cl}}_{4,(m)}$ with $m$ 2MPI components (see Fig.\ref{fig:recursion4pt-cl})
\begin{align}
\label{eq:classical-BS}
\hspace{-7pt}\mathcal{M}^{\text{cl}}_{4,(1)}(p_A,p_B,q) &= \mathcal{K}^{\text{cl}}(p_A,p_B,q) \,, \\
\hspace{-7pt}\mathcal{M}^{\text{cl}}_{4,(m+1)}(p_A,p_B,q) &= \frac{1}{m+1} \int \hat{\mathrm{d}}^4 l \,\mathcal{K}^{\text{cl}}(p_A,p_B,l) \Delta^{\text{cl}}(p_A,p_B,l) \mathcal{M}^{\text{cl}}_{4,(m)}(p_A,p_B,q-l) \,, \,\, \forall m\geq 1 \nonumber 
\end{align}
where the classical two-body propagator is
\begin{align}
\Delta^{\text{cl}}(p_A,p_B,l) = \hat{\delta}(2 p_A \cdot l) \hat{\delta}(2 p_B \cdot l) \,.
\label{eq:classical-prop}
\end{align}
Defining the Fourier transform to the impact parameter space $b$ conjugate the unique momentum transfer $q := q_1 = -q_2$
\begin{align}
\label{eq:def-fourier-impact-conservative}
\widetilde{f}\left(b\right):=\mathcal{F}_{b}[f(q)] \equiv \int \hat{\mathrm{d}}^4 q \, \hat{\delta}\left(2 p_A \cdot q\right) \hat{\delta}\left(2 p_B \cdot q\right) \e^{i q \cdot b/\hbar} f(q)  \;,
\end{align}
the recursion relation (\ref{eq:BS-equation-amplitude-earlier}) is solved, in the classical limit, by~\cite{Adamo:2022ooq}
\begin{align}
\label{eq:4-pt-sol-earlier}
\widetilde{\mathcal{M}}_4^{\mathrm{cl}}\left(p_A, p_B, b\right)= \e^{\tilde{\mathcal{K}}^{\mathrm{cl}}(p_A, p_B, b)} -1  \,, 
\end{align}
where $\mathcal{K}^{\mathrm{cl}}$ is the classical part of the 2MPI interaction kernel and the $-1$ is simply subtraction of the forward scattering contribution.

\begin{figure}[t!]
\centering
\includegraphics[scale=0.8]{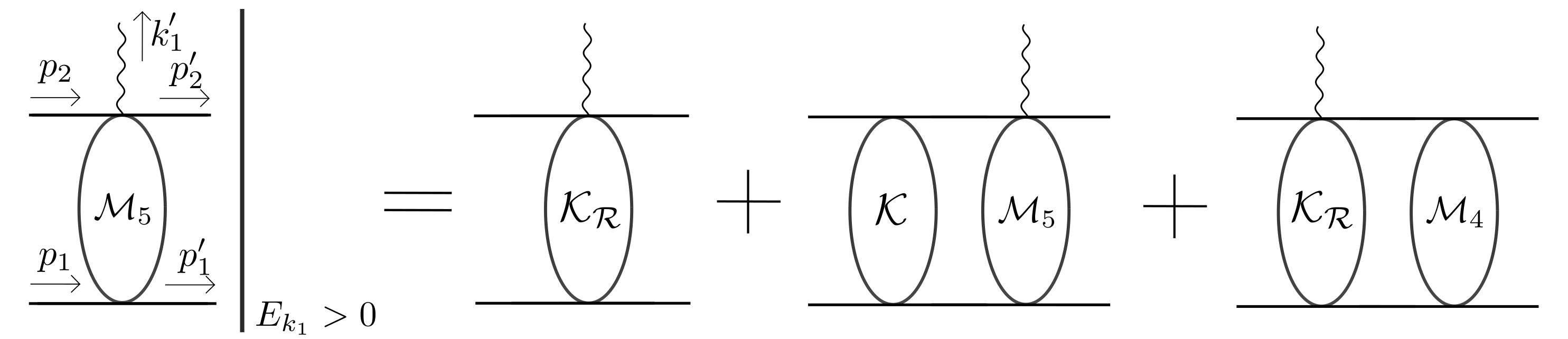}
\caption{The 5-pt amplitude recursion relation simplifies dramatically when restricted to the emission of strictly positive energy gravitons.}
\label{fig:recursion5ptpositive}
\end{figure}

Our task here is to extend this calculation to include radiative contributions. To take the classical limit, we then define a space of classical $(4+N)$-point diagrams
\begin{align}
\mathcal{H}_{4+N,\text{cl}} := \mathcal{H}_{4+N} / \Sigma
\label{eq:space_classical}
\end{align}
as the quotient space of Feynman diagrams contributing to $4+N$-point amplitudes, $\mathcal{H}_{4+N}$, up to the permutation group $\Sigma$ of \emph{all} (virtual and real) graviton emissions. 

\begin{figure}[t!]
\centering
\includegraphics[scale=0.8]{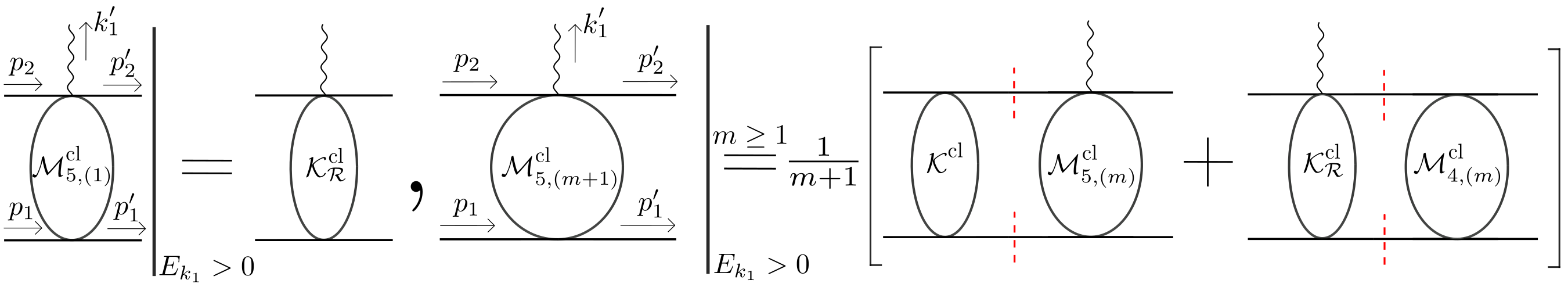}
\caption{The classical 5-pt amplitude recursion relation with positive energy gravitons.}
\label{fig:recursion5ptpositive-cl}
\end{figure}

We focus on the case $N=1$ and restrict to the emission of a positive energy graviton $E_{\mathbf{k_1}^{\prime}}>0$ from here on; the emission of multiple gravitons and the zero-energy sector are discussed in Appendix~\ref{app:SD_radiative}. The 5-point amplitude recursion relation, derived from the Schwinger-Dyson equations in Appendix \ref{app:SD_radiative}, is
\begin{align}
\label{eq:recursion5pt-positive}
\mathcal{M}^{\mu_1 \nu_1}_5(p_1, p_2; p_1^{\prime}, p_2^{\prime}, k_1^{\prime})  &=  \mathcal{K}_{\mathcal{R}}^{\mu_1 \nu_1}(p_1, p_2; p_1^{\prime}, p_2^{\prime}, k_1^{\prime}) \\ 
&+\int \hat{\mathrm{d}}^4 w_1 \mathcal{K}(p_1, p_2; w_1, w_2) \Delta(w_1,w_2) \mathcal{M}^{\mu_1 \nu_1}_5(w_1, w_2; p^{\prime}_1, p^{\prime}_2, k_1^{\prime}) \nonumber \\
&+  \int \hat{\mathrm{d}}^4 w_1 \, \mathcal{K}_{\mathcal{R}}^{\mu_1 \nu_1}(p_1, p_2; w_1, w_2, k_1^{\prime}) \Delta(w_1,w_2)  \mathcal{M}_4( w_1, w_2; p^{\prime}_1, p^{\prime}_2) \,, \nonumber 
\end{align}
which is depicted in Fig.~\ref{fig:recursion5ptpositive}.  From \eqref{eq:space_classical} and \eqref{eq:recursion5pt-positive} we obtain a recursion relation for the classical $5-$pt amplitude. Defining the momentum transfers for each massive particle $q_j := p_j - p_j^{\prime}$, which obey $q_1 + q_2 = k_1^{\prime}$, the recursion relation is (see Fig.~\ref{fig:recursion5ptpositive-cl})
\begin{align}
\label{eq:recursion5pt-positive-cl}
 &\mathcal{M}^{\text{cl},\mu_1 \nu_1}_{5,(1)}(p_A, p_B; q_1, q_2) 
=  \mathcal{K}_{\mathcal{R}}^{\text{cl},\mu_1 \nu_1} (p_A, p_B; q_1, q_2) \;, \\
&\mathcal{M}^{\text{cl},\mu_1 \nu_1}_{5,(m+1)}(p_A, p_B; q_1, q_2) 
\nonumber \\
& \qquad \quad=  \frac{1}{m+1} \int \hat{\mathrm{d}}^4 l \, \Big[ \mathcal{K}^{\text{cl}} (p_A, p_B;l) \Delta^{\text{cl}}(p_A, p_B; l) \mathcal{M}^{\text{cl},\mu_1 \nu_1}_{5,(m)}(p_A, p_B; q_1 - l, q_2 + l) \nonumber \\
& \qquad \qquad \qquad \qquad+  
 \mathcal{K}_{\mathcal{R}}^{\text{cl},\mu_1 \nu_1} (p_A, p_B; q_1 - l, q_2 + l)  \Delta^{\text{cl}}(p_A, p_B; l) \mathcal{M}^{\text{cl}}_{4,(m)}(p_A, p_B; l) \Big] \,, \,\, \forall m\geq 1\,. \nonumber
\end{align}
To proceed, we use the classical recursion relation for the 4-point amplitude (\ref{eq:classical-BS}) to decompose each two-massive particle reducible amplitude $\mathcal{M}^{\text{cl},\mu_1 \nu_1}_{5,(m)}$ and $\mathcal{M}^{\text{cl}}_{4,(m)}$ in terms of the irreducible kernels $\mathcal{K}_{\mathcal{R}}^{\text{cl},\mu_1 \nu_1}$ and $\mathcal{K}^{\text{cl}}$. This allows us to rewrite \eqref{eq:recursion5pt-positive-cl} as, using compact notation,
\begin{align}
\label{eq:recursion5pt-positive-cl-v2}
\mathcal{M}^{\text{cl},\mu_1 \nu_1}_{5,(1)} & 
= \mathcal{K}_{\mathcal{R}}^{\text{cl},\mu_1 \nu_1}  \,, \\
\mathcal{M}^{\text{cl},\mu_1 \nu_1}_{5,(m+1)} &
= \frac{1}{(m+1)!} \Big[\mathcal{K}^{\text{cl},\mu_1 \nu_1}_{\mathcal{R}}  (\Delta^{\text{cl}} \mathcal{K}^{\text{cl}})^m + \mathcal{K}^{\text{cl}}  (\Delta^{\text{cl}} \mathcal{K}^{\text{cl},\mu_1 \nu_1}_{\mathcal{R}}) (\Delta^{\text{cl}}  \mathcal{K}^{\text{cl}})^{m-1} + \ldots  \nonumber \\
&+ \mathcal{K}^{\text{cl}} (\Delta^{\text{cl}}  \mathcal{K}^{\text{cl}})^{j} \underbrace{(\Delta^{\text{cl}} \mathcal{K}^{\text{cl},\mu_1 \nu_1}_{\mathcal{R}})}_{j-\text{th position}} (\Delta^{\text{cl}}  \mathcal{K}^{\text{cl}})^{m-j}  + \ldots + \mathcal{K}^{\text{cl}} (\Delta^{\text{cl}}  \mathcal{K}^{\text{cl}})^{m-1} (\Delta^{\text{cl}} \mathcal{K}^{\text{cl},\mu_1 \nu_1}_{\mathcal{R}}) \Big] \,, \nonumber 
\end{align}
which has a natural interpretation as the sum of all possible single graviton dressings of the two-massive particle irreducible kernel $\mathcal{K}^{\text{cl}}$ in the amplitude recursion. 

Equation~\eqref{eq:recursion5pt-positive-cl-v2} can again be solved by working in impact parameter space, now defined by the following Fourier-like transform over the relative momentum transfers  $(q_1,q_2)$:
\begin{align}
\widetilde{f}\left(b_1,b_2\right) &\equiv \int \hat{\mathrm{d}}^4 q_1 \hat{\mathrm{d}}^4 q_2 \, \hat{\delta}\left(q_1 + q_2 - k_1^{\prime}\right) \nonumber \\
& \qquad \qquad \times\hat{\delta}\left(2 p_A \cdot q_1\right) \hat{\delta}\left(2 p_B \cdot q_2\right) \e^{i (q_1 \cdot b_1 + q_2 \cdot b_2) / \hbar} f(q_1,q_2) \,.
\label{eq:impact_parameter_radiative}
\end{align}
Indeed, we notice that factorisation property~\cite{Brandhuber:2023hhy} 
\begin{align}
&\int \hat{\mathrm{d}}^4 q_1 \hat{\mathrm{d}}^4 q_2 \, \hat{\delta}\left(q_1 + q_2 - k_1^{\prime}\right) \hat{\delta}\left(2 p_A \cdot q_1\right) \hat{\delta}\left(2 p_B \cdot q_2\right) \e^{i (q_1 \cdot b_1 + q_2 \cdot b_2) / \hbar} \nonumber \\
& \qquad \qquad \times \int  \hat{\mathrm{d}}^4 l \, \mathcal{K}^{\text{cl},\mu_1 \nu_1}_{\mathcal{R}}(p_A,p_B; q_1 - l, q_2 + l)  \Delta^{\text{cl}}(p_A,p_B;l) \mathcal{K}^{\text{cl}}(p_A,p_B;l) \nonumber \\
&= \int  \hat{\mathrm{d}}^4 l \, \hat{\delta}\left(2 p_A \cdot l\right) \hat{\delta}\left(2 p_B \cdot l\right) \mathcal{K}^{\text{cl}}(p_A,p_B;l) \e^{i l \cdot (b_1 - b_2) /\hbar} \int \hat{\mathrm{d}}^4 q_{1 R}  \, \hat{\mathrm{d}}^4 q_{2 R} \, \hat{\delta}\left(q_{1 R} + q_{2 R} - k_1^{\prime}\right) \nonumber \\
& \qquad \qquad  \times \hat{\delta}\left(2 p_A \cdot q_{1 R}\right) \hat{\delta}\left(2 p_B \cdot q_{2 R}\right) \e^{i (q_{1 R} \cdot b_1 + q_{2 R} \cdot b_2) / \hbar} \mathcal{K}^{\text{cl},\mu_1 \nu_1}_{\mathcal{R}}(p_A,p_B; q_{1 R}, q_{2 R}) \nonumber \\
& = \mathcal{\widetilde{K}}^{\text{cl}}(p_A,p_B;b_1,b_2)\, \mathcal{\widetilde{K}}^{\text{cl},\mu_1 \nu_1}_{\mathcal{R}}(p_A,p_B;b_1,b_2,k_1^{\prime}) \,,  
\label{eq:factorization-radiative}
\end{align}
where for the intermediate stages of the calculation we introduced the variables
\begin{align}
q_{1 R} := q_1 - l\,, \qquad q_{2 R} := q_2 + l \,.
\end{align}
Iterating the factorization argument \eqref{eq:factorization-radiative}, we solve the non-trivial part of \eqref{eq:recursion5pt-positive-cl-v2}
\begin{align}
\label{eq:recursion5pt-positive-cl-v3}
\mathcal{\widetilde{M}}^{\text{cl},\mu_1 \nu_1}_{5,(1)} 
= \mathcal{\widetilde{K}}_{\mathcal{R}}^{\text{cl},\mu_1 \nu_1}  \,, \qquad \qquad
\mathcal{\widetilde{M}}^{\text{cl},\mu_1 \nu_1}_{5,(m+1)} 
& = \frac{1}{(m)!} \mathcal{\widetilde{K}}^{\text{cl},\mu_1 \nu_1}_{\mathcal{R}} (\mathcal{\widetilde{K}}^{\text{cl}})^m \,, \quad  \forall m\geq 1 
\end{align}
implying the result
\begin{align}\label{eq:5-ptexp}
\mathcal{\widetilde{M}}^{\text{cl},\mu_1 \nu_1}_{5}(p_A,p_B;b_1,b_2,k_1^{\prime}) & 
= \e^{\mathcal{\widetilde{K}}^{\text{cl}}(p_A,p_B;b_1,b_2)}\, \mathcal{\widetilde{K}}^{\text{cl},\mu_1 \nu_1}_{\mathcal{R}}(p_A,p_B;b_1,b_2,k_1^{\prime}) \,.
\end{align}
This is perfectly consistent with the resummation discussed in \cite{Cristofoli:2021jas,DiVecchia:2022nna}, with the difference that the Fourier transforms of the amplitudes are now completely localized on the plane defined by the classical momentum transfers for the two massive particles \cite{Damgaard:2021ipf,Brandhuber:2023hhy} and we have achieved an exact diagonalization of the partial wave basis \cite{Kol:2021jjc}. This shares many similarities with the eikonal but the subtraction terms are different, even at the classical level, see section 6.3.4 of \cite{DiVecchia:2023frv} for a comparison.

We are now ready to turn to the waveform, and for this the reader may proceed directly to Section~\ref{sec:waveform_B2B}. For completeness, though, we include in the following subsection the extension of the above results to higher-point amplitudes, and relate this to the conjectured exponentiation of the classical $S$-matrix.

\subsection{Exponentiation conjecture}

Proceeding as above, we can derive the recursion relation for $4+N$-point amplitudes; this is described in Appendix~\ref{app:SD_radiative}.  To illustrate, we state here the classical recursion relation in the case of $2$ graviton emissions with (positive) energies\footnote{Zero-energy contributions can be included as for the 5-pt amplitude, see Appendix \ref{app:SD_radiative}.} $E_{\mathbf{k_1}^{\prime}}$, $E_{\mathbf{k_2}^{\prime}}$: 
\begin{align}
\label{eq:recursion6pt-positive-cl}
&\mathcal{M}^{\text{cl}, \mu_1 \nu_1 \mu_2 \nu_2}_{6,(1)}  = \mathcal{K}_{\mathcal{R}}^{\text{cl}, \mu_1 \nu_1 \mu_2 \nu_2}  \,, \\
&\mathcal{M}^{\text{cl}, \mu_1 \nu_1 \mu_2 \nu_2}_{6,(m+1)}(p_A, p_B; q_1, q_2, k_1^{\prime}, k_2^{\prime}) \nonumber \\
& \quad =  \frac{1}{m+1} \Big[\frac{1}{2!} \int \hat{\mathrm{d}}^4 l \, \mathcal{K}^{\text{cl}, (\mu_1 \nu_1} (p_A, p_B;l, k_1^{\prime}) \Delta^{\text{cl}}(p_A, p_B; l) \mathcal{M}^{\text{cl},\mu_2 \nu_2)}_{5,(m)}(p_A, p_B; q_1 - l, q_2 + l, k_2^{\prime}) \nonumber \\
& \qquad +  \int \hat{\mathrm{d}}^4 l \, \mathcal{K}_{\mathcal{R}}^{\text{cl}, \mu_1 \nu_1 \mu_2 \nu_2} (p_A, p_B; q_1 - l, q_2 + l, k_1^{\prime}, k_2^{\prime})  \Delta^{\text{cl}}(p_A, p_B; l) \mathcal{M}^{\text{cl}}_{4,(m)}(p_A, p_B; l) \nonumber \\
& \qquad +  \int \hat{\mathrm{d}}^4 l \, \mathcal{K}^{\text{cl}} (p_A, p_B; q_1 - l, q_2 + l)  \Delta^{\text{cl}}(p_A, p_B; l) \mathcal{M}^{\text{cl}, \mu_1 \nu_1 \mu_2 \nu_2}_{6,(m)}(p_A, p_B; l, k_1^{\prime}, k_2^{\prime}) \Big] \,, \,\, \forall m\geq 1\nonumber 
\end{align}
where the $1/2!$ comes from the symmetrization procedure applied to the external gravitons (similarly to what happens for the symmetrization over virtual gravitons~\cite{Adamo:2022ooq}). The solution of this recursion relation in impact parameter space is naturally obtained by promoting the S-matrix to an operator acting in the Hilbert space of real gravitons\footnote{There is a factor of $i/\hbar$ between the kernel and the amplitude, so this operator is  unitary.}
\begin{align}
\label{eq:exp_Smatrix}
& \hat{S}^{\text{cl}} =  e^{\mathcal{\widetilde{K}}^{\text{cl}}(p_A,p_B;b_1,b_2)}  \exp \left[\sum_{\sigma_1} \int \mathrm{d} \Phi(k_1^{\prime})  \hat{\alpha}^{\text{cl}}_{5, \mathcal{R}}(k_1^{\prime}) + \sum_{\sigma_1,\sigma_2} \int \mathrm{d} \Phi(k_1^{\prime},k_2^{\prime}) \hat{\alpha}^{\text{cl}}_{6, \mathcal{R}}(k_1^{\prime},k_2^{\prime}) \right] \,, \nonumber \\
& \hat{\alpha}^{\text{cl}}_{5, \mathcal{R}}(k_1^{\prime}) = \mathcal{\widetilde{K}}^{\text{cl}}_{5,\mathcal{R}}(p_A,p_B;b_1,b_2,k_1^{\prime}) a_{\sigma_1}^{\dagger}(k_1^{\prime})- \mathcal{\widetilde{K}}^{* \text{cl}}_{5,\mathcal{R}}(p_A,p_B;b_1,b_2,k_1^{\prime}) a_{\sigma_1}(k_1^{\prime})\,,   
\end{align}
where $ \mathcal{\widetilde{K}}^{\text{cl}}_{5,\mathcal{R}} \equiv \varepsilon^{\sigma_1}_{\mu \nu}(k_1^{\prime})\mathcal{\widetilde{K}}^{\text{cl} \mu \nu}_{\mathcal{R}}$ and $\hat{\alpha}^{\text{cl}}_{5,\mathcal{R}}$ includes the 6-point 2MPI radiative kernel $\mathcal{K}_{\mathcal{R}}^{\text{cl}, \mu_1 \nu_1 \mu_2 \nu_2}$ and terms quadratic in the creation and annihilation of the external graviton states (in a way that resemble what was recently found in \cite{Fernandes:2024xqr} in the soft limit). Including the emission of $N$ positive energy gravitons is then straightforward (see the Appendix), as this would modify the inelastic operator \eqref{eq:exp_Smatrix} by adding the $4+N$-point 2MPI radiative kernel contributions to the exponent $\sum_{\sigma_1, \dots, \sigma_N} \left[\prod_{j=1}^N \int \mathrm{d} \Phi(k_j^{\prime}) \right] \hat{\alpha}^{\text{cl}}_{4+j, \mathcal{R}}(k_1^{\prime},\dots,k_N^{\prime})$. 

Nevertheless, an explicit calculation \cite{Cristofoli:2021jas,Britto:2021pud} (see also \cite{DiVecchia:2022nna,Herderschee:2023fxh,Georgoudis:2023lgf,Brandhuber:2023hhy}), showed that (for positive graviton energies), at least at tree-level, $\mathcal{K}_{\mathcal{R}}^{\text{cl}, \mu_1 \nu_1 \mu_2 \nu_2} |_{\mathcal{O}(\kappa^{2})} = 0$. This was also recently confirmed by a full calculation of the radiated momentum at 4PM, which does not receive classical contributions from the double graviton insertion \cite{Damgaard:2023ttc}. Therefore, following the conjectural exponentiation of the radiative S-matrix \cite{Cristofoli:2021jas,DiVecchia:2022nna,Britto:2021pud} which we can rephrase in our language as the constraint\footnote{It is worth noticing that the 6-point recursion relation \eqref{eq:recursion6pt-positive-cl} and the analogous higher-point recursion relations are directly solved by a coherent state operator under the constraint \eqref{eq:conjecture-exp}.}
\begin{align}
\mathcal{K}_{\mathcal{R}}^{\text{cl}, \mu_1 \nu_1 \dots \mu_N \nu_N} \stackrel{?}{=} 0 \qquad N \geq 2\,, \qquad  E_{\mathbf{k}_1^{\prime}}>0, \dots, E_{\mathbf{k}_N^{\prime}}>0 \,,
\label{eq:conjecture-exp}
\end{align}
one would then find for positive energy gravitons a coherent state operator 
\begin{align}
\label{eq:exp_Smatrix-final}
\hat{S}^{\text{cl}} \overset{?}{=}  e^{\mathcal{\widetilde{K}}^{\text{cl}}(p_A,p_B;b_1,b_2)}  \exp \left[\sum_{\sigma_1} \int \mathrm{d} \Phi(k_1^{\prime})  \hat{\alpha}^{\text{cl}}_{5, \mathcal{R}}(k_1^{\prime}) \right]\,.
\end{align}
It would thus be interesting to establish whether \eqref{eq:conjecture-exp} holds more rigorously.

The structure \eqref{eq:conjecture-exp} shares similarities with eikonal resummation, but as explained earlier the classical subtraction terms are different than in~\cite{Cristofoli:2021jas,DiVecchia:2022nna}. One advantage is that we expect this to be a more natural basis: for the conservative case there is a direct relation with the radial action~\cite{Kol:2021jjc,Damgaard:2021ipf,Adamo:2022ooq,Brandhuber:2023hhy}, and this provides the most direct extension to the radiative case. For example, it was shown in \cite{Herderschee:2023fxh} (see also \cite{Brandhuber:2023hhy,Georgoudis:2023lgf,Elkhidir:2023dco,Georgoudis:2023eke}) that superclassical terms in the HEFT formalism are irrelevant for the waveform calculation, and the representation ~\eqref{eq:exp_Smatrix-final} makes it manifest. Interestingly, two-massive particle reducible terms can still enter into the final calculation for observables like the waveform which depend not only linearly but also quadratically from the amplitude (and therefore from the classical kernels) through the cut contribution \cite{Caron-Huot:2023vxl}. We leave further discussion of this point to a future analysis: in this paper, we will only work with the tree-level radiative kernel $\mathcal{K}_{\mathcal{R}}^{\text{cl}}$.

\section{Analytic continuation of the Post-Minkowskian waveform}
\label{sec:waveform_B2B}

The example of a linearized Schwarzschild background in Section~\ref{sec:bound_wavefunction} suggests a natural analytic continuation for two-body observables in terms of the binding energy (or equivalently, the rapidity). As shown in Section~\ref{sec:theory-intro}, the need to take residues on the bound state poles in~\eqref{eq:B2B-wavefunction2} came from superclassical iterations, which is avoided if we focus on the classical kernel $\mathcal{K}_{\mathcal{R}}^\text{cl}$ appearing in the exponent of the S-matrix. In this section, we formulate the analytic continuation of the PM tree-level gravitational waveform in the \emph{full} two-body problem, verifying the result through comparison of its PN expansion with the multipoles obtained via the quasi-Keplerian parametrization.

Following~\cite{Cristofoli:2021vyo}, we define the helicity-dependent spectral waveform $W(b_1,b_2,k^{\prime \pm})$, which is entirely determined to tree-level order by the radiative 2MPI kernel\footnote{This equation generalizes to higher loops by including the appropriate KMOC subtraction terms, but we will not discuss loop contributions here. The one-loop waveform calculation has been tackled in~\cite{Brandhuber:2023hhy,Georgoudis:2023lgf,Herderschee:2023fxh,Elkhidir:2023dco}, but some missing contributions were pointed out in~\cite{Caron-Huot:2023vxl}. The one-loop calculations have been corrected in the latest arXiv versions of \cite{Brandhuber:2023hhy,Herderschee:2023fxh} and in the recent works \cite{Bohnenblust:2023qmy,Georgoudis:2023eke,Georgoudis:2023ozp}. It would be interesting to study those results further, given the appearance of tail effects in the next-to-leading PM waveform.}
\begin{align}
W(b_1,b_2,k^{\prime \sigma}) &= \int \mathrm{\hat{d}}^4 q_1 \, \mathrm{\hat{d}}^4 q_2  \, \hat{\delta}(2 p_A \cdot q_1) \hat{\delta}(2 p_B \cdot q_2)\, \e^{i (q_1 \cdot b_1+q_2 \cdot b_2) / \hbar} \mathcal{K}^{\text{cl}}_{5,\mathcal{R}}(q_1,q_2,k^{\prime \sigma}) \,.
\label{eq:spectral_waveform}
\end{align}
In terms of the spectral waveform, the strain at future null infinity, $\mathscr{I}^+$, is given by
\begin{align}
h^{>}(x)  = \frac{\kappa}{4 \pi r}  \int_0^{+\infty} \hat{\mathrm{d}} \omega^{\prime}  \, \Big(W(b_1,b_2,k^{\prime -})\, \e^{-i \omega^{\prime} u} + W(b_1,b_2,k^{\prime +})\, \e^{+i \omega^{\prime} u}\Big)\,,
\label{eq:waveform-KMOC}
\end{align}
where the external graviton momentum is related to the observer location $n=(1,\hat{n})$ via $k^{\prime \, \mu} = \hbar \omega^{\prime} n^{\mu}$, $u = t - r$ is the retarded time, and $r$ is the asymptotic distance of the observer from the two-body system. We note that only the relative difference of the impact parameters $b^{\mu} := b_1^{\mu} - b_2^{\mu}$ is relevant in~\eqref{eq:spectral_waveform}, and at tree-level this is connected to the orbital angular momentum $L$\footnote{At higher orders, there is a more subtle connection between the incoming impact parameter (related to $L$) and the eikonal impact parameters $b_1,b_2$ relevant for the exponentiation in the HEFT variables~\cite{Bini:2023fiz,Georgoudis:2023eke}.} by
\begin{align}
b^{\mu} = (0, \mathbf{b})\,, \qquad |\mathbf{b}|= \frac{L}{P_{\infty}} = \frac{E}{m_A m_B} \frac{L}{\sqrt{y^2 - 1}} \,.
\label{eq:impact_parameter}
\end{align}
Throughout the remainder of this section we adopt the following convenient null tetrad at future null infinity $\mathscr{I}^+$ to paramterize the waveform:
\begin{align}
n^{\mu} = (1, \hat{n}) \,, \quad l^{\mu} \,, \quad m^{\mu} = \varepsilon^{\mu}_{+}(\hat{n}) \,, \quad \bar{m}^{\mu} = (\varepsilon^{\mu}_{+}(\hat{n}))^* = \varepsilon^{\mu}_{-}(\hat{n}) \,,
\label{eq:nulltet}
\end{align}
where $m^{\mu}$, $\bar{m}^{\mu}$ are null vectors normalized by $m \cdot \bar{m} = -1$ while $l^{\mu}$ is an auxiliary null vector satisfying $l \cdot m = l \cdot \bar{m} = 0$ and $n \cdot l = 1$. Using spherical coordinates $(\theta,\phi)$ on the celestial sphere, the spatial components of the tetrad are
\begin{align}
\mathbf{n} &= \big(\sin(\theta) \sin(\phi),\sin(\theta) \cos(\phi), \cos(\theta)\big) \,, \\
\mathbf{m} = \frac{1}{\sqrt{2}} \bigg(\partial_{\theta} \mathbf{n}(\theta,\phi) + &\frac{i}{\sin(\theta)} \partial_{\phi} \mathbf{n}(\theta,\phi)\bigg) \,, \qquad \mathbf{\bar{m}} = \frac{1}{\sqrt{2}} \bigg(\partial_{\theta} \mathbf{n}(\theta,\phi) - \frac{i}{\sin(\theta)} \partial_{\phi} \mathbf{n}(\theta,\phi)\bigg) \,, \nonumber
\end{align}
with the time-components fixed by the tetrad normalization conditions.
 
\subsection{Static and dynamical contributions to the tree-level waveform}
To proceed with the discussion of the waveform, it is necessary to understand the relevance of a decomposition into \emph{static} and \emph{dynamical} terms, along with the relation to the choice of BMS frame for the supertranslation charge.

In Section~\ref{sec:2MPI_kernels}, we derived expression \eqref{eq:5-ptexp} for the 5-point amplitude in terms of the 2MPI radiative kernel $\mathcal{K}_{5,\mathcal{R}}^{\text{cl}}$, valid for the emission of positive energy gravitons, which we identify as a \emph{dynamical} contribution. More generally, $\mathcal{K}_{\mathcal{R}}^{\text{cl}}$ also receives contributions from zero-energy graviton emissions, henceforth called \emph{static}, which are related to the soft behaviour in the $k^{\prime} \to 0$ limit.  With the help of a low-energy cutoff (see e.g.~\cite{DiVecchia:2022piu}), we could formally write
\begin{align}
\mathcal{K}^{\text{cl}}_{5,\mathcal{R}}(p_1, p_2;p_1^{\prime}, p_2^{\prime}, k^{\prime \sigma})
\equiv
\mathcal{K}_{5,\mathcal{R}}^{\text{cl},\text{dyn}}(p_1, p_2;p_1^{\prime}, p_2^{\prime}, k^{\prime  \sigma})
+
\mathcal{K}_{5,\mathcal{R}}^{\text{cl},\text{stat}}(p_1, p_2;p_1^{\prime}, p_2^{\prime}, k^{\prime \sigma})  \,,
\end{align}
though here we identify the static contributions as the (exact) soft limit $k^{\prime} \to 0$ of the corresponding amplitude:
\begin{align}
\mathcal{K}_{5,\mathcal{R}}^{\text{cl},\text{stat}}(p_1, p_2;p_1^{\prime}, p_2^{\prime}, k^{\prime \sigma}) \equiv \lim _{E_{\mathbf{k}^{\prime}} \rightarrow 0}  \mathcal{K}_{5,\mathcal{R}}^{\text{cl}}(p_1, p_2;p_1^{\prime}, p_2^{\prime}, k^{\prime \sigma})  \,.
\end{align}
We observe that to solve the (classical version of the) 5-point recursion relation in Fig.~\ref{fig:recursion5pt} in the zero-energy limit, we can formally redefine our \emph{static} radiative kernel to include the disconnected 5-point contribution
\begin{align}
\mathcal{K}_{5,\mathcal{R}}^{\text{cl},\text{stat}} \to \mathcal{K}_{5,\mathcal{R}}^{\text{cl},\text{stat}} + \left\langle k^{\prime \sigma} p_1^{\prime}|T| p_1\right\rangle \delta_{\Phi}\left(p_2, p_2^{\prime}\right)+\left\langle k^{\prime \sigma} p_2^{\prime}|T| p_2\right\rangle \delta_{\Phi}\left(p_1, p_1^{\prime}\right) \,.
\label{eq:static_kernel}
\end{align}
Combining this with equation~\eqref{eq:spectral_waveform} shows that the waveform receives contribution both at order $G_N$ and $G_N^2$.

First consider the \emph{disconnected} 5-point contribution at order $\mathcal{O}(G_N)$. This term was first computed in~\cite{Strominger:2014pwa} (see also~\cite{Damour:2020tta}), and later reproduced by worldline~\cite{Jakobsen:2021smu,Mougiakakos:2021ckm} and amplitude methods in~\cite{DiVecchia:2022owy} and section 4.6 of~\cite{Gonzo:2022rfk}. Following the latter, since the energy of the graviton has to be exactly zero because of the on-shell 3-point kinematics, Weinberg's soft graviton theorem~\cite{Weinberg:1965nx} can be applied to find, at order $\kappa$,
\begin{align}
\label{eq:kernel-soft}
\mathcal{K}_{5,\mathcal{R}}^{\text{cl},\text{stat}}(p_1, p_2;p_1^{\prime}, p_2^{\prime}, k^{\prime \sigma})
& =
\lim _{E_{\mathbf{k}^{\prime}} \rightarrow 0}
\frac{\kappa}{2} \bigg(\frac{\varepsilon_{\mu \nu}^{\sigma}\left(\hat{n}\right) p_1^{\mu} p_1^{\nu}}{k^{\prime} \cdot p_1-i \epsilon}-\frac{\varepsilon_{\mu \nu}^{\sigma}\left(\hat{n}\right) p_1^{\prime \mu} p_1^{\prime \nu}}{k^{\prime} \cdot p_1^{\prime}+i \epsilon}\bigg) \delta_{\Phi}\left(p_2, p_2^{\prime}\right) + (1 \leftrightarrow 2) \nonumber \\
& = i\, \frac{\kappa}{2} \varepsilon_{\mu \nu}^{\sigma}\left(\hat{n}\right) p_1^{\mu} p_1^{\nu}\, \hat{\delta}\left(k^{\prime} \cdot p_1\right) \delta_{\Phi}\left(p_2, p_2^{\prime}\right) + (1 \leftrightarrow 2) \,, 
\end{align}
where the $\hat{\delta}\left(k^{\prime} \cdot p_j\right)$ is the expected distributional support for zero-energy contributions. We identify this contribution as static, using the language introduced earlier. The corresponding $\mathcal{O}(G_N)$ contribution to the strain is
\begin{align}
h^{> \text{stat}}(u,\hat{n}) \Big|_{\mathcal{O}(G_N)}
= \frac{4 G_N}{r} \left[\frac{m_A^2 (\bar{m} \cdot v_A)^2 }{E_1 - \mathbf{p}_1 \cdot \hat{n}} + \frac{m_B^2 (\bar{m} \cdot v_B)^2 }{E_2 - \mathbf{p}_2 \cdot \hat{n}} \right] \,,
\label{eq:linear-memory}
\end{align}
consistent with~\cite{Damour:2020tta,DiVecchia:2022owy}. This entirely static contribution  can be directly related to the action of a BMS supertranslation as discussed in~\cite{Strominger:2014pwa}. 

A similar static contribution arises through the application of the soft theorem to the \emph{connected} 5-point tree-amplitude:
\begin{align}
\label{eq:kernel-soft2}
\mathcal{K}_{5,\mathcal{R}}^{\text{cl},\text{stat}}(p_1, p_2;&p_1^{\prime}, p_2^{\prime}, k^{\prime \sigma})  \nonumber \\
&= \lim _{E_{\mathbf{k}^{\prime}} \rightarrow 0} \frac{\kappa}{2}  \sum_{j=1}^2 \left(\frac{(\varepsilon^{\sigma}(\hat{n}) \cdot p_j)^2}{k^{\prime} \cdot p_j-i \epsilon}-\frac{(\varepsilon^{\sigma}(\hat{n}) \cdot p_j^{\prime})^2}{k^{\prime} \cdot p_j^{\prime}+i \epsilon}\right) \mathcal{M}^{(0)\text{cl}}_4(p_1, p_2;p_1^{\prime}, p_2^{\prime}) \,. 
\end{align}
Combining \eqref{eq:kernel-soft2} with the dynamical contributions to the 2MPI kernel given by the tree-level connected 5-point amplitude $\mathcal{M}^{(0)\text{cl}}_5(p_1, p_2;p_1^{\prime}, p_2^{\prime}, k^{\prime})$ for $E_{\mathbf{k}^{\prime}} > 0$, we recover the Kovacs-Thorne waveform \cite{Peters:1970mx,Kovacs:1977uw,Kovacs:1978eu,Jakobsen:2021smu} which can be written in a covariant form \cite{DeAngelis:2023lvf}
\begin{align}
	\hspace{-14pt}h^{> \text{stat}+\text{dyn}}&(u,\hat{n}) \Big|_{\mathcal{O}(G_N^2)} = \frac{2 G_N^2 m_A m_B}{r \sqrt{-b^2}} \frac{1}{w_1^2 w_2^2 \sqrt{1+T_{2}^2} \left(y+\sqrt{\left(1+T_{1}^2\right) \left(1+T_{2}^2\right)}+T_{1} T_{2}\right)} \nonumber \\
	&\times \Bigg( \frac{3 w_1 + 2y\left(2 T_1 T_2 w_1-T_2^2 w_2+w_2\right)-\left(2 y^2-1\right) w_1}{y^2-1} f_{1,2}^2  +4 \left(1+T_2^2\right) w_2 f_1 f_2\nonumber \\
	&\quad-\frac{4yT_2 w_2f_1+2\left(2 y^2-1\right) \left[T_1 \left(1+T_2^2\right) w_2 f_1+T_2 (T_1 T_2 w_1+w_2)f_2\right]}{ \sqrt{y^2-1}} f_{1,2} \nonumber \\
	&\quad-4 y \left(1+T_2^2\right) w_2 \left(f_1^2+f_2^2\right) +2 \left(2 y^2-1\right) \left(1+2 T_2^2\right) w_2 f_1 f_2 \Bigg) + \left(1\leftrightarrow 2\right)\,,
 \label{eq:KT_waveform}
\end{align}
where the variables $w_1,w_2,T_1,T_2$ are defined as:
\begin{align}
	w_1 = v_A \cdot n\,,\qquad  w_2 = v_B \cdot n\,, \qquad T_i = \frac{\sqrt{y^2-1} \, (u-b_i\cdot n)}{\sqrt{-(b_1-b_2)^2} \, w_i}\,,
\end{align}
and $f_1$,$f_2$ and $f_{1,2}$ as
\begin{align}
f_{1} = &(\tilde{b} \cdot \bar{m})\, (v_A \cdot n) - (\tilde{b} \cdot n)\, (v_A \cdot \bar{m})\,, \qquad f_{2} = (\tilde{b} \cdot \bar{m})\, (v_B \cdot n) - (\tilde{b} \cdot n)\, (v_B \cdot \bar{m}) \, \nonumber \\
 	f_{1,2} &= (v_A \cdot \bar{m})\, (v_B \cdot n) - (v_A \cdot n)\, (v_B \cdot \bar{m}) \,, \qquad  \tilde{b}^{\mu} = \frac{b_1^{\mu} - b_2^{\mu}}{\sqrt{-(b_1 - b_2)^2}}\,,
\end{align}
keeping in mind that $\bar{m}^{\mu}$ arise from the choice of emitted graviton helicity \eqref{eq:nulltet}.

Before making contact with the corresponding waveform for eccentric bound orbits, we must first clarify the role of static and dynamical contributions to the tree-level waveform. Observe that the static contributions generically give rise to $u$-independent terms, which are essentially determined by the BMS frame chosen in our calculation. This suggests that any relation with the analogous contribution for bound orbits will be subtle: for example, it is clear that at order $G_N$ the ``linear memory'' effect in~\eqref{eq:linear-memory} should not be present for bound orbits, given that the system returns to its original configuration after a period~\cite{Favata:2011qi}.

Consequently, we will restrict our attention to dynamical waveform contributions alone; these -- as we will see below -- can be related to the instantaneous trajectory of the relative motion. Terms such as linear memory, above, which depend on the full dynamical history, would require a different analysis.

\medskip
At this point, inspired by the wavefunction result \eqref{eq:B2B-wavefunction2}, we conjecture the following Post-Minkowskian scattering-to-bound map for the (complex) strain defined in~\eqref{eq:waveform-KMOC}:
\begin{align}
    h^{< \text{dyn}}\!\left(u,\hat{n};\sqrt{1-y^2},L\right) = h^{> \text{dyn}}\!\left(u,\hat{n};\sqrt{y^2 - 1} \to i \sqrt{1- y^2},L\right) \,, \qquad \mathcal{E} < 0\,,
    \label{eq:B2Bwaveform}
\end{align}
in which the rapidity $y$ is related to the incoming energy $E$ through
\begin{align}
    y = v_A \cdot v_B = \frac{E^2 - m_A^2 - m_B^2}{2 m_A m_B} \,,
    \label{eq:rapidity}
\end{align}
and we remind the reader that the impact parameters $b_1$,$b_2$ appearing in $h^>$ need to be expressed in terms of the energy and the orbital angular momentum~\eqref{eq:impact_parameter}. In particular, as explained in Section~\ref{sec:theory-intro}, the quantities discussed here appear in the exponent of the amplitude. This fact, combined with the Fourier transform to impact parameter space, removes the Gamma-function poles present in the wavefunctions, so there is no need to take a residue as in \eqref{eq:B2B-wavefunction2}. Hence, the analytic continuation is implemented directly for the waveform, without the need to extract a residue.

In terms of the convenient variables 
\begin{align}
    p_{\infty} = \sqrt{y^2 - 1} \,, \qquad \tilde{p}_{\infty} = \sqrt{1-y^2}\,,
    \label{eq:def_pinf}
\end{align}
\eqref{eq:B2Bwaveform} can alternatively be written as 
\begin{align}
    \boxed{h^{< \text{dyn}}(u,\hat{n};\tilde{p}_{\infty},L) = h^{> \text{dyn}}(u,\hat{n};p_{\infty} = + i \tilde{p}_{\infty},L) \,, \qquad \mathcal{E} < 0\,.}
    \label{eq:B2Bwaveform2}
\end{align}
It is worth adding a clarifying comment here: the analytic continuation in \eqref{eq:B2Bwaveform2} is a \emph{geometric statement} about the (positive) branch cut in $y$ starting at $y=1$ and extending up to $y=+\infty$. We interpret it as holding also for other functions with a discontinuity around the threshold $y = 1$. Take for example $\sqrt{y-1}$; under the analytic continuation in~\eqref{eq:B2Bwaveform2} this should become $i \sqrt{1-y}$. We stress that for some of the waveforms we will discuss, which are only functions of the binding energy $\mathcal{E}$ and therefore of $p_{\infty}^2$, the other choice of branch leading to $p_{\infty} = - i \tilde{p}_{\infty}$ gives an equivalent answer. We will return to these points when we discuss the analytic continuation of fluxes in Section~\ref{sec:KMOC-bound}. 

We now test this proposal by checking it against the Post-Newtonian expansion of the tree-level dynamical waveform, in both the frequency and time domain. As such we will, for the next sections only, restore powers of the speed of light $c$. 

\subsection{A direct check: the PN expansion}

We will compare the Post-Newtonian expansion of the tree-level Post-Minkowskian waveform with the analogous result provided by the Multipolar Post-Minkowskian (MPM) formalism, following~\cite{Bini:2023fiz}. The latter has the advantage of being defined for both scattering and bound trajectories, which will be very convenient for checking our conjectured B2B map~\eqref{eq:B2Bwaveform2} for the waveform. 

In the MPM formalism, one usually chooses the incoming center-of-mass frame of the two-body system, for convenience. At tree-level, since radiation reaction effects are negligible for such frame choice \cite{Bini:2023fiz}, we define our time vector and impact parameters as 
\begin{align}
    e_0^{\mu} &= \frac{p_1^{\mu} + p_2^{\mu}}{E} \sim (1,\mathbf{0})\,, \\
    \frac{1}{E} \left(E_1 b^{\mu}_1 + E_2 b^{\mu}_2 \right) &= 0\,, \qquad b_1 \cdot e_0 = b_2 \cdot e_0 = 0\,.
\end{align}
Explicitly, our choice of velocities $v_A$, $v_B$ and impact parameters $b_1$,$b_2$, $b = b_1 - b_2$ is
\begin{align}
    v_A &= \frac{1}{m_A} (E_1,0,P_{\infty},0) \,,
    \qquad
    v_B = \frac{1}{m_B} (E_2,0,-P_{\infty},0) \,,
    \nonumber \\
    b_1^{\mu} &= b \Big(0,\frac{E_2}{E},0,0\Big) \,,
    \qquad
    \qquad b_2^{\mu} = - b \Big(0,\frac{E_1}{E},0,0\Big)  \,,
    \qquad
    \, b^{\mu} = (0,b,0,0) \,,
    \nonumber \\
    E_1 &= \frac{1}{m_A} (y m_B + m_A)\,,
    \qquad E_2 =\frac{1}{m_B} (y m_A + m_B)\,, \qquad E = E_1 + E_2\,.
    \label{eq:scattering_com}
\end{align}
The Post-Newtonian expansion is an expansion in $1/c$, which is achieved for the conservative case by defining 
\begin{align}
    y = \sqrt{1 + p_{\infty}^2}\,, \qquad j = \frac{L c}{G_N m_A m_B} \,,
    \label{eq:PN-def}
\end{align}
so that PN-counting is given by $p_{\infty} \propto \mathcal{O}(1/c)$ and $j \propto \mathcal{O}(c)$ with a finite Newtonian eccentricity
\begin{align}
    e_{\text{N}} = \sqrt{1 + p_{\infty}^2 j^2} \,. 
    \label{eq:eccentricity_Newton}
\end{align}
For the radiative case, the frequency of the emitted gravitational wave must scale as $\omega \sim p_{\infty} \propto \mathcal{O}(1/c)$ \cite{Bini:2023fiz}. This implies that, when performing the PN expansion of the time-domain waveform in powers of $p_{\infty}$, it is convenient to define a new time variable\footnote{The  physical motivation for this change of variable will become clear later by studying the analytic solution of the Kepler equation for the relative time-dependent trajectories, where the ($j-$rescaled) mean motion $\tilde{n}^{>}$ appears together with the time $t$.}
\begin{align}
    \tilde{u}^{>} := \frac{c p_{\infty}}{b} u \,,
    \label{eq:time_PN}
\end{align}
such that $\tilde{u}^{>}$ is kept fixed (together with the Newtonian eccentricity $e_{\text{N}}$). We then obtain 
\begin{align}
   h^{>}\left(u = \frac{b}{p_{\infty} c} \tilde{u}^{>} ,\theta,\phi\right) = \frac{4 G_N}{c^4} \left(W^{>}_{\text{N}} + \frac{1}{c} W^{>}_{\text{0.5PN}} + \frac{1}{c^2} W^{>}_{\text{1PN}} + \dots  \right)\,,
\end{align}
with some PN coefficients $W^{>}_{\text{N}}$, $W^{>}_{\text{0.5PN}}$, $W^{>}_{\text{1PN}}$, etc.

It suffices, to analyse the analytic continuation of the waveform, to focus only on the Newtonian term. To further simplify the result we consider the equatorial plane by setting $\theta = \pi/2$, which yields
\begin{align}
   W^{>\text{dyn}}_{\text{N}}(\tilde{u}^{>};p_{\infty},j) &= -\frac{m_A m_B p_{\infty}}{4 j \left[1+(\tilde{u}^{>})^2\right]^{3/2}} \Big[ \left((\tilde{u}^{>})^2 +3\right) \cos (2 \phi ) \nonumber \\
   & \qquad\qquad\qquad\qquad+ \left(1+(\tilde{u}^{>})^2\right) +2 \left((\tilde{u}^{>})^3+2 \tilde{u}^{>}\right) \sin (2 \phi )\Big] \,,
  \label{eq:Newtonian_time_hypwaveform}
\end{align}
where we have neglected static terms\footnote{There is also a static contribution
\begin{align}
   W^{>\text{stat}}_{\text{N}}(p_{\infty},j) = -\frac{m_A m_B p_{\infty}}{2 j} \sin(2 \phi) \,,
\end{align}
which we will recover later from the computation of the quadrupole with the hyperbolic trajectory. These terms can be included in the matching with the PN waveform, but we will not discuss them here.} and the time $\tilde{u}^{>}$ can be expressed in terms of Newtonian value of the ($j-$rescaled) mean motion $\tilde{n}^{>}$ as
\begin{align}
   \tilde{u}^{>} = \tilde{n}^{>}_{\text{N}}  u\,, \qquad \tilde{n}^{>}_{\text{N}} = \frac{c p_{\infty}}{b} \,.
   \label{eq:mean-motion-rescal}
\end{align}
We now show that we can recover this result within the post-Newtonian framework, given the identification of the PN coefficients with the even and odd parity $2^{\ell}$ radiative multipoles $U_{\ell}(u,\theta,\phi)$ and $V_{\ell}(u,\theta,\phi)$ defined in the MPM formalism\footnote{Note that the retarded time $T_r = u - 2 (G_N E /c^5) \log(r/(c t_0))$ which usually appears in the PN literature~\cite{Blanchet:2013haa} as the argument of multipole, after the matching with the source, includes a logarithmic shift proportional to the total energy of the system $E$ with an arbitrary time-scale $t_0$. Such a shift is irrelevant at tree-level, so we can safely approximate $T_r \sim u$ here.}, i.e. at tree-level \cite{Bini:2023fiz}
\begin{align}
    h^{>}(u,\theta,\phi) &= \frac{4 G_N}{c^4} \Big(U_2^{>}(u,\theta,\phi) + \frac{1}{c} (U_3^{>}(u,\theta,\phi) + V_2^{>}(u,\theta,\phi)) \nonumber \\
    & \qquad + \frac{1}{c^2} (U_4^{>}(u,\theta,\phi) + V_3^{>}(u,\theta,\phi)) + \dots \Big)\,,
    \label{eq:waveform_multipoles}
\end{align}
which are defined in terms of symmetric-trace-free (STF) Cartesian tensors of order $\ell$ in the 3d space (for the notation see \cite{Blanchet:2013haa} and references therein)
\begin{align}
    U_{\ell}(u,\theta,\phi) &= \left(\bar{m}^i \bar{m}^j\right)  \left[\frac{1}{l!} n^{i_1} n^{i_2} \dots n^{i_{l-2}} U_{i j i_1 i_2 \dots i_{l-2}}(u,\theta,\phi)\right] \,, \nonumber \\
    V_{\ell}(u,\theta,\phi) &= \left(\bar{m}^i \bar{m}^j\right) \left[ \frac{1}{l!} \frac{2 l}{l+1} n^{c} n^{i_1} n^{i_2} \dots n^{i_{l-2}} \epsilon_{c d i} V_{j d i_1 i_2 \dots i_{l-2}}(u,\theta,\phi) \right] \,.
\end{align}
We now focus on the quadrupole contribution, which at Newtonian order gives 
\begin{align}
\label{eq:quadrupole-def}
W^{>}_{\text{N}} &= U_2^{>} = \frac{1}{2!} \text{STF}_{i j} \frac{\mathrm{d}^2}{\mathrm{d} t^2} \left(\mu x^i x^j \right) \Bigg|_{t = u} + \mathcal{O}\left(\frac{1}{c^2}\right) \,,
\end{align}
where we work in the incoming center of mass system with the \emph{relative trajectory}
\begin{align}
x^i(t) = x^i_1(t) - x^i_2(t) = r \, n^i_{\text{CM}} = r(t) (\cos(\varphi(t)), \sin(\varphi(t)), 0) \,,
\end{align}
and $r(t)$, $\varphi(t)$ are the (dimensionless) quasi-Keplerian radius and phase of the equatorial orbit.  

The quasi-Keplerian parametrization of the relative motion is reviewed in detail in appendix \ref{app:quasiKepler}, but for our calculation here we need the simpler version
\begin{align}
r^{>} &= a^{>} (e^{>}_r \cosh(\mathsf{v}) - 1) \,, \nonumber \\
\varphi^{>} &= k^{>} \Theta^{>} + \mathcal{O}\left(\frac{1}{c^2}\right)\,, \nonumber \\
l^{>} &= n^{>} t =  e^{>}_t \sinh(\mathsf{v}) - \mathsf{v} + \mathcal{O}\left(\frac{1}{c^2}\right) \,, \label{eq:kepler1} \\
\Theta^{>} &= 2 \arctan \left(\sqrt{\frac{e^{>}_{\phi} + 1}{e^{>}_{\phi} - 1}} \tanh\left(\frac{\mathsf{v}}{2}\right)\right) + \mathcal{O}\left(\frac{1}{c^2}\right) \,,
\label{eq:orbital_elements_hyp_1PN}
\end{align}
which is a function of several variables we are now going to define. First, the trajectory is naturally parametrized by the (hyperbolic) eccentric anomaly $\mathsf{v}$ while the dependence on time $t$ is \emph{implicit} through the transcendental Kepler equation (\ref{eq:kepler1}), in which $n^{>}$ is the (hyperbolic) mean motion. It is thus difficult to find an exact map $\mathsf{v}=\mathsf{v}(t)$. Then, we have the semi-latus rectum $a^{>}$, the hyperbolic periastron precession $k^{>}$ and the three eccentricities $e^{>}_r$,$e^{>}_t$ and $e^{>}_{\phi}$ which have an explicit expression in terms of the symmetric mass ratio $\nu$, the orbital angular momentum $j$ and the binding energy $\mathcal{E}$\footnote{This is equivalent to the usual definition of $\bar{E} = \mathcal{E}$ in \cite{Bini:2017wfr}.}.

Given that from the post-Minkowskian approach we expect an explicit dependence on time, it must be possible to solve analytically for some time-dependent trajectory $x^i(t)$ in the PM expansion. This was shown for example in \cite{Wagoner1976} (see also section VI of \cite{Bini:2017wfr}), by expanding at large eccentricities the Kepler equation. We take here a similar path by expanding at large orbital angular momentum $j$, which has the advantage of being well-defined both the hyperbolic and the elliptic case. We then define the ($j$-rescaled) mean motion,
\begin{align}
\tilde{n}^{>} = \frac{n}{j p_{\infty}} \,,
\end{align}
which generalizes the Newtonian result in \eqref{eq:mean-motion-rescal}. We then observe that the Kepler equation 
\begin{align}
\tilde{n}^{>} t = \frac{1}{j p_{\infty}} \left[ e^{>}_t \sinh(\mathsf{v}) - \mathsf{v}  + \mathcal{O}\left(\frac{1}{c^2}\right) \right]\,,
\end{align}
can be solved perturbatively as an expansion in $1/j$, i.e. we have at Newtonian order
\begin{align}
\mathsf{v} = \arcsinh\left(\tilde{n}^{>} t\right) +  \frac{1}{j} \left[\frac{\arcsinh(\tilde{n}^{>} t)}{p_{\infty} \sqrt{1+(\tilde{n}^{>} t)^2}}\right] + \mathcal{O}\left(\frac{1}{j^2},\frac{1}{c^2}\right)\,.
\label{eq:eccentricity-anomaly_time}
\end{align}
This allows to obtain analytic time dependence for the orbital elements $r,\phi$ which will enter into the evaluation of the quadrupole. Indeed, by plugging \eqref{eq:eccentricity-anomaly_time} into \eqref{eq:orbital_elements_hyp} we get
\begin{align}
r^{>}(t) &= -\frac{1}{p_{\infty}^2}+\frac{\tilde{n}^{>}  t \arcsinh(\tilde{n}^{>}  t)}{p_{\infty}^2 \sqrt{(\tilde{n}^{>} t)^2+1}}+\frac{\left(1+2 p_{\infty}^2 j^2\right) \left((\tilde{n}^{>} t)^2+1\right)+\arcsinh^2(\tilde{n}^{>}  t)}{2 j p_{\infty}^{3/2} \sqrt{(\tilde{n}^{>} t)^2+1}} \,, \nonumber \\
\varphi^{>}(t) &= \arctan (\tilde{n}^{>} t)+\frac{\tilde{n}^{>} t \sqrt{1+(\tilde{n}^{>} t)^2}+\log \left(\sqrt{1+(\tilde{n}^{>} t)^2}+\tilde{n}^{>} t\right)}{p_{\infty} j \left(1+(\tilde{n}^{>} t)^2\right)}  \,.
\label{eq:hyperbolic_trajectories}
\end{align}
We now use \eqref{eq:hyperbolic_trajectories} to evaluate the quadrupole, which we can conveniently rewrite as 
\begin{align}
\hspace{-7pt}\frac{\mathrm{d} U_2^{>}}{\mathrm{d} t} = -\text{STF}_{i j} \left[\frac{G_N m_A m_B }{(r^{>}(t))^4} \left(4 \, r^{>}(t) x^i(t) \frac{\mathrm{d} x^j(t)}{\mathrm{d} t}-3 \frac{\mathrm{d} r^{>}(t)}{\mathrm{d} t} x^i(t) x^j(t) \right) \right] + \mathcal{O}\bigg(\frac{1}{c^2}\bigg) \,.
\end{align}
A direct calculation shows that, as anticipated, we recover \eqref{eq:Newtonian_time_hypwaveform}
\begin{align}
U_2^{>\text{dyn}}(u,\phi) = W^{>\text{dyn}}_{\text{N}}(u,\phi) \,,
\end{align}
where we have retained only time-dependent contributions when choosing the boundary conditions of the time integration\footnote{The contribution from the static terms, as mentioned earlier, is in agreement with the PN expansion of the tree-level waveform $U_2^{>\text{stat}}(u,\phi) = W^{>\text{stat}}_{\text{N}}(u,\phi)$ but we will not discuss this further in this paper.}. 

We can also check that, in Fourier domain, we recover equation (4.13) of~\cite{Bini:2023fiz}. We first define the waveform in frequency domain as 
\begin{align}
    \tilde{h}(\omega,\theta,\phi) = \int_{-\infty}^{+\infty} \mathrm{d} u \, \e^{i \omega u} h(u,\theta,\phi) \,. 
\end{align}
Then, we need to compute integrals of the form
\begin{align}
    \int_{-\infty}^{+\infty} \mathrm{d} \tilde{u} \, \frac{\tilde{u}^{p}\, \e^{i \omega \tilde{u}}}{\left(\tilde{u}^2+1\right)^{\beta}} \quad p \in \mathbb{N}_{>} \,,
\end{align}
which can be all easily obtained from the integral (valid for $\Re \beta > 0$)
\begin{align}
    \int_{-\infty}^{+\infty} \mathrm{d} \tilde{u} \, \frac{ \e^{i \omega \tilde{u}}}{\left(1+\tilde{u}^2\right)^{\beta}}  = \frac{\sqrt{\pi } 2^{\frac{3}{2}-\beta } \omega^{\beta -\frac{1}{2}} K_{\frac{1}{2}-\beta }(\omega)}{\Gamma (\beta )} \,,
\end{align}
and its $p$-order frequency derivatives. The result is
\begin{align}
U_2^{>\text{dyn}}(\Omega^{>},\theta=\pi/2,\phi) &= - \frac{G_N m_A m_B}{2 p_{\infty}} \Big[K_0(\Omega^{>}) ((\cos (2 \phi )+1)+2 i \Omega^{>} \sin (2 \phi ))\nonumber \\
& \qquad\qquad\qquad\qquad+K_1(\Omega^{>}) (2 \Omega^{>} \cos (2 \phi )+2 i \sin (2 \phi ))) \Big]\,,
\end{align}
where $K_0$,$K_1$ are modified Bessel functions of the second kind and $\Omega^{>}$ is defined as
\begin{align}
\Omega^{>}= \frac{\omega}{\tilde{n}^{>}_{\text{N}}} = \frac{\omega b}{c p_{\infty}} \,.
\end{align}
We now turn to the analogous calculation in the elliptic bound case, using the PN formalism.

\subsection{Boundary to bound dictionary of quasi-Keplerian orbits}

In a seminal work~\cite{Damour:1985}, Damour and DeRuelle found the first instance of an analytic continuation between the orbital elements defined in \eqref{eq:hyperbolic_trajectories} at 1PN, which was also recently revisited up to 3PN order with a modified prescription in \cite{Cho:2018upo}. The relation between Post-Newtonian waveforms for elliptic and hyperbolic orbits was developed at the level of the multipoles already early on in the PN literature, for example see the results of Junker and Sch\"afer \cite{Junker:1992kle} (and \cite{Wagoner1976,Turner1977,Blanchet:1989cu}) which we will discuss further in section \ref{sec:1PN_B2B}. Here we generalize the discussion by deriving a Post-Minkowskian map from tree-level scattering waveforms to bound waveforms in time-domain. The quasi-Keplerian parametrization for elliptic orbits is, up to 1PN, 
\begin{align}
r^{<} &= a^{<} (1 - e^{<}_r \cos(\mathsf{u})) \,, \nonumber \\
\varphi^{<} &= k^{<} \Theta^{<} + \mathcal{O}\left(\frac{1}{c^2}\right)\,, \nonumber \\
l^{<} &= n^{<} t =  \mathsf{u} - e^{<}_t \sin(\mathsf{u}) + \mathcal{O}\left(\frac{1}{c^2}\right) \,, \nonumber \\
\Theta^{<} &= 2 \arctan \bigg[\sqrt{\frac{e^{<}_{\phi} + 1}{1 - e^{<}_{\phi}}} \tan\left(\frac{\mathsf{u}}{2}\right)\bigg] + \mathcal{O}\bigg(\frac{1}{c^2}\bigg) \,,
\label{eq:orbital_elements_ell1PN}
\end{align}
where $\mathsf{u}$ here is the elliptic eccentric anomaly and the other variables are the elliptic analogue of \eqref{eq:orbital_elements_hyp}. As shown in appendix \ref{app:quasiKepler}, up to 1PN there is a straightforward analytic continuation between the orbital elements for the hyperbolic and the elliptic case
\begin{align}\label{B2Bmainbody}
     n^{>} &\to -i n^{<}\,, e_t^{>} \to e_t^{<}\,, e^{>}_r \to e_r^{<}\,, e_{\phi}^{>} \to e_{\phi}^{<}\,, \mathsf{v} \to i \mathsf{u} \,, a^{>} \to -a^{<}\,, k^{>} \to k^{<} \,,
    \end{align}
which can be completely understood in terms of the binding energy $\mathcal{E}$ \cite{Damour:1985}. 

Working at Newtonian order, we can solve the elliptic Kepler equation at large\footnote{The reader may note that this expansion in large (imaginary) eccentricity \eqref{eq:eccentricity_Newton} is unphysical for bound orbits. As we will show in Section~\ref{sec:1PN_B2B}, though,  the analytic continuation back to the physical regime where $0<e_t^{<}<1$ is unique and well-defined, at least up to 1PN.} $j$,
\begin{align}
i \mathsf{u} = -\arcsinh\left(\tilde{n}^{<} t\right) -  \frac{1}{j} \left[\frac{\arcsinh(\tilde{n}^{<} t)}{i \tilde{p}_{\infty} \sqrt{1+(\tilde{n}^{<} t)^2}}\right] + \mathcal{O}\left(\frac{1}{j^2},\frac{1}{c^2}\right)\,.
\label{eq:eccentricity-anomaly_time-ell}
\end{align}
which allows us to compute analytic form of the elliptic trajectory \eqref{eq:orbital_elements_ell1PN} 
\begin{align}
r^{<}(t) &= \frac{1}{\tilde{p}_{\infty}^2}-\frac{\tilde{n}^{<}  t \arcsinh(\tilde{n}^{<}  t)}{\tilde{p}_{\infty}^2 \sqrt{(\tilde{n}^{<} t)^2+1}}+i \frac{\left(1-2 \tilde{p}_{\infty}^2 j^2\right) \left((\tilde{n}^{<} t)^2+1\right)+\arcsinh^2(\tilde{n}^{<}  t)}{2 j \tilde{p}_{\infty}^{3/2} \sqrt{(\tilde{n}^{<} t)^2+1}} \,, \nonumber \\
\varphi^{<}(t) &= -\arctan (\tilde{n}^{<} t)+i \frac{\tilde{n}^{<} t \sqrt{1+(\tilde{n}^{<} t)^2}+\log \left(\sqrt{1+(\tilde{n}^{<} t)^2}+\tilde{n}^{<} t\right)}{\tilde{p}_{\infty} j \left(1+(\tilde{n}^{<} t)^2\right)}  \,.
\label{eq:elliptic_trajectories}
\end{align}
which means that the frequency-domain quadrupole \eqref{eq:quadrupole-def} is
\begin{align}
U_2^{<\text{dyn}}(\Omega^{<},\theta=\pi/2,\phi)
&= -i \frac{G_N m_A m_B }{2 \tilde{p}_{\infty}}
\Big[K_0(\Omega^{<})-2 i (\Omega^{<} K_0(\Omega^{<})+K_1(\Omega^{<})) \sin (2 \phi )\nonumber \\
& \qquad \qquad \qquad +(K_0(\Omega^{<})+2 \Omega^{<} K_1(\Omega^{<})) \cos (2 \phi )\Big]\,,
\end{align}
where the modified Bessel functions are functions of the positive frequency $\Omega^{<}$
\begin{align}
\Omega^{<} = -\Omega^{>} = \frac{\omega L}{c^2 \mu \tilde{p}_{\infty}^2}\,.
\end{align}
In time domain, we obtain the following expression
\begin{align}
  W^{<\text{dyn}}_{\text{N}}(\tilde{u}^{<};\tilde{p}_{\infty},j) &=-i \frac{m_A m_B \tilde{p}_{\infty}}{4 j \left[1+(\tilde{u}^{<})^2\right]^{3/2}} \Big[ \left((\tilde{u}^{<})^2 +3\right) \cos (2 \phi ) \nonumber \\
  & \qquad\qquad\qquad\qquad+ \left(1+(\tilde{u}^{<})^2\right)  -2 \left((\tilde{u}^{<})^3+2 \tilde{u}^{<}\right) \sin (2 \phi )\Big] \,,
  \label{eq:Newtonian_time_ellwaveform}
\end{align}
where $\tilde{u}^{<}$ is now related to the Newtonian value of the $j$-rescaled (elliptic) mean motion
\begin{align}
  \tilde{u}^{<} = \tilde{n}^{<}_{\text{N}} u\,, \qquad \tilde{n}^{<}_{\text{N}} = - \tilde{n}^{>}_{\text{N}} = \frac{c^2 \mu \tilde{p}_{\infty}^2}{L}\,.
\end{align}
The final result is the following remarkable map
\begin{align}
  W^{<\text{dyn}}_{\text{N}}(u;\tilde{p}_{\infty},j) &= W^{>\text{dyn}}_{\text{N}}(u;p_{\infty} = + i \tilde{p}_{\infty},j) \,, \quad \mathcal{E} < 0 \,.
\end{align}
We now discuss how the generic PN bound waveform at tree-level order arise explicitly from the expansion of the analytical continuation \eqref{eq:B2Bwaveform} of the Post-Minkowskian result in \eqref{eq:KT_waveform}, working for simplicity in the center of mass frame. We first perform the change of variable to the time $\tilde{u}^{<}$ by setting\footnote{Notice that, compared to \eqref{eq:time_PN}, there is an overall sign in the definition of $\tilde{u}^{<}$ compared to $\tilde{u}^{>}$ since we want the change of variable to be positive definite in the bound kinematic region.}
\begin{align}
  u = \tilde{u}^{<} \frac{L E}{m_A m_B \tilde{p}_{\infty}^2 c^2}\,,
  \label{eq:time_PN_bound}
\end{align}
and we then evaluate all the scalar products with the bound analogues of \eqref{eq:scattering_com}, i.e.
\begin{align}
    v_A &= \frac{1}{m_A} (E_1,0,i \tilde{P}_{\infty},0) \,,
    \qquad
    v_B = \frac{1}{m_B} (E_2,0,-i \tilde{P}_{\infty},0) \,,  \quad b^{\mu} = \frac{L E}{m_A m_B (i \tilde{p}_{\infty})} (0,1,0,0)\,,
    \nonumber \\
     b_1^{\mu} &= \frac{L E}{m_A m_B (i \tilde{p}_{\infty})} \Big(0,\frac{E_2}{E},0,0\Big) \,,
     \qquad
     b_2^{\mu} = - \frac{L E}{m_A m_B (i \tilde{p}_{\infty})} \Big(0,\frac{E_1}{E},0,0\Big)  \,,
    \label{eq:bound_com} \\
    E_1 & = \frac{1}{m_A} (\sqrt{1-\tilde{p}_{\infty}^2} m_B + m_A)\,, \qquad E_2 = \frac{1}{m_B} (\sqrt{1-\tilde{p}_{\infty}^2} m_A + m_B)\,, \qquad E = E_1 + E_2\,. \nonumber
\end{align}
Equipped with \eqref{eq:time_PN_bound} and \eqref{eq:bound_com}, we can finally expand in powers of $\tilde{p}_{\infty}$ to obtain 
\begin{figure}[t!]
\centering
\begin{minipage}{.5\textwidth}
  \centering
  \includegraphics[width=\linewidth]{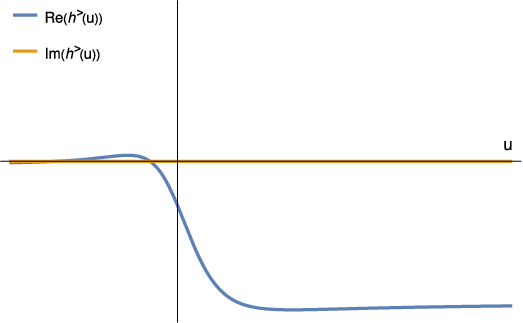}
\end{minipage}%
\begin{minipage}{.5\textwidth}
  \centering
  \includegraphics[width=\linewidth]{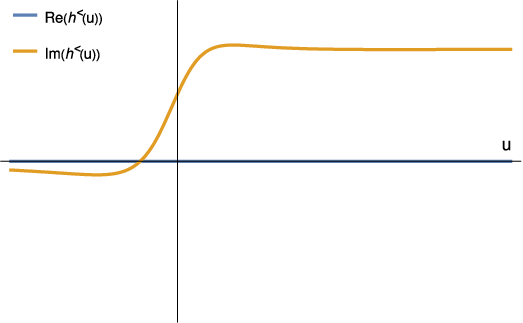}
\end{minipage}
\caption{A comparison between the dynamical tree-level scattering (left) and bound (right) waveforms in the center-of-mass frame and on the equatorial plane, where we have adopted the proposed, one-sided, analytic continuation in energy. As discussed in Sec.~\ref{sec:1PN_B2B} the lack of periodicity in the bound case requires a resummation. 
}
\label{fig:waveform_B2B}
\end{figure}
\begin{align}
   h^{<\text{dyn}}\left(\tilde{u}^{<} \frac{L E}{m_A m_B \tilde{p}_{\infty}^2 c^2},\theta,\phi\right) = \frac{4 G_N}{c^4} \left(W^{<\text{dyn}}_{\text{N}} + \frac{1}{c} W^{<\text{dyn}}_{\text{0.5PN}} + \frac{1}{c^2} W^{<\text{dyn}}_{\text{1PN}} + \dots  \right)\,,
\end{align}
\noindent where now $W^{<\text{dyn}}_{\text{N}}$ agrees exactly with \eqref{eq:Newtonian_time_ellwaveform}.

Examples of the scattering and bound tree-level dynamical waveforms $h^{\text{dyn}}(u)$ are given in Fig.~\ref{fig:waveform_B2B}. A question which immediately arises is: why does the bound waveform fail to be periodic? This is an artefact of the large eccentricity expansion used to truncate Kepler's equation which, while natural for scattering orbits, is unphysical in the bound case. This suggests that to convert the PM scattering waveforms into reasonable bound waveforms a resummation is required. To clarify and resolve this, we turn to, and exploit, the full PN results in the Newtonian limit. 

\subsection{1PN scattering and bound waveforms: effective resummation}
\label{sec:1PN_B2B}

We now compute the Newtonian waveform directly, without performing the large eccentricity Post-Minkowskian expansion. For the scattering case, using \eqref{eq:orbital_elements_hyp_1PN} in the center of mass frame and restricting to the equatorial plane $\theta=\pi/2$, we obtain
\begin{align}
   h_{\text{N}}^{>}\left(\mathsf{v}(u),\frac{\pi}{2},\phi\right) & = \frac{4 G_N}{c^4} \frac{G_N m_A m_B \mathcal{E}}{4 (e_\text{N}\cosh (\mathsf{v})-1)^2} \Big[   \left(4-3 e_{\text{N}}^2\right) \cos (2 \phi -2 \varphi^{>}_\text{N}(\mathsf{v})) \nonumber \\ 
   & +e_\text{N}\Big(4 \sqrt{e_\text{N}^2-1} \sinh (\mathsf{v}) \sin (2 \phi -2 \varphi^{>}_\text{N}(\mathsf{v}))+e_\text{N}\cosh (2 \mathsf{v}) \cos (2 \phi -2 \varphi^{>}_\text{N}(\mathsf{v})) \nonumber \\
   & -2 e_\text{N}\cosh ^2(\mathsf{v})+4 \cosh (\mathsf{v}) \sin ^2(\phi -\varphi^{>}_\text{N}(\mathsf{v}))\Big)\Big]\,,
   \label{eq:Newtonianhyp}
\end{align}
where $e_\text{N}$ is the Newtonian eccentricity in \eqref{eq:eccentricity_Newton},
\begin{align}
   \varphi^{>}_\text{N}(\mathsf{v}) = 2 \arctan\bigg[\sqrt{\frac{e_\text{N}+1}{e_\text{N}-1}} \tanh \left(\frac{\mathsf{v}}{2}\right)\bigg]
\end{align}
and the hyperbolic Kepler equation at Newtonian level becomes
\begin{align}
   u = \frac{1}{n^{>}_\text{N}} \left[ e_\text{N} \sinh(\mathsf{v}) - \mathsf{v} \right]\,.
   \label{eq:Kepler_hyp_Newton}
\end{align}
For the elliptic case, we obtain instead
\begin{align}
   h_{\text{N}}^{<}\left(\mathsf{u}(u),\frac{\pi}{2},\phi\right) & = \frac{4 G_N}{c^4} \frac{G_N m_A m_B \mathcal{E}}{4 (e_\text{N} \cos (\mathsf{u})-1)^2} \Big[ \left(4-3 e_\text{N}^2\right) \cos (2 \phi -2 \varphi^{<}_\text{N}(\mathsf{u}))    \nonumber \\ 
   &   +e_\text{N}\Big(4 \sqrt{1-e_\text{N}^2} \sin (\mathsf{u}) \sin (2 \phi -2 \varphi^{<}_\text{N}(\mathsf{u}))+e_\text{N}\cos (2 \mathsf{u}) \cos (2 \phi -2 \varphi^{<}_\text{N}(\mathsf{u})) \nonumber \\
   & -2 e_\text{N}\cos ^2(\mathsf{u})+4 \cos (\mathsf{u}) \sin ^2(\phi -\varphi^{<}_\text{N}(\mathsf{u}))\Big)\Big]\,,
   \label{eq:Newtonianell}
\end{align}
where 
\begin{align}
   \varphi^{<}_\text{N}(\mathsf{u}) = 2 \arctan\bigg[\sqrt{\frac{e_\text{N}+1}{1-e_\text{N}}} \tan \left(\frac{\mathsf{u}}{2}\right)\bigg] \;,
\end{align}
and the elliptic Kepler equation at Newtonian level becomes
\begin{align}
   u = \frac{1}{n^{<}_\text{N}} \left[\mathsf{u} - e_\text{N} \sin(\mathsf{u})\right]\,.
   \label{eq:Kepler_ell_Newton}
\end{align}
We first emphasise that under the analytic continuation \eqref{B2Bmainbody} the scattering and bound Newtonian waveforms are directly mapped to each other, together with the corresponding Kepler equations \eqref{eq:Kepler_hyp_Newton} and \eqref{eq:Kepler_ell_Newton}. Next, we have checked that~\eqref{eq:eccentricity-anomaly_time} and \eqref{eq:eccentricity-anomaly_time-ell} recover our previous results \eqref{eq:Newtonian_time_hypwaveform} and \eqref{eq:Newtonian_time_ellwaveform} upon using the large angular momentum expansion to make \emph{analytic} contact with the PM waveform. 

\begin{figure}[t!]
\centering
\begin{minipage}{.5\textwidth}
  \centering
  \includegraphics[width=\linewidth]{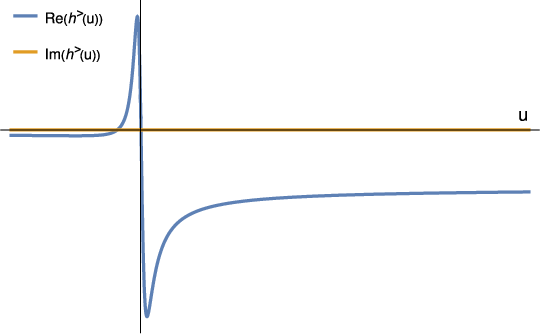}
\end{minipage}%
\begin{minipage}{.5\textwidth}
  \centering
  \includegraphics[width=\linewidth]{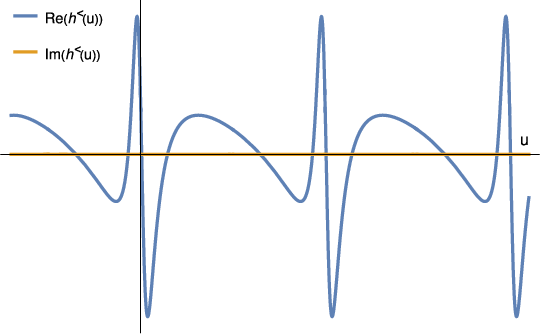}
\end{minipage}
\caption{A comparison between the Newtonian scattering (left) and bound (right) waveforms in the center-of-mass frame, without performing the large eccentricity expansion.}
\label{fig:waveform_Newtonian_B2B}
\end{figure}

Comparing these familiar Newtonian waveforms in Fig.~\ref{fig:waveform_Newtonian_B2B} with their large eccentricity counterparts in Fig.\ref{fig:waveform_B2B}, we see that the expected periodic behaviour in the bound case is restored. The difference lies entirely in how the Kepler equation is treated; solving this numerically is effectively resumming the large-eccentricity expansion, such that periodic behaviour is restored in the bound case. If analytic results are required, then it calls for the development of methods by which to more explicitly resum PM waveforms, similar to the Firsov resummation presented in~\cite{Kalin:2019rwq} for calculations of binding energy in the conservative case or to the inclusion of PN contributions in the fluxes~\cite{Saketh:2021sri,Cho:2021arx}. We leave this very interesting problem to future work.

We comment also on the 1PN waveform calculation by Junker and Sch\"afer \cite{Junker:1992kle}, where the 1PN multipoles \eqref{eq:waveform_multipoles} were computed with the quasi-Keplerian parametrization used in this work. Although it was never explicitly observed in that paper, we have checked that the multipoles for bound and scattering orbits provided in their eq.(58-76) and eq.(79-97), as a function of the eccentric anomalies $\mathsf{u}$ and $\mathsf{v}$, do indeed map into each under \eqref{B2Bmainbody}. Given that the corresponding Kepler equations in equations (78) and (100) of \cite{Junker:1992kle} are also mapped into each other, we can claim that up to 1PN order we have 
\begin{align}
   \label{eq:B2Bwaveform3}
    h^{< \text{dyn}}_{\text{1PN}}(\mathsf{u}(u),\hat{n};\tilde{p}_{\infty},L) = h^{> \text{dyn}}_{\text{1PN}}(\mathsf{v}(u),\hat{n};+ i \tilde{p}_{\infty},L) \,, \qquad \mathcal{E} < 0\,,
\end{align}
which is valid independent of the PM expansion (i.e.~also beyond tree-level). This raises the question of whether we can extend our map to one-loop order, which we return to in the conclusions.

\section{Insights into the analytic continuation of the fluxes}
\label{sec:KMOC-bound}

In the previous section we explored the analytic continuation of the two-body waveform, confirming our intuition from Section~\ref{sec:bound_wavefunction} that we can map the scattering case to the bound case, at least at tree-level and for dynamical contributions. In this section we extend this analytic continuation to more inclusive observables, calculated as expectation values of operators ${\hat{\mathcal{O}}}$ in time-evolved states\footnote{We are ultimately interested in the classical on-shell reduction of such in-in quantities, see e.g.~\cite{Kosower:2018adc}, and appendix A of~\cite{Britto:2021pud} and \cite{Damgaard:2023vnx} for more details on this point.}, $\bra{\text{in}}S^\dagger {\hat{\mathcal O}}S \ket{\text{in}}$. Examples are the total radiated energy $\Delta E^{>}_{\text{rad}}$ and angular momentum $\Delta J^{>}_{\text{rad}}$.

Working in the one-body model of Section~\ref{sec:theory-intro}, we can compute such expectation values of both scattering and bound observables by inserting the completeness relation \eqref{eq:completeness-4pt-back-earlier}, as this includes a sum over both scattering and bound states. Focusing on the total radiated energy $\Delta E_{\text{rad}}$ to illustrate, and using \eqref{eq:completeness-4pt-back-earlier} we obtain for the scattering case the familiar structure
\begin{align}
\Delta E^{>}_{\text{rad}} &= \sum_{\sigma} \int \mathrm{d} \Phi(k') \, \mathrm{d} \Phi(p') \, \hbar \, \omega^{\prime} \bra{\overline{\Psi}^{>}_{\mathbf{p}}} T^{\dagger} \ket{\overline{\mathcal{U}}_{\mathbf{p'}} k^{\prime \sigma}} \bra{\overline{\mathcal{U}}_{\mathbf{p'}} k^{\prime \sigma}} T \ket{\overline{\Psi}^{>}_{\mathbf{p}}} \nonumber \\
&= \sum_{\sigma} \int \mathrm{d} \Phi(k') \, \hbar \, \omega^{\prime} \, |\bar{W}^{>}(k^{\prime \sigma})|^2 \Big|_{k^{\prime \, \mu} = \omega' n^{\mu}}\,
\label{eq:rad_momentum_scatt}
\end{align}
where $\overline{\Psi}^{>}_{\mathbf{p}}$ is the incoming scattering wavefunction for a massive particle of momentum $p$ moving in the background and the spectral waveform is related to the time-domain waveform in the one-body model through
\begin{align}
\int_{-\infty}^{+\infty} \mathrm{d} u \, \e^{i \omega' u} \varepsilon^{\mu \nu}_{\sigma }(\hat{n}) \bar{h}^{>}_{\mu \nu}(u, \hat{n}) = \frac{\kappa}{4 \pi r} \bar{W}^{>}(k^{\prime \sigma})\Big|_{k^{\prime \, \mu} = \omega' n^{\mu}}\,, \quad \omega'>0 \,.
\end{align}
We can then rewrite \eqref{eq:rad_momentum_scatt} equivalently in terms of the scattering waveform\footnote{The expression in terms of the matrix elements can be easily made explicit, as in \eqref{eq:spectral_waveform} for the two-body case, but we feel it is unnecessary here and it makes the argument a bit longer.}
\begin{align}
\Delta E^{>}_{\text{rad}} &= \frac{1}{32 \pi G_N} \int_{\mathcal{C}_{u}^{>}} \mathrm{d} u\, \int_{\mathcal{S}^2}  \mathrm{d} \Omega \, r^2  (\partial_u \bar{h}^{>}_{\mu \nu}(u, \hat{x})) (\partial_u \bar{h}^{> \mu \nu}(u, \hat{x}))\,,
\end{align}
which a well-known expression \cite{Maggiore:2007ulw,Damour:2020tta} where the $u$-contour of integration is $\mathcal{C}_{u}^{>} = \mathbb{R}$.  

We now consider the energy radiated in transitions between bound energy levels, 
%
\begin{align}
\Delta E^{<}_{\text{rad}} &= \sum_{\sigma} \sum_{\{n'\}} \, \int \mathrm{d} \Phi(k^{\prime}) \, \hbar \, \omega^{\prime} \bra{\overline{\Psi}^{<}_{\{n\}}} T^{\dagger} \ket{\mathcal{\bar{B}}_{\{n'\}} k^{\prime \sigma}} \bra{\mathcal{\bar{B}}_{\{n'\}} k^{\prime \sigma}} T \ket{\overline{\Psi}^{<}_{\{n\}}} \nonumber \\
&= \frac{1}{32 \pi G_N} \int_{\mathcal{C}_u^{<}} \mathrm{d} u\,\int_{\mathcal{S}^2}  \mathrm{d} \Omega \, r^2  (\partial_u \bar{h}^{<}_{\mu \nu}(x)) (\partial_u \bar{h}^{< \mu \nu}(x))\,
\label{eq:rad_momentum_bound}
\end{align}
where $\overline{\Psi}^{<}_{\{n\}}$ is the incoming bound state wavefunction with a definite set of quantum numbers $\{n\}$, $\bar{h}^{< \mu \nu}(x)$ is the time-domain bound waveform and $\mathcal{C}_R^{<}$ is the retarded time contour of integration for the bound case which we will discuss shortly.  

Given that the initial and final massive bound states have a discrete energy spectrum, energy conservation (for a static background like Schwarzschild) imposes that the emitted frequencies for the radiated graviton must be discrete as well $E_{\mathbf{k}',\{m'\}} = \hbar \omega_{k',\{m'\}}$. In Fourier space, this is equivalent to an integral over a finite-size length (i.e., a period $T_u$)
\begin{align}
\sum_{\{m'\}} \tilde{f}(\omega_{k',\{m'\}}) \stackrel{\text{Fourier space}}{\leftrightarrow} \int_0^{T_u} \mathrm{d} u f(u)\,,
\end{align}
which means that the $u$-contour of integration in the bound case is $\mathcal{C}_u^{<} = [0, T_u]$. The meaning of these radiative observables for bound states (at least in our simple one-body model) has therefore to be interpreted as the averaged radiated energy over a periodic bound orbit, simply as a consequence of the discreteness of the emitted frequency.

Given the general expressions \eqref{eq:rad_momentum_scatt} and \eqref{eq:rad_momentum_bound}, we can now ask: what is the relation between $\Delta E^{>}_{\text{rad}}$ and $\Delta E^{<}_{\text{rad}}$ in the full two-body case? Extrapolating our findings from Section~\ref{sec:waveform_B2B}, we know that the integrands \emph{do} analytically continue into each other 
\begin{align}
(\partial_u h^{<}_{\mu \nu}(u, \hat{n})) (\partial_u h^{< \mu \nu}(u, \hat{n}))  = (\partial_u h^{>}_{\mu \nu}(u, \hat{n})) (\partial_u h^{> \mu \nu}(u, \hat{n})) \Big|_{p_{\infty} = \pm i \tilde{p}_{\infty}} \,,
\end{align}
where we emphasize that both branch cut prescriptions are allowed at tree-level and at 1PN order. It is essential, though, to consider the integral over the retarded time and the angles, to which we turn now. Let us define the Post-Minkowskian energy flux as
\begin{align}
\mathcal{F}_{E}(u,\mathcal{E},L) = \frac{1}{32 \pi G_N} \int_{\mathcal{S}^2}  \mathrm{d} \Omega \, r^2  (\partial_u h_{\mu \nu}(x)) (\partial_u h^{\mu \nu}(x))\,,
\end{align}
so that with a change of variable to the time $t = u + r$  (at fixed large $r$) we obtain 
%
\begin{align}
\Delta E^{>}_{\text{rad}} &=  \int_{-\infty}^{+\infty} \mathrm{d} t\, \mathcal{F}^{>}_{E}(t,\mathcal{E},L)\,, \quad \Delta E^{<}_{\text{rad}} =  \int_{0}^{T_{\text{orb}}} \mathrm{d} t\, \mathcal{F}^{<}_{E}(t,\mathcal{E},L)\,,
\end{align}
where $T_{\text{orb}}$ is the orbital period.

At this point, we can change the variable from $t$ to the radial coordinate $r$ in the adiabatic approximation so that we get \cite{Bini:2020hmy,Cho:2021arx,Saketh:2021sri}
\begin{align}
\Delta E^{>}_{\text{rad}} &= 2  \int_{r_{\text{min}}(\mathcal{E},L)}^{+\infty} \mathrm{d} r\, \frac{\mathcal{F}_{E}(r,\mathcal{E},L)}{p_r^{>}(r,\mathcal{E},L)}\,, \quad \Delta E^{<}_{\text{rad}} = 2  \int_{r_-(\mathcal{E},L)}^{r_+(\mathcal{E},L)} \mathrm{d} r\, \frac{\mathcal{F}_{E}(r,\mathcal{E},L)}{p_r^{<}(r,\mathcal{E},L)}\,,
\end{align}
where $r_{\text{min}}$ is the radial turning point of the (hyperbolic) scattering motion and $r_{\pm}$ are the radial turning points of the bound motion. It was noticed in~\cite{Cho:2021arx,Saketh:2021sri} that in the isotropic gauge 
\begin{align}
p_r(r,\mathcal{E},L) = p_r(r,\mathcal{E},-L) \,, \qquad \mathcal{F}_{E}(r,\mathcal{E},L) = + \mathcal{F}_{E}(r,\mathcal{E},-L)\,,
\end{align}
which means that $p_r$ and $\mathcal{F}_{E}(r,\mathcal{E},L)$ depend only on $L^2$. Therefore, using the fact that the $r_{ \pm}(\mathcal{E}, L)=r_{\text{min}}(\mathcal{E}, \mp L)$ we recover the analytic continuation
\begin{align}
\Delta E_{\mathrm{rad}}^{<}(\mathcal{E}, L)=\Delta E_{\mathrm{rad}}^{>}(\mathcal{E}, L)-\Delta E_{\mathrm{rad}}^{>}(\mathcal{E},-L), \quad \mathcal{E}<0 \,,
\label{eq:B2B-energy}
\end{align}
for the radiated energy.

Following similar steps for the angular momentum with the Post-Minkowskian flux, in the center of mass frame we obtain the expression \cite{Maggiore:2007ulw}
\begin{align}
\mathcal{F}_{L}^{k}(u,\mathcal{E},L) = \frac{1}{32 \pi G_N} \epsilon^{k i j} \int_{\mathcal{S}^2}  \mathrm{d} \Omega \, r^2  h_{\mu \nu}(u, \hat{x})  \left[\eta^{\nu \rho} \hat{x}_{[i} \partial_{j]} \partial_u+2 \delta_{[i}^\nu \delta_{j]}^\rho \right]  h^{\mu}_{\rho}(u, \hat{x}) \,.
\end{align}
As $\Delta L^k$ is a pseudo-vector and $\mathcal{F}^{k}_{L} \propto L^k$, it follows that
\begin{align}
\mathcal{F}^{k}_{L}(r,\mathcal{E},L) = - \mathcal{F}^{k}_{L}(r,\mathcal{E},-L) \,,
\end{align}
so that we recover the known analytic continuation for the total angular momentum 
\begin{align}
\Delta J^{<}_{\text{rad}}(\mathcal{E},L) &= \Delta J^{>}_{\text{rad}}(\mathcal{E},L) + \Delta J^{>}_{\text{rad}}(\mathcal{E},-L)\,, \quad \mathcal{E} < 0\,.
\label{eq:B2B-angularmomentum}
\end{align}
Interestingly, the analytic continuations \eqref{eq:B2B-energy} and \eqref{eq:B2B-angularmomentum} can be alternatively written purely in terms of the energy (or in terms of the eccentricity \cite{Cho:2021arx}), as we did for the waveform. Given that $L = b P_{\infty}$ and $P_{\infty} \propto p_{\infty}$, the analytic continuation in $p_{\infty}$ is essentially determined by $L$ if we impose $b \to i b$ as in \cite{Kalin:2019inp,Kalin:2019rwq}. Starting with the radial momentum, we find in the isotropic gauge the new relations
\begin{align}
p_r^{<}(r,\tilde{p}_{\infty},L) = p_r^{>}(r,- i\tilde{p}_{\infty},L) = -p_r^{>}(r,+i\tilde{p}_{\infty},L) \,,
\end{align}
while for the radial roots we have\footnote{The analytic continuation of the radial action is then given by 
\begin{align}
I_r^{<}(\tilde{p}_{\infty},L)  = \int_{r_{\text{min}}(-i \tilde{p}_{\infty},L)}^{+\infty} p_r^{<}(-i \tilde{p}_{\infty},L) - \int_{r_{\text{min}}(i \tilde{p}_{\infty},L)}^{+\infty} p_r^{<}(i \tilde{p}_{\infty},L) = I_r^{>}(-i\tilde{p}_{\infty},L) + I_r^{>}(+i\tilde{p}_{\infty},L)\,,    
\end{align}
as first found by an alternative computation in \cite{DiVecchia:2023frv}.}
\begin{align}
r_{-}(\tilde{p}_{\infty}, L) = r_{\text{min}}(-i \tilde{p}_{\infty},L) \,, \quad 
r_{+}(\tilde{p}_{\infty}, L) = r_{\text{min}}(+i \tilde{p}_{\infty},L) \,.
\end{align}
Therefore, using the fact that the fluxes obey
\begin{align}
\mathcal{F}^{<}_E(r,\tilde{p}_{\infty},L) = \mathcal{F}^{>}_E(r,\mp i \tilde{p}_{\infty},L) \,, \quad \mathcal{F}^{< k}_L(r,\tilde{p}_{\infty},L) = \mathcal{F}^{> k}_L(r,\mp i \tilde{p}_{\infty},L) \,,
\end{align}
we find the analytic continuation of the total radiated energy and angular momentum
\begin{align}
\Delta E^{<}_{\text{rad}}(\tilde{p}_{\infty},L) &= \Delta E^{>}_{\text{rad}}(-i \tilde{p}_{\infty},L) + \Delta E^{>}_{\text{rad}}(+i \tilde{p}_{\infty},L) \,, \,\mathcal{E} < 0\,, \nonumber \\
\Delta J^{<}_{\text{rad}}(\tilde{p}_{\infty},L) &= \Delta J^{>}_{\text{rad}}(-i \tilde{p}_{\infty},L) + \Delta J^{>}_{\text{rad}}(+i \tilde{p}_{\infty},L) \,, \,\mathcal{E} < 0\,.
\label{eq:B2B_fluxes_y}
\end{align}
While the analytic continuation for the angular momentum in this form was first observed in \cite{Heissenberg:2023uvo}, the one for the energy loss is new. 

For this reason, we explicitly checked, using the results in \cite{Herrmann:2021lqe,Herrmann:2021tct}, that 
\begin{align}
\Delta E^{<}_{\text{rad}}(\tilde{p}_{\infty},L) = &\Delta E^{>}_{\text{rad}}(\{\sqrt{y^2 - 1} \to -i \sqrt{1- y^2},\sqrt{y - 1} \to -i \sqrt{1 - y}\},L) \nonumber \\
&+\Delta E^{>}_{\text{rad}}(\{\sqrt{y^2 - 1} \to +i \sqrt{1- y^2},\sqrt{y - 1} \to +i \sqrt{1 - y}\},L)\,, 
\label{eq:B2B_E3PM}
\end{align}
holds up to 3PM order. Notice that this is one of the cases where $\sqrt{y -1}$ also appears (in the arcsinh term) besides $\sqrt{y^2 -1}$, and the same branch cut prescription must be applied for both square roots to get the correct answer as discussed after \eqref{eq:B2Bwaveform2}. It is interesting to notice that the sum over the two branch cut prescriptions of $p_{\infty} = \sqrt{y^2 - 1}$ (i.e. $\pm \sqrt{1-y^2}$) in \eqref{eq:B2B_fluxes_y} appear only as a consequence of the time integration of the fluxes, while for the waveform \eqref{eq:B2Bwaveform2} one needs, in general, only a single branch cut prescription.

\section{Summary and conclusions}
\label{sec:conclusion}

Now, more than ever, it is imperative to understand the connection between scattering and bound observables for the classical gravitational two-body problem. In this paper, we have derived and tested an analytic continuation of the tree-level (and 1PN) waveform in terms of the binding energy or, rather, of the rapidity. We began by working in the time domain due to the simpler analytic structure. {Inspired by the matching obtained with the PN multipoles in~\cite{Bini:2023fiz} (there in the frequency domain)}, we defined an equivalent time-domain expansion for the dynamical contribution of the PN waveform in the center-of-mass frame. This correctly reproduced the hyperbolic multipoles computed independently through the quasi-Keplerian parametrization. Using the Damour and DeRuelle~\cite{Damour:1985} analytic continuation of the orbital elements from hyperbolic to elliptic, we also computed the analogous bound version of the multipoles, and successfully compared against the analytic continuation of the tree-level scattering waveform. {The lack of periodicity in this result was simply due to the large-eccentricity expansion used to truncate Kepler's equation in the PM expansion; upon resumming this expansion periodicity is recovered, and the analytic continuation between scattering and bound remains manifest. As a further check, we confirmed that our analytic continuation holds for the 1PN waveform computed in~\cite{Junker:1992kle}.}

{We also discussed the analytic continuation of inclusive observables at 3PM order. While these formulae have been derived before in terms of analytic continuation in the angular momentum (or eccentricity), we provided a new perspective based entirely on analytic continuation in the binding energy. In particular, we showed how the analytic continuation of the total radiated energy and angular momentum (expressed as integrals over the associated PM fluxes constructed from the time-domain waveform) is modified by the retarded-time integral. This leads to a sum over two different branch cut prescriptions (unlike in the case of the waveform) such that our map directly recovers the results of e.g.~\cite{Heissenberg:2023uvo} in the case of angular momentum.}

Tables~\ref{tab:B2B2} and~\ref{tab:B2B1} summarize the current status of the boundary-to-bound dictionary, including results derived here and from the literature, in both the conservative and radiative domains~\cite{Kalin:2019inp,Kalin:2019rwq,Bini:2020hmy,Cho:2021arx,Saketh:2021sri,Jakobsen:2022zsx,Adamo:2022ooq,DiVecchia:2023frv,Heissenberg:2022tsn,Heissenberg:2023uvo,Gonzo:2023goe}. Table~\ref{tab:B2B2} contains results for the scattering angle $\chi$ (the bound analogue of which is the periastron advance $\Delta \Phi$), the time-domain dynamical waveform $h^{\text{dyn}}(u)$ and the total radiated energy and angular momentum $\Delta E_{\text{rad}}$, $\Delta J_{\text{rad}}$. (Note that we consider the total radiated angular momentum $\Delta J_{\text{rad}}$ in the center of mass frame to avoid frame dependence ambiguities~\cite{DiVecchia:2022owy,Manohar:2022dea,Riva:2023xxm}.) We remind the reader of the consequences of the branch cut prescription when performing these analytic continuations, see the explicit example for the radiated energy at 3PM~\eqref{eq:B2B_E3PM}.

\begin{table}[t!]
    \centering
    \begin{tabular}{l|c} \toprule
    Bound observable
    & Scattering observable
    \\ \midrule
    $\Delta \Phi(\tilde{p}_{\infty},L,a,\tidal{c_X})$  & $\chi(-i \tilde{p}_{\infty},L,a,\tidal{c_X}) +  \chi(+i \tilde{p}_{\infty},L,a,\tidal{c_X})$ \\ \midrule
    $\Delta E_{\text{rad}}^{<}(\tilde{p}_{\infty},L,a,\tidal{c_X})$  & $\Delta E_{\text{rad}}^{>}(-i \tilde{p}_{\infty},L,a,\tidal{c_X}) +  \Delta E_{\text{rad}}^{>}(+i \tilde{p}_{\infty},L,a,\tidal{c_X})$ \\ \midrule
    $\Delta J_{\text{rad}}^{<}(\tilde{p}_{\infty},L,a,{c_X})$  & $\Delta J_{\text{rad}}^{>}(-i \tilde{p}_{\infty},L,a,{c_X}) +  \Delta J_{\text{rad}}^{>}(+i \tilde{p}_{\infty},L,a,{c_X})$ \\ \midrule
    $h^{< \text{dyn}}(u;\tilde{p}_{\infty},L,\tidal{a, c_X})$  & $h^{> \text{dyn}}(u;+i \tilde{p}_{\infty},L, \tidal{a, c_X})$ \\ \bottomrule
\end{tabular}
\caption{Summary of analytic continuation for the observables between bound and scattering orbits. \tidal{Blue highlighting} indicates a conjectured extension (see the text) to additional dependence on aligned spin $a$ and leading tidal effects $c_X$.}
\label{tab:B2B2}
\end{table}

\begin{table}[t!]
    \centering
    \begin{tabular}{l|c} \toprule
    Bound observable 
    &
    Scattering observable 
    \\ \midrule
    $\Delta \Phi(\mathcal{E},L,a,\tidal{c_X})$  & $\chi(\mathcal{E},L,a,\tidal{c_X}) +  \chi(\mathcal{E},-L,-a,\tidal{c_X})$ \\ \midrule
    $\Delta E_{\text{rad}}^{<}(\mathcal{E},L,a,\tidal{c_X})$  & $\Delta E_{\text{rad}}^{>}(\mathcal{E},L,a,\tidal{c_X}) -  \Delta E_{\text{rad}}^{>}(\mathcal{E},-L,-a,\tidal{c_X})$ \\ \midrule
    $\Delta J_{\text{rad}}^{<}(\mathcal{E},L,a,{c_X})$  & $\Delta J_{\text{rad}}^{>}(\mathcal{E},L,a,{c_X}) +  \Delta J_{\text{rad}}^{>}(\mathcal{E},-L,-a,{c_X})$ \\ 
    \bottomrule
\end{tabular}
\caption{Summary of analytic continuation for the bound and scattering observables in terms of the variables $\mathcal{E},L,a,c_X$. \tidal{Blue text} again indicates a conjectured extension.}
\label{tab:B2B1}
\end{table}

Along with the variables $p_{\infty}$ and $L$ we make conjectures for the extension of known results to the spinning case with aligned-spin\footnote{For an extension to the mis-aligned spin case, the first steps have been taken in~\cite{Gonzo:2023goe} by considering generic Kerr geodesics with precession, but we will not discuss this here.} discussed in~\cite{Kalin:2019rwq} and to leading tidal couplings $c_X=c_{E_i^2},c_{B_i^2}$ given by the mass/electric-type and current/magnetic-type effects (see \cite{Mougiakakos:2022sic,Heissenberg:2022tsn} for the explicit definitions). Dependencies in black are verified, while those \tidal{highlighted in blue text} remain to be verified. We have checked that the proposed analytic continuation holds with the addition of aligned spin for some observables, as indicated, but the inclusion of leading tidal effects in the scattering angle and radiated energy remains to be verified, as does the inclusion of both aligned spin and leading tidal effects of the time-domain waveform $h^{\text{dyn}}$. Analogous results in terms of the variables $(\mathcal{E},L,a,c_X)$ are summarised in Table~\ref{tab:B2B1}.

It is worth stressing that the the stated relations are expected to be valid when there is local dependence on the trajectory of the motion; that is, at least up to \emph{3PM order for conservative and radiative inclusive observables} and up to \emph{tree-level order for the waveform}.  Indeed, one of the lessons we have learnt here is that it will be essential to make direct contact with the classical trajectories of the system in order to define a suitable map from scattering to bound observables. At higher PM orders, the presence of non-local in time effects in the gravitational case\footnote{This problem is absent in  classical electromagnetism, and therefore the boundary-to-bound map should extend more easily to higher orders in perturbation theory.} appearing at 4PM spoils the naive application of the rules described above, as nicely explained in~\cite{Cho:2021arx}. This implies that the analytic continuation of the waveform will suffer similar problems at one-loop order, which definitely deserves a deeper investigation.

There are many other avenues for future investigation. First, it would be interesting to check our proposed analytic continuation for the tree-level dynamical waveform including spin~\cite{Riva:2022fru,Jakobsen:2021lvp,Aoude:2023dui,DeAngelis:2023lvf,Brandhuber:2023hhl,Bohnenblust:2023qmy} or tidal effects~\cite{Jakobsen:2022psy,AccettulliHuber:2020dal,Mougiakakos:2022sic,Fucito:2023afe}, as well as at one-loop order \cite{Brandhuber:2023hhy,Georgoudis:2023lgf,Herderschee:2023fxh,Elkhidir:2023dco,Caron-Huot:2023vxl,Bohnenblust:2023qmy,Georgoudis:2023eke,Georgoudis:2023ozp} where issues related to tail effects should appear. This issue needs to be resolved in order to extend the scattering-to-bound map to higher PM orders~\cite{Cho:2021arx}; making closer contact with the classical trajectory may help in this respect, as in~\cite{Pound:2021qin}. A pressing problem is also to understand to the resummation at large eccentricity, which is needed to describe the inspiral phase of the two-body problem. It would also be interesting to return to the static terms and their inclusion in the bound state waveforms, perhaps by better addressing their dependence on the choice of BMS frame \cite{Veneziano:2022zwh,Georgoudis:2023eke,Riva:2023xxm,Favata:2011qi}. Finally, we look forward to extending our dictionary in other directions, to e.g.~the local energy and angular momentum flow recently discussed in~\cite{Gonzo:2023cnv}.

\paragraph{Acknowledgments} We thank Stefano de Angelis, Carlo Heissenberg, David Kosower, Luke Lippstreu, Donal O'Connell, Karthik Rajeev, Rodolfo Russo, Matteo Sergola and Canxin Shi for useful discussions. We especially thank Kallia Petraki for explaining the bound state formalism developed in \cite{Petraki:2015hla} to us, and its relevance for the dark matter community. We also thank the participants of the GRAMPA workshop for many useful discussions during the last stages of preparation of this work. RG would like to thank FAPESP grant 2021/14335-0 where part of this work was performed during August 2023. TA \& AI are supported by the STFC consolidator grant ST/X000494/1 `Particle Theory at the Higgs Centre.' TA is supported by a Royal Society University Research Fellowship, the Leverhulme Trust grant RPG-2020-386 and the Simons Collaboration on Celestial Holography MP-SCMPS-00001550-11.

\appendix

\section{Schwinger-Dyson equations for the $4+N$-point Green's function}
\label{app:SD_radiative}
We derive here the Schwinger-Dyson equations for the $4+N$-pt Green's function for the scattering of two massive scalars and the emission of $N$ gravitons. We begin with $N=0$, i.e.~the conservative sector of the two body problem, which is controlled by the 4-pt Green's function of the scalar fields. The Schwinger-Dyson equations imply that this obeys, working in momentum space,
\begin{align}
G\left(p_1, p_2; p_1^{\prime}, p_2^{\prime}\right)&= G_0(p_1, p_2; p_1^{\prime}, p_2^{\prime})  \nonumber \\
&+\int \hat{\mathrm{d}}^4 r_1 \, \hat{\mathrm{d}}^4 s_1 \,  G(s_1, s_2; p_1^{\prime}, p_2^{\prime})\mathcal{K}(r_1, r_2; s_1, s_2) G_0(p_1, p_2; r_1 r_2)  \;,\label{eq:BS-eq}
\end{align}
in which it is understood that $p_1^{\prime}+p_2^{\prime} = p_1 + p_2 = r_1+ r_2 = s_1+s_2$ by momentum conservation, $\mathcal{K}$ is the interaction kernel given by connected two-massive-particle irreducible (2MPI) diagrams, and $G_0(p_1, p_2; p_1^{\prime}, p_2^{\prime})$ is the disconnected Green's function
\begin{align}
	G_0(p_1, p_2; p_1^{\prime}, p_2^{\prime})=(2 \pi)^4 \delta^4\left(p_1^{\prime}-p_1\right) \Delta_1\left(p_1\right) \Delta_2\left(p_2\right)\,, \qquad  \Delta_j(p) = \frac{i}{p^2 - m_j^2 + i \epsilon} \;,
\end{align}
which is simply the product of massive scalar (tree level, free-theory) propagators for particles 1 and 2. We now make the external legs in the Green's function explicit using the ansatz (here and below $\Delta(p_1, p_2) \equiv \Delta_1(p_1) \Delta_2(p_2)$),
\begin{align}
 G\left(p_1, p_2; p_1^{\prime}, p_2^{\prime}\right)
	= G_0\left(p_1, p_2; p_1^{\prime}, p_2^{\prime}\right)
	+  \Delta(p_1^{\prime}, p_2^{\prime}) \mathcal{M}_4(p_1, p_2; p_1^{\prime}, p_2^{\prime}) \Delta(p_1, p_2) \;,
\end{align}
where $\mathcal{M}_4$ is the \emph{off}-shell 4-pt amplitude. Inserting~\eqref{eq:BS-eq} generates, upon the standard LSZ reduction, the following recusion relation for the on-shell amplitude $\mathcal{M}_4$: 
\begin{align}
\mathcal{M}_4(p_1, p_2; p_1^{\prime}, p_2^{\prime}) &=
\mathcal{K}_4(p_1, p_2; p_1^{\prime}, p_2^{\prime}) \nonumber \\
&+ \int\! \hat{\mathrm{d}}^4 s_1 \,  \mathcal{K}(p_1, p_2; s_1, s_2)  \Delta(s_1,s_2) \mathcal{M}_4(s_1, s_2; p_1^{\prime}, p_2^{\prime}) \,,
\label{eq:BS-equation-amplitude}
\end{align}
as illustrated in Fig.\ref{fig:recursion4pt-earlier}. This recovers eq.(3.1) of~\cite{Adamo:2022ooq}.

We now turn to the radiative sector, that is $N\geq 1$.  The generalization of two-body recursion relations for form factors, with the insertion of (conserved) current operators, was first derived in~\cite{Gross:1987bu}. At the level of Green's functions, recursion relations for vector (or tensor) currents require some gauge-fixing, as was first discussed in~\cite{Kvinikhidze:1997,Haberzettl:1997jg} and further clarified in the series of works~\cite{Kvinikhidze:1998xn,Kvinikhidze:1999xp}.  In the gravitational case, this ``gauging'' procedure -- which is equivalent to deriving the Schwinger-Dyson equations -- amounts to dressing the $4$-pt Green's function by attaching $N$ gravitons on each line in the diagrammatic representation.

Beginning with the case of one graviton emission, the Schwinger-Dyson equation for the 5-pt Green's function is
\begin{align}
\label{eq:G5-first}
G^{\mu_1 \nu_1}&\left(p_1, p_2; p_1^{\prime}, p_2^{\prime}, k_1^{\prime}\right)=G_0^{\mu_1 \nu_1}\left(p_1, p_2; p_1^{\prime}, p_2^{\prime}, k_1^{\prime}\right) \nonumber \\
& \quad +\int \hat{\mathrm{d}}^4 r_1 \, \hat{\mathrm{d}}^4 s_1  G_0^{\mu_1 \nu_1}\left(p_1, p_2; r_1, r_2, k_1^{\prime}\right) \mathcal{K}\left(r_1, r_2; s_1, s_2\right) G\left(s_1, s_2; p_1^{\prime}, p_2^{\prime} \right) \nonumber \\
& \quad +\int \hat{\mathrm{d}}^4 t_1 \, \hat{\mathrm{d}}^4 u_1 \, G_0\left(p_1, p_2; t_1, t_2 \right) \hat{\mathcal{K}}_R^{\mu_1 \nu_1}
\left(t_1, t_2; u_1, u_2, k_1^{\prime} \right) G\left(u_1, u_2; p_1^{\prime}, p_2^{\prime}\right) \nonumber \\
& \quad +\int \hat{\mathrm{d}}^4 v_1 \, \hat{\mathrm{d}}^4 w_1 G_0\left(p_1, p_2; v_1, v_2 \right) \mathcal{K}\left(v_1, v_2; w_1, w_2 \right) G^{\mu_1 \nu_1}\left(w_1, w_2; p_1^{\prime}, p_2^{\prime}, k_1^{\prime} \right) \,,  
\end{align}
where we denote with ${\hat{\mathcal{K}}}_R$ the graviton dressed 2MPI radiative kernel, and the momentum conservation relations $p_1+p_2=p_1^{\prime}+p_2^{\prime}+k_1^{\prime}$, $p_1+p_2=t_1+t_2=v_1+v_2=w_1+w_2$, and $p_1^{\prime}+p_2^{\prime}=r_1+r_2=s_1+s_2=u_1+u_2$ are implicit. In a compact notation (\ref{eq:G5-first}) may be written
\begin{align}
G^{\mu_1 \nu_1}=G_0^{\mu_1 \nu_1}+G_0^{\mu_1 \nu_1} \mathcal{K} G+G_0 \hat{\mathcal{K}}_R^{\mu_1 \nu_1} G+G_0 \mathcal{K} G^{\mu_1 \nu_1} \,.
\label{eq:recursion_DS1}
\end{align}
\begin{figure}[t!]
\centering
\includegraphics[scale=0.75]{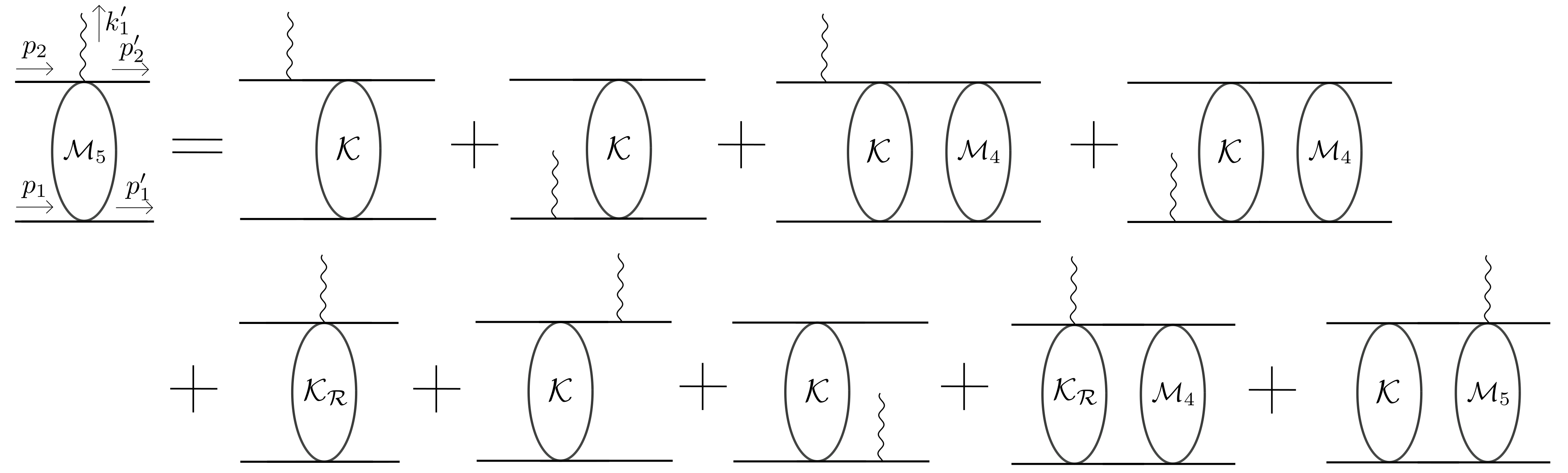}
\caption{The 5-pt amplitude recursion relation which follows from Schwinger-Dyson equations, which naturally include contributions from emission of zero-energy gravitons.}
\label{fig:recursion5pt}
\end{figure}
We proceed to construct from this expression a recursion relation for the 5-pt amplitude $\mathcal{M}^{\mu\nu}_5$. As for the conservative sector, we redefine our Green's function in terms of matrix elements and external free propagators, 
\begin{align}
\label{eq:M5-ansatz}
G\left(p_1, p_2; p_1^{\prime}, p_2^{\prime}, k_1^{\prime}\right) & = \varepsilon^{\mu_1 \nu_1}_{\sigma} (k_1^{\prime}) \Delta_{\mu_1 \nu_1 \alpha_1 \beta_1} (k_1^{\prime})  \mathcal{M}^{\alpha_1 \beta_1}_5(p_1, p_2; p_1^{\prime}, p_2^{\prime}, k_1^{\prime}) \Delta(p_1,p_2) \Delta(p_1^{\prime}, p_2^{\prime})  \nonumber \\
&+ \varepsilon^{\mu_1 \nu_1}_{\sigma} (k_1^{\prime})  \Delta_{\mu_1 \nu_1 \alpha_1 \beta_1 } (k_1^{\prime})  \Big[ \hat{\delta}^4 (p_1 - p_1^{\prime}) \Delta_1(p_1) \mathcal{M}_3^{\alpha_1 \beta_1} (p_2; p_2^{\prime}, k_1^{\prime}) \Delta_2(p_2,p_2^{\prime}) \nonumber \\
& \qquad\qquad \qquad\qquad\qquad+ (p_1, p_2; p_1^{\prime}, p_2^{\prime}) \leftrightarrow (p_2, p_1; p_2^{\prime}, p_1^{\prime})  \Big] \;,
\end{align}
in which $\Delta_{\mu_1 \nu_1 \alpha_1 \beta_1} (k_1^{\prime})$ is the free graviton propagator and we note in the second and third lines the presence of disconnected contributions. Combining this ansatz with (\ref{eq:recursion_DS1}) we obtain a lengthy relation for $\mathcal{M}_5^{\mu\nu}$ shown diagrammatically in Fig.\ref{fig:recursion5pt}, with the amputated radiative kernel $\mathcal{K}^{\mu_1 \nu_1}_{R}$ defined by
\begin{align}
\hat{\mathcal{K}}^{\mu_1 \nu_1}_{R} := \Delta^{\mu_1 \nu_1}_{\quad \, \alpha_1 \beta_1} (k_1^{\prime}) \mathcal{K}_{\mathcal{R}}^{\alpha_1 \beta_1} \,.
\end{align}
We observe that all 3-pt amplitudes $\mathcal{M}_3^{\mu_1 \nu_1}$ contributing to Fig.\ref{fig:recursion5pt} have support only on $k_1^{\prime} = 0$, i.e. zero-energy gravitons, and moreover that $\mathcal{M}^{\mu_1 \nu_1}_5(p_1^{\prime} p_2^{\prime} k_1^{\prime}; w_1 w_2)$ includes, in general, disconnected contributions. If we focus on the kinematic region where the energy of the graviton is strictly positive $E_{\mathbf{k_1}^{\prime}}>0$, the amplitude recursion relation obtained from (\ref{eq:recursion_DS1}) and (\ref{eq:M5-ansatz}) simplifies considerably, to
\begin{align}
\mathcal{M}^{\mu_1 \nu_1}_5(p_1, p_2; p_1^{\prime}, p_2^{\prime}, k_1^{\prime}) \Big|_{E_{\mathbf{k_1}^{\prime}}>0} &=  \mathcal{K}_{\mathcal{R}}^{\mu_1 \nu_1}(p_1, p_2; p_1^{\prime}, p_2^{\prime}, k_1^{\prime}) \\ 
&+\int \hat{\mathrm{d}}^4 w_1 \mathcal{K}(p_1, p_2; w_1, w_2) \Delta(w_1,w_2) \mathcal{M}^{\mu_1 \nu_1}_5(w_1, w_2; p^{\prime}_1, p^{\prime}_2, k_1^{\prime}) \nonumber \\
&+  \int \hat{\mathrm{d}}^4 w_1 \, \mathcal{K}_{\mathcal{R}}^{\mu_1 \nu_1}(p_1, p_2; w_1, w_2, k_1^{\prime}) \Delta(w_1,w_2)  \mathcal{M}_4( w_1, w_2; p^{\prime}_1, p^{\prime}_2) \,. \nonumber 
\end{align}
which matches exactly \eqref{eq:recursion5pt-positive} of the main text.

These results may be generalized to $N>1$ graviton emissions by iterating the gauging procedure. We define the set $\Sigma_N$ of all gaugings of the conservative Schwinger-Dyson equations by dressing with $N$ pairs of indices $\{\mu_i \nu_i\}_{i=1,\dots,N}$. There are a total of $|\Sigma_N| = 1 + 3^N$ terms\footnote{Combinatorically, this is the number of ways ($=3^N$) of assigning $N$ distinct objects (pairs of indices) to $3$ distinct boxes (i.e.~the combination $G_0 \mathcal{K} G$) plus the single contribution from the free term $G_0$.} on the RHS of such equations -- indeed we can check that there are $4 = 1 + 3$ terms for the $N=1$ case in \eqref{eq:recursion_DS1}. Schematically, the Schwinger-Dyson equation for $4+N$-pt Green's functions becomes
\begin{align}
G^{\mu_1 \nu_1 \mu_2 \nu_2 \dots \mu_N \nu_N}&=G_0^{\mu_1 \nu_1 \mu_2 \nu_2 \dots \mu_N \nu_N}+
\sum_{\substack{(\sigma_1, \sigma_2,\sigma_3) \in \Sigma_N \\ |\sigma_1| + |\sigma_2| +|\sigma_3| = N}} G_0^{\sigma_1} \hat{\mathcal{K}}_R^{\sigma_2} G^{\sigma_3} \,,
\label{eq:recursion_DSn}
\end{align}
where $\hat{\mathcal{K}}_R^{\varnothing} \equiv \mathcal{K}$ and $\sigma_1,\sigma_2,\sigma_3$ are (possibly empty) distinct collections of pairs of indices $\{\mu_i \nu_i\}_{i=1,\dots,N}$ such that $|\sigma_1| + |\sigma_2| +|\sigma_3| = N$. From equation \eqref{eq:recursion_DSn}, following similar steps to the one adopted for the $N=1$ case, it would be straightforward to obtain the generic amplitude recursion relation for $N$ graviton emissions.

\section{Decomposition of Green's functions into scattering and bound states}
\label{sec:appendixA}

We revisit in this appendix the traditional Bethe-Salpeter approach for radiative transitions in the two-body problem, as originally proposed by Mandelstam \cite{Mandelstam:1955sd} (see also \cite{Gross:1987bu,Faustov:1970hn,Faustov:1974qt}). Following \cite{Petraki:2015hla}, and referring to \cite{Silagadze:1998ri} for further details, we first demonstrate how the conservative Green's functions both of the one-body model and the full two-body case admit a non-perturbative decomposition in terms of scattering and bound wavefunctions. We then show how to rigorously define the radiative matrix elements for the bound state formation by using the Schwinger-Dyson equations combined with the corresponding wavefunctions for scattering and bound states. We work here in units where $\hbar = c = 1$.

\subsection{2-pt Green's function in the one-body model}

We begin with the 2-pt Green's function, which encodes the classical geodesic equation of a particle in the conservative piece of the metric ${g}^{\text{\tiny OB}}_{\mu\nu}$, 
\begin{align}
G^{\text{\tiny OB}}(x_1; y_1) = \bra{\Omega^{\text{\tiny OB}}} T \bar{\Phi}(y_1) \bar{\Phi}^{\dagger}(x_1) \ket{\Omega^{\text{\tiny OB}}} \,,
\label{eq:2-ptGreenEOB}
\end{align}
which may be computed perturbatively as a solution of the Schwinger-Dyson equation 
\begin{align}
\label{eq:2-ptGreenEOB-recursion}
G^{\text{\tiny OB}}\left(p_1; p_1^{\prime}\right)&=G_{0}^{\text{\tiny OB}}\left(p_1 ; p_1^{\prime}\right) +\int \!\hat{\mathrm{d}}^4 r_1 \hat{\mathrm{d}}^4 s_1  \, G_{0}^{\text{\tiny OB}}\left(p_1; r_1\right) \mathcal{K}^{\text{\tiny OB}}\left(r_1; s_1\right) G^{\text{\tiny OB}}\left(s_1; p_1^{\prime}\right) \,,  
\end{align}
where $\mathcal{K}^{\text{\tiny OB}}$ is the effective one-body potential determined via matching~\cite{Neill:2013wsa,Cheung:2018wkq,Bern:2019crd,Kalin:2019rwq,Kalin:2019inp}. 

Assuming the existence of a Hilbert space decomposition of the (self-adjoint) Hamiltonian of our one-body model into the corresponding scattering and bound eigenvectors, we now insert the completeness relation \eqref{eq:completeness-4pt-back-earlier} into \eqref{eq:2-ptGreenEOB}, which trivially yields a decomposition of the Green's function into bound and scattering contributions:
\begin{align}
G^{\text{\tiny OB}}(x_1; y_1) = G^{> \text{\tiny{OB}}}(x_1; y_1) + \sum_{\{n\}} G^{< \text{\tiny{OB}}}_{\{n\}}(x_1; y_1) \,,
\end{align}
where 
\begin{align}
G^{>\text{\tiny{OB}}}(x_1; y_1) &= \int \frac{\mathrm{\hat{d}}^3 P}{2 E^{>}_{\mathbf{P}}}  \bra{\Omega^{\text{\tiny OB}}} \bar{\Phi}(y_1) \ket{\mathcal{\bar{U}}_{\mathbf{P}}}\bra{\mathcal{\bar{U}}_{\mathbf{P}}} \bar{\Phi}^{\dagger}(x_1) \ket{\Omega^{\text{\tiny OB}}} \theta(y_1^0-x_1^0) \;,
\nonumber \\
G^{<\text{\tiny{OB}}}_{\{n\}}(x_1; y_1)  &= \left\langle\Omega^{\text{\tiny OB}}\left|\bar{\Phi}(y_1) \right| \mathcal{\bar{B}}_{\{n\}}\right\rangle \bra{\mathcal{\bar{B}}_{\{n\}}}\bar{\Phi}^{\dagger}(x_1) \ket{\Omega^{\text{\tiny OB}}} \theta(y_1^0-x_1^0) \,,
\nonumber 
\end{align}
in which we immediately recognise the essential ingredients of the OB wavefunctions defined in \eqref{eq:EOB-wavefunctions-earlier}.  In each case we transform the theta function into a momentum integral using the standard representation
\begin{align}
\theta(z)= i \int_{-\infty}^{\infty} \!\mathrm{\hat{d}} P^0 \,\frac{\e^{-i (P^0 - E) z}}{P^0 - E+i \epsilon}\,,
\label{eq:theta_function}
\end{align}
and appropriate choice of $E$. The Fourier transform from $x_1$, $y_1$ to momenta $p_1$, $p_1'$ gives for the scattering part of the Green's function 
\begin{align}
G^{>\text{\tiny{OB}}}(p_1; p_1^{\prime}) &= \int\!\mathrm{d}^4 x_1 \mathrm{d}^4 y_1 \, e^{i\left(p_1 x_1 - p_1^{\prime} y_1\right)} G^{>\text{\tiny{OB}}}(x_1; y_1) \nonumber \\
&= i \int \!\mathrm{\hat{d}}^4 P \int\!\mathrm{d}^4 x_1 \mathrm{d}^4 y_1 \, \e^{i\left(p_1 x_1 - p_1^{\prime} y_1\right)}  \bar{\Psi}_{\mathbf{P}}^{>}(y_1) \bar{\Psi}_{\mathbf{P}}^{* >}(x_1) \frac{e^{-i (P^0 - E^{>}_{\mathbf{P}}) (y_1^0-x_1^0)}}{(2 E^{>}_{\mathbf{P}}) (P^0 - E^{>}_{\mathbf{P}}+i \epsilon)} \,. 
\label{eq:GEOB_exactdecomposition}
\end{align}
Therefore, around the on-shell pole, $P^0 \to E^{>}_{\mathbf{P}}$, we obtain
\begin{align}
\label{eq:scattering-poleEOB}
G^{>\text{\tiny{OB}}} &
\sim i  \int \frac{\mathrm{\hat{d}}^3 P}{(2 E^{>}_{\mathbf{P}})}  \frac{\bar{\Psi}_{\mathbf{P}}^{>}(p_1^{\prime}) \bar{\Psi}_{\mathbf{P}}^{* >}(p_1)}{P^0 - E^{>}_{\mathbf{P}}+i \epsilon} \,.
\end{align}
Analogously, the bound state contribution behaves, near the bound state pole $ E^{<}_{\{n\}}$, as
\begin{align}
\label{eq:bound-poleEOB}
G^{< \text{\tiny{OB}}}_{\{n\}} & \sim 
i \frac{\bar{\Psi}_{n}^{<}(p_1^{\prime}) \bar{\Psi}_{\{n\}}^{* <}(p_1)}{P^0 - E^{<}_{\{n\}}+i \epsilon} \,.
\end{align}

\subsection{$3$-pt Green's function in the one-body model}

For the generic case with radiation, we need to study the $2+N$-pt Green's function in the one-body model
\begin{align}
\label{eq:n-ptGreenEOB}
G^{\text{\tiny OB}}_{\mu_1 \nu_1 \mu_2 \nu_2 \dots \mu_N \nu_N}&(x_1; y_1,z_1,z_2,\dots,z_N) \\
&= \bra{\Omega^{\text{\tiny OB}}} T \bar{h}_{\mu_1 \nu_1}(z_1) \bar{h}_{\mu_2 \nu_2}(z_2) \dots \bar{h}_{\mu_N \nu_N}(z_N) \bar{\Phi}(y_1) \bar{\Phi}^{\dagger}(x_1)  \ket{\Omega^{\text{\tiny OB}}} \,, \nonumber 
\end{align}
where the field operators are now defined on background of the conservative piece of the metric ${g}^{\text{\tiny OB}}_{\mu\nu}$ determined by the matching. As explained in the main text, this is a (self-)consistent approximation of the full radiative dynamics, but in general we might be neglecting backreaction effects (i.e., recoil). Focusing on the emission of a single graviton we obtain the Schwinger-Dyson equation
\begin{align}
\label{eq:3-ptGreenEOB-pert}
G^{\text{\tiny{OB}} \, \mu_1 \nu_1}\left(p_1; p_1^{\prime}, k_1^{\prime} \right)&=G_0^{\mu_1 \nu_1}\left(p_1; p_1^{\prime}, k_1^{\prime}\right) \\
&+\int \hat{\mathrm{d}}^4 r_1 \, \hat{\mathrm{d}}^4 s_1  G_0^{\text{\tiny{OB}} \, \mu_1 \nu_1}\left(p_1; r_1, k_1^{\prime}\right) \mathcal{K}^{\text{\tiny OB}}\left(r_1; s_1\right) G^{\text{\tiny OB}}\left(s_1; p_1^{\prime}\right) \nonumber \\
& +\int \hat{\mathrm{d}}^4 t_1 \, \hat{\mathrm{d}}^4 u_1 \, G_0^{\text{\tiny OB}}\left(p_1; t_1\right) \hat{\mathcal{K}}_R^{\text{\tiny{OB}} \, \mu_1 \nu_1}\left(t_1; u_1, k_1^{\prime}\right) G^{\text{\tiny OB}}\left(u_1; p_1^{\prime}\right) \nonumber \\
& +\int \hat{\mathrm{d}}^4 v_1 \, \hat{\mathrm{d}}^4 w_1 G^{\text{\tiny OB}}_0\left(p_1; v_1\right) \mathcal{K}^{\text{\tiny OB}}\left(v_1; w_1\right) G^{\text{\tiny{OB}} \, \mu_1 \nu_1}\left(w_1; p_1^{\prime}, k_1^{\prime}\right) \,, \nonumber
\end{align}
where $\hat{\mathcal{K}}_R^{\text{\tiny{OB}} \, \mu_1 \nu_1}$ is the graviton dressed one-particle irreducible (1MPI) radiative kernel. 

At this point, using the Bethe-Salpeter approach, we can then combine \eqref{eq:2-ptGreenEOB}, \eqref{eq:2-ptGreenEOB-recursion} and \eqref{eq:3-ptGreenEOB-pert} into the following equation
\begin{align}
\label{eq:radiative-captureEOB}
& G^{\text{\tiny{OB}} \, \mu_1 \nu_1}\left(p_1; p_1^{\prime}, k_1^{\prime}\right) \\
&=\int \hat{\mathrm{d}}^4 t_1 \, \hat{\mathrm{d}}^4 u_1 \, G^{\text{\tiny OB}}\left(p_1; t_1\right) \left[\Gamma_0^{\text{\tiny{OB}} \, \mu_1 \nu_1}\left(t_1; u_1, k_1^{\prime} \right) + \hat{\mathcal{K}}_R^{\text{\tiny{OB}} \, \mu_1 \nu_1}\left(t_1; u_1, k_1^{\prime}\right) \right] G^{\text{\tiny OB}}\left(u_1; p_1^{\prime}\right) \,, \nonumber 
\end{align}
where $\Gamma_0^{\text{\tiny{OB}} \, \mu_1 \nu_1} := (G_0^{\text{\tiny OB}})^{-1} G_0^{\text{\tiny{OB}} \, \mu_1 \nu_1} (G_0^{\text{\tiny OB}})^{-1}$. 

The interesting feature of \eqref{eq:radiative-captureEOB} is that we can combine the standard LSZ reduction of the LHS 3-pt Green's function with the matrix element obtained by taking the on-shell poles of the 2-pt Green's functions on the RHS. On one hand, using the definition of the scattering and bound OB wavefunctions  \eqref{eq:EOB-wavefunctions-earlier} and of the graviton OB wavefunction
\eqref{eq:EOB-graviton-earlier}, the LSZ reduction of the 3-pt classical OB Green's function \eqref{eq:n-ptGreenEOB} gives

\begin{align}
G^{\mu_1 \nu_1}\left(p_1; p_1^{\prime}, k_1^{\prime}\right) &\sim \int \frac{\hat{\mathrm{d}}^3 K}{2 E^{>}_{\mathbf{K}^{\prime}} } \,  \frac{i \bar{H}_{\mathbf{K}^{\prime}}^{> \mu_1 \nu_1}(k_1^{\prime})}{(K^0-E^{>}_{\mathbf{K}^{\prime}}+i \epsilon)}  \, \frac{i \bar{\Psi}^{* <}_{\{n\}}(p_1)}{(P^0-E^{<}_{\{n\}}+i \epsilon)} \nonumber \\
& \times \int \frac{\hat{\mathrm{d}}^3 P}{2 E^{>}_{\mathbf{P}} } \,  \frac{i \bar{\Psi}^{>}_{\mathbf{P}}(p_1^{\prime})}{(P^0-E^{>}_{\mathbf{P}}+i \epsilon)} \left\langle\mathcal{\bar{B}}_{n} ; \bar{h}_{\mathbf{K}^{\prime}} |S| \mathcal{\bar{U}}_{\mathbf{P}}\right\rangle \,.
\label{eq:LSZ-3-ptGreenLHS}
\end{align}
On the other hand, using the decomposition of the classical OB 2-pt Green's function \eqref{eq:GEOB_exactdecomposition}, we can write the RHS of \eqref{eq:radiative-captureEOB} around the scattering and bound poles as
\begin{align}
\hspace{-10pt} G^{\mu_1 \nu_1}\left(p_1; p_1^{\prime}, k_1^{\prime}\right) &\sim \int \mathrm{\hat{d}}^4 r_1 \, \mathrm{\hat{d}}^4 t_1 \, \mathrm{\hat{d}}^4 u_1  \left[\int \frac{\hat{\mathrm{d}}^3 K}{2 E^{>}_{\mathbf{K}^{\prime}} }  \frac{i \bar{H}_{\mathbf{K}^{\prime}}^{> \mu_1 \nu_1}(k_1^{\prime}) \bar{H}_{\mathbf{K}^{\prime}}^{* > \alpha_1 \beta_1}(r_1) }{(K^0-E^{>}_{\mathbf{K}^{\prime}}+i \epsilon)}\right] \left[\frac{i \bar{\Psi}^{<}_{\{n\}}(p_1^{\prime}) \bar{\Psi}^{*<}_{\{n\}}(u_1) }{(P^0-E^{<}_{\{n\}}+i \epsilon)}\right] \nonumber \\
& \quad \times \, \left[\int \frac{\hat{\mathrm{d}}^3 P}{2 E^{>}_{\mathbf{P}} } \frac{i \bar{\Psi}^{> }_{\mathbf{P}}(t_1) \bar{\Psi}^{* > }_{\mathbf{P}}(p_1)}{(P^0-E^{>}_{\mathbf{P}}+i \epsilon)} \right] \bar{M}_{3,\alpha_1 \beta_1} \left(t_1; u_1, r_1 \right) \,.
\label{eq:LSZ-3-ptGreenRHS}
\end{align}
Finally combining \eqref{eq:LSZ-3-ptGreenLHS} and \eqref{eq:LSZ-3-ptGreenRHS} we obtain
\begin{align}
\left\langle\mathcal{\bar{B}}_{\{n\}} ; \bar{h}_{\mathbf{K}^{\prime}} |S| \mathcal{\bar{U}}_{\mathbf{P}}\right\rangle &= \int
\hat{\mathrm{d}}^4 r\,
\hat{\mathrm{d}}^4 s\,
\hat{\mathrm{d}}^4 t\, 
\bar{H}^{* >}_{\mathbf{K}^{\prime}, \alpha \,\beta}(t)  \bar{\Psi}_{\{n\}}^{* <}(r) \bar{\Psi}^{>}_{\mathbf{P}}(s) \bar{M}^{\alpha \,\beta}_{3} (r, t; s) \,.
\label{eq:master-eq}
\end{align}

\subsection{4-pt Green's function for the two-body dynamics}

We follow here a similar strategy as in the one-body model. We start by inserting the completeness relation \eqref{eq:completeness-4pt} into the 4-pt Green's function defined as
\begin{align}
G(x_1, x_2; y_1, y_2) = \bra{\Omega} T \phi_1(y_1) \phi_2(y_2) \phi_1^{\dagger}(x_1) \phi^{\dagger}_2(x_2) \ket{\Omega} \,,
\end{align}
to obtain the following decomposition into scattering ($>$) and bound ($<$) contributions:
\begin{align}
G(x_1, x_2; y_1, y_2) &=  G^{>}(x_1, x_2; y_1, y_2) + \sum_{\{n\}} G^{<}_{\{n\}}(x_1, x_2; y_1, y_2)  \,, \\
G^{<}_{\{n\}}(x_1, x_2; y_1, y_2) &= \left\langle\Omega\left|T \phi_1(y_1) \phi_2(y_2)\right| \mathcal{B}_{\{n\}}\right\rangle \bra{\mathcal{B}_{\{n\}}}T \phi_1^{\dagger}(x_1) \phi^{\dagger}_2(x_2) \ket{\Omega} \theta\left[F\right] \,, \nonumber \\
G^{>}(x_1, x_2; y_1, y_2) &= \int \frac{\mathrm{\hat{d}}^3 P}{2 E^{>}_{\mathbf{P},\mathbf{Q}}}
\frac{\mathrm{\hat{d}}^3 Q}{2 \epsilon^{>}_{\mathbf{P},\mathbf{Q}}} \bra{\Omega} T \phi_1(y_1) \phi_2(y_2)\ket{\mathcal{U}_{\mathbf{P},\mathbf{Q}}}\bra{\mathcal{U}_{\mathbf{P},\mathbf{Q}}} T \phi_1^{\dagger}(x_1) \phi^{\dagger}_2(x_2) \ket{\Omega} \theta\left[F\right] \,, \nonumber 
\end{align}
with
\begin{align}
F(x_1^0,x_2^0,y_1^0,y_2^0) = \min \left(y_1^0, y_2^0\right)-\max &\left(x_1^0, x_2^0\right)  \,. 
\end{align}
We focus now on the scattering contribution, since the other one is completely analogous. Again using the representation \eqref{eq:theta_function} of the theta function, along with the definition of the two-body wavefunctions in position space \eqref{eq:two-bodywavefunction} and the coordinate variables $X,Y,x,y$ in \eqref{eq:coord_variables} we can equivalently write the scattering Green's function as 
\begin{align}
G^{>}&= i  \int \frac{\mathrm{\hat{d}}^3 P}{2 E^{>}_{\mathbf{P},\mathbf{Q}}} \int \frac{\mathrm{\hat{d}}^3 Q}{2 \epsilon^{>}_{\mathbf{P},\mathbf{Q}}} \, \Psi_{\mathbf{P}, \mathbf{Q}}^{>}(y) \Psi_{\mathbf{P}, \mathbf{Q}}^{* >}(x) \e^{-i E_{\mathbf{P}} (Y^0 - X^0)} \e^{i \mathbf{P} \cdot (\mathbf{Y} - \mathbf{X})} \int \mathrm{\hat{d}} P^0  \frac{\e^{-i (P^0 - E^{>}_{\mathbf{P}, \mathbf{Q}}) F}}{P^0 - E^{>}_{\mathbf{P}, \mathbf{Q}}+i \epsilon} \nonumber \\
&= i \int  \mathrm{\hat{d}}^4 P \int \frac{\mathrm{\hat{d}}^3 Q}{(2 \epsilon^{>}_{\mathbf{P},\mathbf{Q}})} \, \Psi_{\mathbf{P}, \mathbf{Q}}^{>}(y) \Psi_{\mathbf{P}, \mathbf{Q}}^{* >}(x) \e^{i P \cdot (X - Y)} \frac{1}{(2 E^{>}_{\mathbf{P},\mathbf{Q}})}  \frac{\e^{-i (P^0 - E^{>}_{\mathbf{P}, \mathbf{Q}}) \tilde{F}}}{P^0 - E^{>}_{\mathbf{P}, \mathbf{Q}}+i \epsilon} \,,
\label{eq:Gscattering_decomposition}
\end{align}
where we have used the relative coordinates introduced in \eqref{eq:coord_variables} and we have defined
\begin{align}
\tilde{F}(x_1^0,x_2^0,y_1^0,y_2^0)&= F(x_1^0,x_2^0,y_1^0,y_2^0) - (Y^0 - X^0) \;.\nonumber
\end{align}
At this point, we can perform the Fourier transform of \eqref{eq:Gscattering_decomposition} to get
\begin{align}
\label{eq:Gscattering_exactdecomposition}
G^{>}&(P / 2+q, P / 2-q; P / 2 + q^{\prime}, P / 2 - q^{\prime}) \\
&= \int \mathrm{d}^4 (X-Y) \mathrm{~d}^4 x \mathrm{~d}^4 y \, \e^{i\left(P (X-Y) +q x - q^{\prime} y\right)} G^{>}(X+x/2, X-x/2;Y+y/2, Y-y/2) \nonumber \\
&= i \int \frac{\mathrm{\hat{d}}^3 Q}{(2 \epsilon^{>}_{\mathbf{P},\mathbf{Q}})}\, \int\mathrm{~d}^4 x \mathrm{~d}^4 y \, \e^{i (q^{\prime} y - q x)}  \Psi_{\mathbf{P}, \mathbf{Q}}^{>}(y) \Psi_{\mathbf{P}, \mathbf{Q}}^{* >}(x) \frac{\e^{-i (P^0 - E^{>}_{\mathbf{P}, \mathbf{Q}}) \tilde{F}}}{(2 E^{>}_{\mathbf{P},\mathbf{Q}}) (P^0 - E^{>}_{\mathbf{P}, \mathbf{Q}}+i \epsilon)} \,, \nonumber
\end{align}
which means that the behaviour near the on-shell pole $P^0 \to E^{>}_{\mathbf{P},\mathbf{Q}}$ is
\begin{align}
G^{>}&\stackrel{P^0 \to E^{>}_{\mathbf{P},\mathbf{Q}}}{\sim} i \int \frac{\mathrm{\hat{d}}^3 Q}{(2 \epsilon^{>}_{\mathbf{P},\mathbf{Q}})}  \frac{\Psi_{\mathbf{P}, \mathbf{Q}}^{>}(q^{\prime}) \Psi_{\mathbf{P}, \mathbf{Q}}^{* >}(q)}{(2 E^{>}_{\mathbf{P},\mathbf{Q}}) (P^0 - E^{>}_{\mathbf{P}, \mathbf{Q}}+i \epsilon)} \,.
\end{align}
The analogue of \eqref{eq:Gscattering_exactdecomposition} for the bound contribution is
\begin{align}
\label{eq:Gbound_exactdecomposition}
G^{<}_{\{n\}}(P / 2&+q, P / 2-q; P / 2 + q^{\prime}, P / 2 - q^{\prime}) \\
&=  i  \int\mathrm{~d}^4 x \mathrm{~d}^4 y \, \e^{i (q^{\prime} y - q x)}  \Psi_{\{n\}}^{<}(y) \Psi_{\{n\}}^{* <}(x) \frac{\e^{-i (P^0 - E^{<}_{\{n\}}) \tilde{F}}}{P^0 - E^{<}_{\{n\}}+i \epsilon} \,, \nonumber
\end{align}
which on the pole $P^0 \to E^{<}_{\{n\}}$ becomes
\begin{align}
G^{<}_{\{n\}}& \sim i \frac{\Psi_{n}^{<}(q^{\prime}) \Psi_{\{n\}}^{* <}(q)}{P^0 - E^{<}_{\mathbf{P}, \{n\}}+i \epsilon} \,.
\end{align}

\subsection{$5$-pt Green's function for the two-body dynamics}

We start by considering the 5-pt Green's function \eqref{eq:green-intro} and we notice that, as in the one-body case, we can recast the Schwinger-Dyson recursion \eqref{eq:recursion_DS1} in the form
\begin{align}
\label{eq:radiative-capture-5pt}
G^{\mu_1 \nu_1}  &= (\text{id} - G_0 \mathcal{K})^{-1} \left[G_0^{\mu_1 \nu_1}+G_0^{\mu_1 \nu_1} \mathcal{K} G+G_0 \hat{\mathcal{K}}_R^{\mu_1 \nu_1} G\right]  \\
&= G \left(G_0^{-1} G_0^{\mu_1 \nu_1} G_0^{-1}+\hat{\mathcal{K}}_R^{\mu_1 \nu_1}\right) G \nonumber
\end{align}
where we have used the Bethe-Salpeter equation \eqref{eq:BS-eq}. Physically, this means that the 5-pt Green's function can be thought as a sum of a one-body $\Gamma_0^{\mu_1 \nu_1} := G_0^{-1} G_0^{\mu_1 \nu_1} G_0^{-1}$ and a two-body $\hat{\mathcal{K}}_R^{\mu_1 \nu_1}$ current 
\begin{multline}
G^{\mu_1 \nu_1}\!\left(p_1, p_2; p_1^{\prime}, p_2^{\prime}, k_1^{\prime}\right) =\int \hat{\mathrm{d}}^4 t_1 \, \hat{\mathrm{d}}^4 u_1 \, G\left(p_1, p_2; t_1, t_2 \right) \Big[\Gamma_0^{\mu_1 \nu_1}\left(t_1, t_2; u_1, u_2, k_1^{\prime}\right)  \\
\left.+ \hat{\mathcal{K}}_R^{\mu_1 \nu_1}\left(t_1, t_2; u_1, u_2, k_1^{\prime}\right) \right] G\left(u_1, u_2; p_1^{\prime}, p_2^{\prime}\right) \,, 
\end{multline}
where the $5$-pt Green's function in momentum space is defined as (see the coordinates \eqref{eq:coord_variables})
\begin{align}
& G^{\mu_1 \nu_1}\left(P / 2+q, P / 2-q; P^{\prime} / 2 + q^{\prime}, P^{\prime} / 2 - q^{\prime},  k_1^{\prime}\right) \, \nonumber \\
& \qquad \qquad \,=\int \mathrm{d}^4 X \,\mathrm{d}^4 Y \,\mathrm{d}^4 x \,\mathrm{d}^4 y \,\mathrm{d}^4 z_1  \,\e^{-i (P X + q x -P^{\prime} Y^{\prime} - q^{\prime} y^{\prime} - k_1^{\prime} z_1)} \nonumber \\
& \qquad \qquad \qquad \qquad \qquad \times G^{\mu_1 \nu_1}\left(X+x/2, X-x/2; Y+y/2, Y-y/2, z_1\right) \,.  
\label{eq:Fouriertransform-5pt-v2}
\end{align}
We would like now to isolate the the energy poles corresponding to the incoming two-body scattering state, the outgoing two-body bound state and the outgoing graviton 
\begin{align}
P^0 \to E^{>}_{\mathbf{P},\mathbf{Q}} \,, \qquad (P')^0 \to E^{<}_{\{n\}} \,, \qquad (k_1^{\prime})^{0} \to E_{\mathbf{k}_1^{\prime}} = |\mathbf{k_1}^{\prime}|\,,
\end{align}
where we recall that $P = p_1 + p_2$ and $P' = p_1^{\prime} + p_2^{\prime}$. On one hand, we can perform the LSZ reduction on the position space Green's function on the LHS of \eqref{eq:Fouriertransform-5pt-v2}
\begin{align}
& \int \mathrm{d}^4 X \e^{-i P X} \int \mathrm{d}^4 z_1 \e^{i k_1^{\prime} z_1} \int \mathrm{d}^4 Y e^{i P^{\prime} Y} G^{\mu_1 \nu_1}\left(X+x/2, X-x/2; Y+y/2, Y-y/2, z_1\right) \nonumber \\
& \qquad \qquad \qquad \sim\left[\frac{i\left\langle\Omega|h^{\mu_1 \nu_1}(0)| h_{\mathbf{k_1}^{\prime}}\right\rangle}{2 E_{\mathbf{k_1}^{\prime}} ((k_1^{\prime})^0-E_{\mathbf{k_1}^{\prime}}+i \epsilon)}\right]\left[\frac{i\left\langle\Omega\left|T \phi_1\left(y/2\right) \phi_2\left(-y/2\right)\right| \mathcal{B}_{\{n\}}\right\rangle}{(P')^0-E^{<}_{\{n\}}+i \epsilon}\right] \nonumber \\
& \qquad \qquad \qquad \times \int \frac{\mathrm{\hat{d}}^3 Q}{2 \epsilon^{>}_{\mathbf{P},\mathbf{Q}}} \,\frac{i\bra{\mathcal{U}_{\mathbf{P}, \mathbf{Q}}} T \phi_1^{\dagger}\left(x/2\right) \phi_2^{\dagger}\left(-x/2\right)\ket{\Omega}}{2 E^{>}_{\mathbf{P}, \mathbf{Q}} (P^0-E^{>}_{\mathbf{P}, \mathbf{Q}}+i \epsilon)} \left\langle\mathcal{B}_{\{n\}} ; h_{\mathbf{k_1}^{\prime}} |S| \mathcal{U}_{\mathbf{P}, \mathbf{Q}}\right\rangle \,,
\label{eq:LSZ-5-ptGreen}
\end{align}
where we have used the Poincar\'e invariance of the vacuum state\footnote{Using the translation operator, we have identities of the form $\left\langle\Omega\left|T \phi_1\left(Y + y/2\right) \phi_2\left(Y -y/2\right)\right| \mathcal{B}_{\{n\}}\right\rangle = \left\langle\Omega\left|T \phi_1\left(y/2\right) \phi_2\left(-y/2\right)\right| \mathcal{B}_{\{n\}}\right\rangle$. } and in the last line we have isolated the matrix element of interest for the bound state formation
\begin{align}
\left\langle\mathcal{B}_{\{n\}} ; h_{\mathbf{k_1}^{\prime}} |S| \mathcal{U}_{\mathbf{P}, \mathbf{Q}}\right\rangle \,.
\end{align}
Using the two-body scattering and bound state wavefunctions \eqref{eq:two-bodywavefunction} we can express the full momentum space Green's function \eqref{eq:LSZ-5-ptGreen} around the pole as
\begin{align}
\label{eq:LSZ-5-ptGreenLHS}
& G^{\mu_1 \nu_1}\left(P / 2+q, P / 2-q; P^{\prime} / 2 + q^{\prime}, P^{\prime} / 2 - q^{\prime},  k_1^{\prime}\right) \\
& \qquad \qquad \qquad \qquad   \sim\left[\frac{i \varepsilon^{\mu_1 \nu_1}(k_1^{\prime})}{2 E_{\mathbf{k_1}^{\prime}} (k_1^0-E_{\mathbf{k_1}^{\prime}}+i \epsilon)}\right] \left[\frac{i \Psi^{<}_{\{n\}}(q^{\prime}) }{(P')^0-E^{<}_{\{n\}}+i \epsilon}\right] \nonumber \\
& \qquad \qquad \qquad \qquad \times \int \frac{\mathrm{\hat{d}}^3 Q}{2 \epsilon^{>}_{\mathbf{P},\mathbf{Q}}} \,\frac{i \Psi_{\mathbf{P}, \mathbf{Q}}^{* >}(q)}{2 E^{>}_{\mathbf{P}, \mathbf{Q}} (P^0-E^{>}_{\mathbf{P}, \mathbf{Q}}+i \epsilon)} \left\langle\mathcal{B}_{\{n\}} ; h_{\mathbf{k_1}^{\prime}} |S| \mathcal{U}_{\mathbf{P}, \mathbf{Q}}\right\rangle \,.  \nonumber
\end{align}
On the other hand, we can consider the RHS of \eqref{eq:radiative-capture-5pt} and perform the decomposition of the 4-pt Green's function as in \eqref{eq:Gscattering_exactdecomposition} and \eqref{eq:Gbound_exactdecomposition}, obtaining the transition element 
\begin{align}
\label{eq:5-ptGreenRHS}
& G^{\mu_1 \nu_1}\left(P / 2+q, P / 2-q; P^{\prime} / 2 + q^{\prime}, P^{\prime} / 2 - q^{\prime},  k_1^{\prime}\right) \\
& \quad \sim \int \hat{\mathrm{d}}^4 r_1 \, \hat{\mathrm{d}}^4 s_1 \, \left[\frac{i \varepsilon_{\sigma}^{\mu_1 \nu_1}(k_1^{\prime}) \varepsilon^{*}_{\sigma \alpha_1 \beta_1}(k_1^{\prime}) }{2 E_{\mathbf{k_1}^{\prime}} (k_1^0-E_{\mathbf{k_1}^{\prime}}+i \epsilon)}\right] \left[\frac{i \Psi^{<}_{\{n\}}(q^{\prime}) \Psi^{* <}_{\{n\}}(r_1) }{(P')^0-E^{<}_{\{n\}}+i \epsilon}\right] \nonumber \\
& \quad \times \int \frac{\mathrm{\hat{d}}^3 Q}{2 \epsilon^{>}_{\mathbf{P},\mathbf{Q}}} \,\frac{i \Psi^{>}_{\mathbf{P}, \mathbf{Q}}(s_1) \Psi^{*>}_{\mathbf{P}, \mathbf{Q}}(q)}{2 E^{>}_{\mathbf{P}, \mathbf{Q}} (P^0-E^{>}_{\mathbf{P}, \mathbf{Q}}+i \epsilon)} M^{\alpha_1 \beta_1}_5\left(P/2 + s_1, P/2 - s_1; P^{\prime}/2 + r_1, P^{\prime}/2 - r_1,  k_1^{\prime}\right) \,, \nonumber 
\end{align}
where 
\begin{align}
M^{\alpha_1 \beta_1}_5 (\cdot) := \hat{\delta}^4(P^{\prime} +k_1^{\prime} - P) \mathcal{M}^{\alpha_1 \beta_1}_5 (\cdot) \,.
\end{align}
Comparing \eqref{eq:LSZ-5-ptGreenLHS} and \eqref{eq:5-ptGreenRHS} and identifying the residues, we finally get (see also \cite{Petraki:2015hla})
\begin{align}
\left\langle\mathcal{B}_{\{n\}} ; h^{\sigma}_{\mathbf{k_1}^{\prime}} |S| \mathcal{U}_{\mathbf{P}, \mathbf{Q}}\right\rangle &= \varepsilon^{*\sigma }_{\alpha_1 \beta_1}(k_1^{\prime}) \int \hat{\mathrm{d}}^4 r_1 \, \hat{\mathrm{d}}^4 s_1 \Psi_{\{n\}}^{* <}(r_1) \Psi^{>}_{\mathbf{P}, \mathbf{Q}}(s_1) \hat{\delta}^4(P^{\prime} +k_1^{\prime} - P) \nonumber \\
&  \times \mathcal{M}^{\alpha_1 \beta_1}_5 \left(P/2 + s_1, P/2 - s_1; P^{\prime}/2 + r_1, P^{\prime}/2 - r_1,  k_1^{\prime}\right) \,.
\end{align}
This is a very interesting result, which in principle allows to determine our transition elements for the bound state formation in a general quantum field theory setup. Please note that the matrix element $\mathcal{M}^{\alpha_1 \beta_1}_5 \left(P/2 + s_1, P/2 - s_1; P^{\prime}/2 + r_1, P^{\prime}/2 - r_1,  k_1^{\prime}\right)$ is defined for generic momenta (i.e., not on-shell), and this is essential as it appears in the convolution with the external wavefunctions. The scattering-scattering and bound-bound transition elements are obtained similarly.

\section{Orbital elements in the quasi-Keplerian parametrization}
\label{app:quasiKepler}

We review here some definition of the orbital elements in the quasi-Keplerian parametrization \cite{Damour:1985,Damour:1988mr,Blanchet:2013haa} used in the recent work \cite{Bini:2020nsb,Bini:2020hmy}, restoring also the powers of $c$. We work with dimensionless variables here, so we normalize the physical radius $r^{\text{phys}}$ and time $t^{\text{phys}}$ as $r = (c^2 r^{\text{phys}})/(G_N (m_A + m_B))$ and $t = (c^3 t^{\text{phys}})/(G_N (m_A + m_B))$. Starting with the hyperbolic case, we can parametrize the relative motion as
\begin{align}
r^{>} &= a^{>} (e^{>}_r \cosh(\mathsf{v}) - 1) \,, \nonumber \\
\varphi^{>} &= k^{>} (\Theta^{>} + f^{>}_{\phi} \sin (2 \Theta^{>}) + g^{>}_{\phi} \sin (3 \Theta^{>}))  + \mathcal{O}\left(\frac{1}{c^4}\right)\,, \nonumber \\
l^{>} &= n^{>} t =  e^{>}_t \sinh(\mathsf{v}) - \mathsf{v} + f^{>}_{t} \Theta^{>} + g^{>}_{t} \sin (\Theta^{>}) + \mathcal{O}\left(\frac{1}{c^4}\right) \,, \nonumber \\
\Theta^{>} &= 2 \arctan \left(\sqrt{\frac{e^{>}_{\phi} + 1}{e^{>}_{\phi} - 1}} \tanh\left(\frac{\mathsf{v}}{2}\right)\right) + \mathcal{O}\left(\frac{1}{c^4}\right) \,,
\label{eq:orbital_elements_hyp}
\end{align}
where the gauge-invariant quantities $k^{>}$ and $n^{>}$ are the same in all coordinates, while the eccentricities $e^{>}_{t}$, $e^{>}_{r}$ and $e^{>}_{\phi}$ depend on the coordinate system. Working in the harmonic gauge, we have up to 2PN accuracy the following relations
\allowdisplaybreaks
\begin{align}
n^{>} &=(2 \mathcal{E})^{3/2} \left(1  +\frac{1}{4 c^2} \mathcal{E} (15-\nu ) + \frac{1}{32 c^4} \mathcal{E}^2 \left(11 \nu ^2+30 \nu +555\right)\right) \,, \nonumber \\
a^{>} &= \frac{1}{2 \mathcal{E}} \left[1 + \frac{1}{2 c^2} \mathcal{E} (7-\nu )+\frac{1}{4 c^4} \mathcal{E}^2 \left(1+\nu ^2-\frac{8 (7 \nu -4)}{\mathcal{E} j^2}\right) \right] \,, \nonumber \\
k^{>}&=1 + \frac{3}{c^2 j^2}+\frac{3}{4 c^4 j^4} \left(5 (7-2 \nu )-(2 \mathcal{E}) j^2 (2 \nu -5)\right)\,, \nonumber \\
(e^{>}_r)^2&=1 +2 \mathcal{E} j^2 +\frac{1}{c^2} \mathcal{E} \left(2 (\nu -6)-5 \mathcal{E} j^2 (3-\nu )\right) \nonumber \\
&+ \frac{1}{c^4} \mathcal{E}^2 \left(\mathcal{E} j^2 \left(4 \nu ^2-45 \nu +80\right)+\frac{8 (7 \nu -4)}{\mathcal{E} j^2}+\left(\nu ^2+74 \nu +30\right)\right)\,, \nonumber \\
(e^{>}_t)^2&=1 + 2 \mathcal{E} j^2 + \frac{1}{c^2} \mathcal{E} \left(4 (1-\nu )-\mathcal{E} j^2 (7 \nu -17)\right) \nonumber \\
&+ \frac{1}{c^4} \mathcal{E}^2 \left(\mathcal{E} j^2 \left(16 \nu ^2-47 \nu +112\right)+\frac{4 (7 \nu -4)}{\mathcal{E} j^2}+2 \left(5 \nu ^2+18 \nu +3\right)\right) \,, \nonumber \\
(e^{>}_{\phi})^2&=1 + 2 \mathcal{E} j^2 + \frac{1}{c^2} \mathcal{E} \left(\mathcal{E} j^2 (-(15-\nu ))-12\right) \nonumber \\
&+ \frac{1}{4 c^4} \mathcal{E}^2 \left(2 \mathcal{E} j^2 \left(3 \nu ^2-31 \nu +160\right)+\frac{15 \nu ^2+91 \nu -416}{2 \mathcal{E} j^2}+2 \left(9 \nu ^2+17 \nu -20\right)\right)\,, \nonumber \\
f_t^{>}&= \frac{1}{c^4} \frac{3 (2 \mathcal{E})^{3/2} (5-2 \nu )}{2 j} \,,  \nonumber \\
g_t^{>}&=-\frac{1}{c^4} \frac{\nu  (\nu -15) (2 \mathcal{E})^{3/2} \sqrt{1+2 \mathcal{E} j^2}}{8 j} \,, \nonumber \\
f_{\phi}^{>}&=\frac{1}{c^4} \frac{\left(-3 \nu ^2+19 \nu +1\right) \left(1+2 \mathcal{E} j^2\right)}{8 j^4}\,,  \nonumber \\
g_{\phi}^{>}&=\frac{1}{c^4} \frac{\nu  (1-3 \nu ) \left(1+2 \mathcal{E} j^2\right)^{3/2}}{32 j^4}\,.
\end{align}
For the elliptic case, instead, we can parametrize the relative motion as
\begin{align}
r^{<} &= a^{<} (1 - e^{<}_r \cos(\mathsf{u})) \,, \nonumber \\
\varphi^{<} &= k^{<} (\Theta^{<} + f^{<}_{\phi} \sin (2 \Theta^{<}) + g^{<}_{\phi} \sin (3 \Theta^{<})  + \mathcal{O}\left(\frac{1}{c^4}\right)\,, \nonumber \\
l^{<} &= n^{<} t =  \mathsf{u} - e^{<}_t \sin(\mathsf{u}) + f^{<}_t \sin(\Theta^{<}) + g^{<}_t (\Theta^{<} - \mathsf{u}) + \mathcal{O}\left(\frac{1}{c^4}\right) \,, \nonumber \\
\Theta^{<} &= 2 \arctan \left(\sqrt{\frac{e^{<}_{\phi} + 1}{1 - e^{<}_{\phi}}} \tan\left(\frac{\mathsf{u}}{2}\right)\right) + \mathcal{O}\left(\frac{1}{c^4}\right) \,,
\label{eq:orbital_elements_ell}
\end{align}
where up to 2PN accuracy in harmonic gauge we have
\allowdisplaybreaks
\begin{align}
n^{<} &=(-2 \mathcal{E})^{3/2} \left[1 +\frac{(-2 \mathcal{E})}{8 c^2} (\nu -15) + \frac{(-2 \mathcal{E})^2 }{128 c^4} \left(\frac{192 (2 \nu -5)}{\sqrt{-2 \mathcal{E}} j}+11 \nu ^2+30 \nu +555\right)\right] \,, \nonumber \\
a^{<} &= \frac{1}{(-2 \mathcal{E})} \left[1 +\frac{(-2 \mathcal{E})}{4 c^2} (\nu -7)+\frac{(-2 \mathcal{E})^2 }{16 c^4} \left(\frac{16 (7 \nu -4)}{(-2 \mathcal{E}) j^2}+\nu ^2+1\right) \right] \,, \nonumber \\
k^{<}&=1 -\frac{3}{c^2 j^2}+\frac{1}{4 c^4}\left[\frac{15 (7-2 \nu )}{j^4}+\frac{3 (-2 \mathcal{E}) (2 \nu -5)}{j^2}\right]\,, \nonumber \\
(e^{>}_r)^2&=1 +2 \mathcal{E} j^2 -\frac{(-2 \mathcal{E})}{c^2} \left[\frac{5}{4} (3-\nu )(-2 \mathcal{E}) j^2+\nu -6\right] \nonumber \\
&+ \frac{(-2 \mathcal{E})^2}{8 c^4}  \left[- \left(4 \nu ^2-45 \nu +80\right)(-2 \mathcal{E}) j^2+\frac{32 (7 \nu -4)}{(-2 \mathcal{E}) j^2}+2 \left(\nu ^2+74 \nu +30\right)\right]\,, \nonumber \\
(e^{>}_t)^2&=1 + 2 \mathcal{E} j^2 -\frac{(-2 \mathcal{E})}{4 c^2} \left[(7 \nu -17)(-2 \mathcal{E}) j^2+8 (1-\nu )\right] \nonumber \\
&+ \frac{(-2 \mathcal{E})^2}{8 c^4} \Big[- (16 \nu^2 -47 \nu +112)(-2 \mathcal{E}) j^2-\frac{16 (7 \nu -4)}{(-2\mathcal{E}) j^2} \nonumber \\
& \qquad \qquad \qquad \qquad \qquad +4 (5 \nu^2 +18 \nu+3)+\frac{24 (2 \nu -5) (1+2 \mathcal{E} j^2)}{\sqrt{-2 \mathcal{E}} j}\Big] \,, \nonumber \\
(e^{>}_{\phi})^2&=1 + 2 \mathcal{E} j^2 - \frac{(-2 \mathcal{E}) }{c^2} \left(\frac{1}{4} (-2 \mathcal{E}) j^2 (15-\nu )-6\right) \nonumber \\
&- \frac{(-2 \mathcal{E})^2}{16 c^4} \left[\left(3 \nu ^2-31 \nu +160\right)(-2 \mathcal{E}) j^2 +\frac{15 \nu ^2+91 \nu -416}{(-2 \mathcal{E}) j^2}-2 \left(9 \nu ^2+17 \nu -20\right)\right]\,, \nonumber \\
f_t^{<}&= -\frac{1}{c^4} \frac{\nu  (\nu -15) (-2 \mathcal{E})^{3/2} \sqrt{1+2 \mathcal{E} j^2}}{8 j} \,,  \nonumber \\
g_t^{<}&=-\frac{1}{c^4} \frac{3 (-2 \mathcal{E})^{3/2}  (2 \nu -5)}{2 j} \,, \nonumber \\
f_{\phi}^{<}&=\frac{1}{c^4} \frac{\left(-3 \nu ^2+19 \nu +1\right) \left(1+2 \mathcal{E} j^2\right)}{8 j^4}\,,  \nonumber \\
g_{\phi}^{<}&=-\frac{1}{c^4} \frac{\nu  (3 \nu -1) \left(1+2 \mathcal{E} j^2\right)^{3/2}}{32 j^4}\,.
\end{align}
It is easy to verify that up to 1PN accuracy there is a straightforward analytic continuation between the orbital elements for the hyperbolic and the elliptic case determined purely in terms of the binding energy \cite{Damour:1985}
\begin{align}
     n^{>} &\to -i n^{<}\,, e_t^{>} \to e_t^{<}\,, e^{>}_r \to e_r^{<}\,, e_{\phi}^{>} \to e_{\phi}^{<}\,, \mathsf{v} \to i \mathsf{u} \,, a^{>} \to -a^{<}\,, k^{>} \to k^{<} \,,
     \label{eq:B2B-orbital}
    \end{align}
while $n^{>} \neq -i n^{<}$ and $e_t^{>} \neq e_t^{<}$ at 2PN (and beyond). Interestingly, it might be possible to modify the quasi-Keplerian parametrization to get an analytic continuation up to 3PN \cite{Cho:2018upo}, but this is beyond the scope of this work.

\bibliographystyle{JHEP} 
\bibliography{references}
\end{document}